\documentclass[aps,pra,twocolumn,superscriptaddress,letterpaper]{revtex4-1}
\usepackage{graphicx}
\usepackage{amsmath}
\usepackage{amssymb}
\usepackage{esint}
\usepackage{verbatim}
\usepackage{xcolor}
\usepackage{soul}
\usepackage{siunitx}
\usepackage{hyperref}
\newcommand{\be}{\begin{equation}}
\newcommand{\ee}{\end{equation}}
\newcommand{\ba}{\begin{array}}
\newcommand{\ea}{\end{array}}
\newcommand{\bqa}{\begin{eqnarray}}
\newcommand{\eqa}{\end{eqnarray}}

\renewcommand{\Re}{\text{Re}}

\newcommand{\ncav}{n_{\text{c}}}

\newcommand{\gzero}{g_{\text{0}}}

\newcommand{\gammaOM}{\gamma_{\text{OM}}}
\newcommand{\Qo}{Q_{\text{opt}}}

\newcommand{\kappaE}{\kappa_{\text{e}}}

\newcommand{\QoE}{Q_{\text{opt,e}}}

\newcommand{\gammai}{\gamma_{\text{i}}}

\newcommand{\aopt}{a_\text{o}}
\newcommand{\amech}{a_\text{m}}

\begin{document}
% \begin{bibunit} % references for the Main Text?

\title{Topological phonon transport in an optomechanical system}
%\title{Topological phonon transport in micro-optomechanical system}
%\title{Topological phonon transport in mesoscopic optomechanical system}

\author{Hengjiang Ren}
\thanks{These authors contributed equally to this work.}
\affiliation{Thomas J. Watson, Sr., Laboratory of Applied Physics, California Institute of Technology, Pasadena, California 91125, USA}
\affiliation{Kavli Nanoscience Institute, California Institute of Technology, Pasadena, California 91125, USA}
\affiliation{Institute for Quantum Information and Matter, California Institute of Technology, Pasadena, California 91125, USA}
\author{Tirth Shah}
\thanks{These authors contributed equally to this work.}
\affiliation{Max Planck Institute for the Science of Light, Staudtstrasse 2, 91058 Erlangen, Germany}
\affiliation{Department of Physics, Friedrich-Alexander Universit\"at Erlangen-N\"urnberg, Staudtstrasse 7, 91058 Erlangen, Germany}
\author{Hannes Pfeifer}
\altaffiliation[Current Address: ]{Institut f\"{u}r Angewandte Physik, Universit\"{a}t Bonn, Wegelerstraße 8, 53115 Bonn, Germany}
\affiliation{Max Planck Institute for the Science of Light, Staudtstrasse 2, 91058 Erlangen, Germany}
\author{Christian Brendel}
\affiliation{Max Planck Institute for the Science of Light, Staudtstrasse 2, 91058 Erlangen, Germany}
\author{Vittorio Peano}
\affiliation{Max Planck Institute for the Science of Light, Staudtstrasse 2, 91058 Erlangen, Germany}
\author{Florian Marquardt}
\affiliation{Max Planck Institute for the Science of Light, Staudtstrasse 2, 91058 Erlangen, Germany}
\affiliation{Department of Physics, Friedrich-Alexander Universit\"at Erlangen-N\"urnberg, Staudtstrasse 7, 91058 Erlangen, Germany}
\author{Oskar Painter}
\affiliation{Thomas J. Watson, Sr., Laboratory of Applied Physics, California Institute of Technology, Pasadena, California 91125, USA}
\affiliation{Kavli Nanoscience Institute, California Institute of Technology, Pasadena, California 91125, USA}
\affiliation{Institute for Quantum Information and Matter, California Institute of Technology, Pasadena, California 91125, USA}
\affiliation{AWS Center for Quantum Computing, Pasadena, California 91125, USA.}
\email{opainter@caltech.edu}

\date{\today}

%\begin{abstract} 
%\end{abstract}

\maketitle

{\bf
Recent advances in cavity-optomechanics~\cite{Aspelmeyer2014} have now made it possible to use light not just as a passive measuring device of mechanical motion~\cite{deGroot2019}, but also to manipulate the motion of mechanical objects down to the level of individual quanta of vibrations (phonons).  At the same time, microfabrication techniques have enabled small-scale optomechanical circuits capable of on-chip manipulation of mechanical and optical signals~\cite{massel2012multimode,mian2015synchronization,xu2016topological,kharel2019high_frequency,ruesink2016nonreciprocity,peterson2017demonstration,bernier2017nonreciprocal,fang2017generalized,xu2019nonreciprocal,mathew_synthetic_2020}. Building on these developments, theoretical proposals have shown that larger scale optomechanical arrays can be used to modify the propagation of phonons, realizing a form of topologically protected phonon transport~\cite{peano2015topological,brendel2017pseudomagnetic,brendel2018snowflake,mathew_synthetic_2020,sanavio2020}. Here, we report the observation of topological phonon transport within a multiscale optomechanical crystal structure consisting of an array of over $800$ cavity-optomechanical elements. Using sensitive, spatially resolved optical read-out~\cite{teufel2009nanomechanical,wilson2015measurement} we detect thermal phonons in a $0.325-0.34$~GHz band traveling along a topological edge channel, with substantial reduction in backscattering.  This represents an important step from the pioneering macroscopic mechanical systems work~\cite{Susstrunk2015,nash2015topological,lu2017observation,Miniaci_PRX_2018,yu_elastic_2018} towards topological phononic systems at the nanoscale, where hypersonic frequency ($\gtrsim$~GHz) acoustic wave circuits consisting of robust delay lines~\cite{hafezi2011robust} and non-reciprocal elements~\cite{cha2018experimental,ma2020nanomechanical,nassar2020nonreciprocity} may be implemented.  Owing to the broadband character of the topological channels, the control of the flow of heat-carrying phonons, albeit at cryogenic temperatures, may also be envisioned.
}

% revised main-text-intro:
% emphasis on topology/general
% and then on multiscale OMC
% (officially put here 12.8./F.M.)
Topology deals with features invariant to smooth deformations. The band structure for waves in a periodic medium may display such topological features, and this can have immediate consequences for transport along boundaries, e.g. producing protected edge states~\cite{hasan2010colloquium}. In recent years, these conceptual insights, first acquired for electrons, were quickly expanded to cover arbitrary waves~\cite{aidelsburger_artificial_2018}. This includes, in particular, mechanical vibrations~\cite{peano2015topological,Susstrunk2015,nash2015topological,lu2017observation,cha2018experimental,ma2020nanomechanical,mousavi_topologically_2015,Miniaci_PRX_2018,yu_elastic_2018,deng2020acoustic}, with their potential for far-reaching applications in signal processing and other domains when implemented in compact chip-scale acoustic devices. 
A very promising approach to lower the footprint for excitation and read-out, and to boost the sensitivity to high-frequency vibrations, is to use radiation pressure forces in so-called optomechanical crystals (OMCs)~\cite{Eichenfield2009,Safavi-Naeini2010b,Safavi-Naeini2014PRL,ren2020two}. OMCs are patterned structures that can be engineered to yield large radiation-pressure coupling between cavity photons and phonons.

Here, we demonstrate the optomechanical detection of topological phonon transport in a multiscale OMC fabricated into the surface of a silicon microchip. In contrast to standard single-scale devices, the multiscale OMC consists of a superlattice structure, superimposing two patterns with very different but commensurate lattice spacings. This multiscale approach adds an extra degree of flexibility, decoupling the engineering of photonic and phononic modes.  In our design, at the larger scale is a phononic crystal.  Embedded within each unit cell of the phononic crystal is a smaller scale photonic crystal, which hosts a high-$Q$ optical nanocavity for optical read-out of phonons.  Local changes within the OMC lattice of the phononic crystal unit cell are used to create topologically distinct mechanical domains, the boundary of which host phononic helical edge states based on the Valley Hall effect~\cite{Martin_2008_PRL,ju_topological_2015}. The optomechanical arrays in this work consist of over $800$ phononic unit cells, each with a corresponding optical mode for single-site resolution of phonon transport.

%%%% END OF INTRO %%%%%

\begin{figure*}
\begin{center}
\includegraphics[width=2\columnwidth]{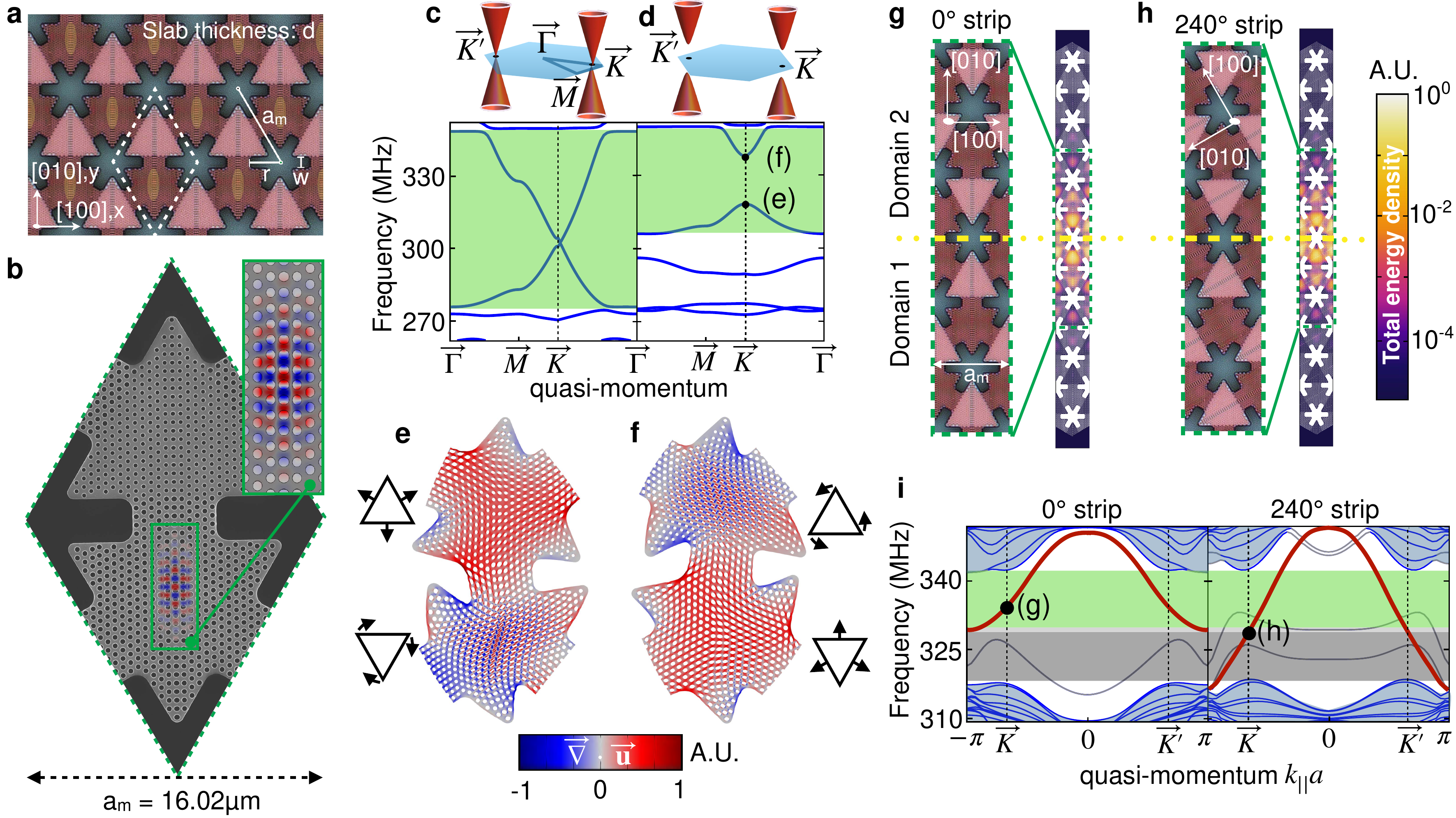}
\caption{
\textbf{Design of the multiscale optomechanical crystal for topological phononics.} 
\textbf{a}, Optical microscope image showing the snowflake triangular lattice (unit cell dashed) with parameters $(d, r, w, \amech)$ = $(0.22, 5.77, 2.34, 16.02)$~\si{\micro\metre}.
%Throughout the paper, the axes are aligned with the lattice vectors of the silicon crystal, see SI.
The  axes are aligned with the silicon crystal.
\textbf{b}, Focused Ion Beam (FIB) image of unit cell geometry with the simulated photonic-crystal cavity mode profile ($E_\text{[100]}$ component of the electric field; red/blue indicates sign).
%(simulated optical $\Qo = 1.2 \times 10^6$). Inset: FEM simulated mode profile of the fundamental optical resonance at $\omegac/2\pi = 194$~THz 
\textbf{c} and \textbf{d}, Simulated phononic band structures with $M_y$ mirror symmetry intact and broken (design in \textbf{b}), respectively. Inset: Sketches of the Dirac cones.
\textbf{e,f}, Snapshots of the mechanical mode deformation (colours indicate the local volume change, $\vec{\nabla} \cdot \vec{u}$; red corresponding to expansion and blue compression). The arrows in the associated pictograms indicate the dynamics of the motion. 
%The mode \textbf{f} in the upper band has a higher unit-cell vacuum optomechanical coupling rate ($g_0 = 2\pi\times33.6$kHz) than the lower-band mode ($g_0 = 2\pi\times8.4$ kHz), since it produces breathing motion (indicated by pictogram) in the lower triangle that hosts the cavity.
\textbf{g} and \textbf{h}, Optical microscope images and simulated mechanical mode profiles for two strip configurations, each comprising two topologically distinct domains (domain 1 as in \textbf{a,b}). The domain wall (dashed) has slope $0^{\circ}$ (horizontal, \textbf{g}) or $240^{\circ}$ (slanted, \textbf{h}) relative to the [100] axis.
%$x$ axis.
\textbf{i}, 1-D band structures calculated for the horizontal and slanted configurations. The red lines indicate the topological edge state dispersion, the grey lines are the additional edge state modes localized at the top and bottom boundaries of the geometry (away from the domain wall), the blue parts are bulk modes. The color shading inside the bulk band gap identifies different transport regimes in systems where the two types of domain walls are connected.
%When the two type of strips are combined into a closed loop the resulting edge states will be qualitatively different in the three frequency ranges highlighted in i dark grey (edge states localized within the slanted edges), (ii) green (edge states extended over the whole domain wall no significant backscattering), and (iii) light grey (extended edge states but significant backscattering), cf.~Fig2. (d-f). 
}
\label{fig1}
\end{center}
\end{figure*}

%%% SECTION Multiscale 191

Images of a fabricated multiscale OMC structure are shown in Figs.~\ref{fig1}a,b. In our design, a triangular lattice of snowflake-shaped holes with lattice spacing $\amech=16.02$ \si{\micro\metre} is superimposed onto another triangular lattice of cylindrical holes with a much smaller spacing $\aopt=450$~nm.  This hole pattern has been etched into the thin ($220$~nm thickness) silicon device layer of a silicon-on-insulator (SOI) microchip.  After releasing the underlying buried oxide layer, this produces an array of connected triangular silicon membranes forming the phononic crystal, each hosting a photonic crystal defined by the smaller holes (see Fig.~\ref{fig1}b and inset).  The snowflake pattern is adopted from a well-known single-scale OMC design \cite{Safavi-Naeini2014PRL,ren2020two} and has also been proposed theoretically as a platform for topological phononics \cite{brendel2017pseudomagnetic,brendel2018snowflake}. In this work we have increased the snowflake lattice spacing by a factor of $\sim 30$, enabling every triangular membrane to harbor an optical nanocavity consisting of a localized defect in the triangular photonic crystal hole pattern. The purpose of using a cavity is to boost the optomechanical interaction (see App.~\ref{sec:opt_cav_design}). In Fig.~\ref{fig1}a and b, such a cavity is present only in the downward-pointing triangular membranes, with the upward-pointing triangular membranes having an unperturbed photonic crystal pattern.  Although the two lattices (phononic and photonic) are at vastly different scales, the patterning of the photonic crystal within each triangular membrane does (weakly) influence the phononic properties, providing an extra knob to trim the mechanical properties.

%%% SECTION Valley Hall, general 132
%{\it Valley Hall physics.} 
%Recap of valley Hall Physics
We employ these tuning knobs of the multiscale design to realize a structure supporting robust helical edge states based on the Valley Hall effect~\cite{Martin_2008_PRL}. The Valley Hall effect is relevant for a wide range of systems that support Dirac cones, including electronic \cite{Martin_2008_PRL,ju_topological_2015}, photonic \cite{zeng_electrically_2020}, and mechanical systems \cite{lu2017observation, Miniaci_PRX_2018}. In this context, valley refers to the quasi-momentum region around a Dirac cone. In a time-reversal-symmetric system, the Dirac cones, and thus the corresponding valleys, come in pairs mapped onto each other by the operation of time reversal. Thus, the valley can be viewed as a binary degree of freedom akin to the spin. In the Valley Hall effect, valley-polarized edge excitations propagate in opposite directions, analogous to spin-polarized edge states in the Spin Hall effect.

%Here, we are interested in a scenario where inter-valley transitions are strongly suppressed as they involve large quasimomentum transfer and, thus, the two valley are effectively decoupled. In such a setting

%%% SECTION Bulk bandstructure in our design 202
As we are pursuing an optomechanical approach to the detection of mechanical edge excitations, we focus here on the vibrational modes that couple to light, the in-plane modes which are even under the mirror operator $M_z$ ($z\mapsto-z$). For these modes, the snowflake phononic crystal supports a pair of Dirac cones well-isolated from the remaining bands~\cite{brendel2017pseudomagnetic,brendel2018snowflake}. In our experiment, the Dirac cones have a center frequency of approximately $0.3$~GHz, with linear Dirac-like dispersion across a bandwidth of $70$~MHz (see Fig. \ref{fig1}c). These cones are protected by a symmetry under $M_y$ (see App.~\ref{sec:dirac_equation_derivation}). We open the bulk band gap that will host the helical edge states by breaking this symmetry. Decreasing the size of the photonic-crystal holes in the upward-pointing triangles by a factor of $0.78$ produces a band gap of width $18$~MHz (see Fig \ref{fig1}d). The underlying vibrational Bloch waves, calculated using finite-element method (FEM) simulations (see App.~\ref{sec:FEM_sim}), are shown in Fig.~\ref{fig1}e-f. A comparatively large unit-cell vacuum optomechanical coupling ($g_0 = 2\pi\times33.7$kHz) is produced for the higher-frequency mode in Fig.~\ref{fig1}f because it displays breathing motion around the optical cavity. A detailed discussion of the optomechanical coupling is provided in App.~\ref{sec:opt_cav_design}. 
% and Supplementary Videos 1-2

%%% SECTION Valley Chern number and domain walls 148

\begin{figure*}[htp!]
\begin{center}
\includegraphics[width=2\columnwidth]{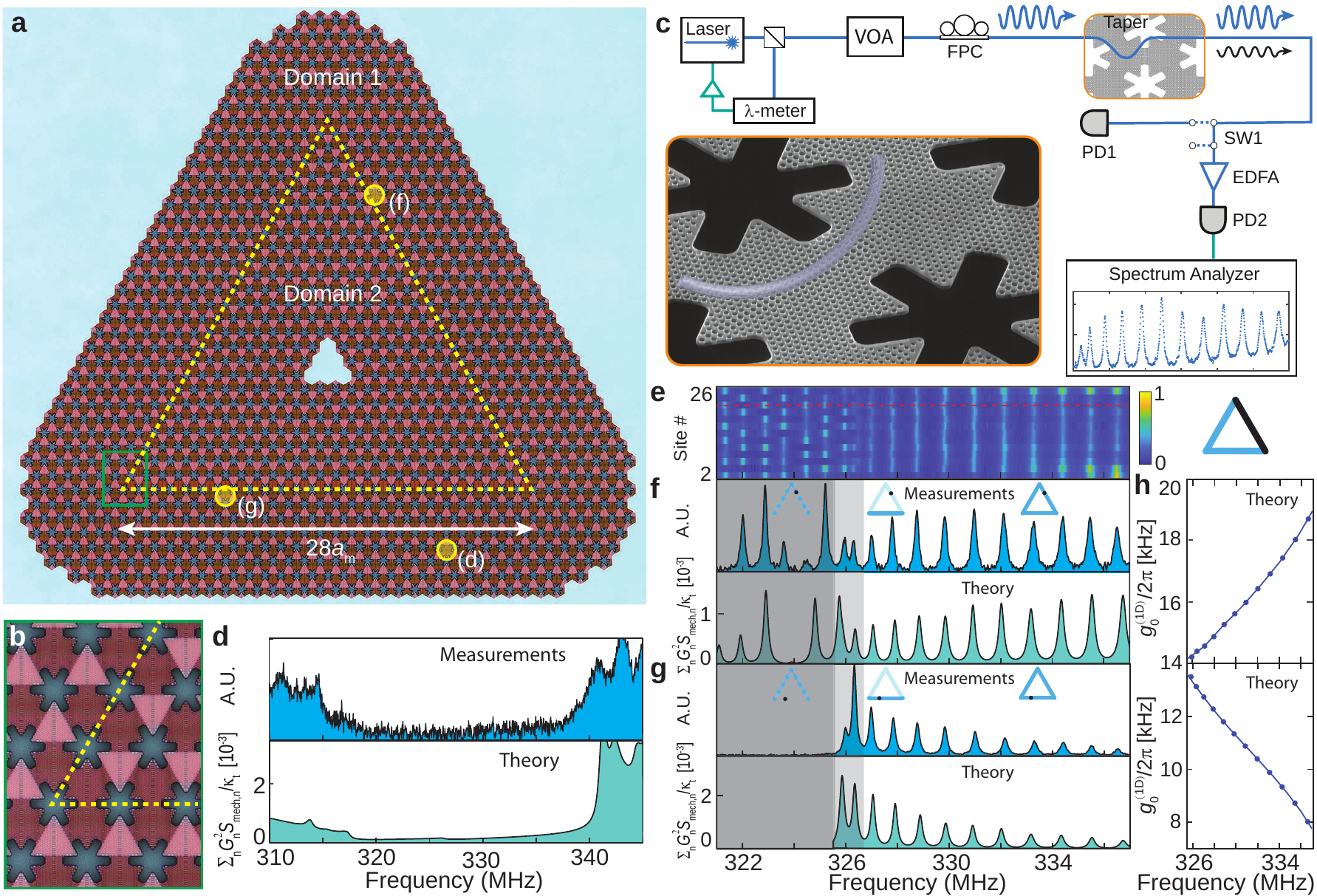}
\end{center}
\caption{
\textbf{Characterization of topological edge states using optomechanical read-out.}
\textbf{a}, Optical microscope image of triangular topological mechanical cavity (Domain wall: dashed line. Read-out cavities for the measurements in \textbf{d}, \textbf{f},  and \textbf{g}: yellow dots). %The local density of states of standing waves formed in this triangular geometry is read out optically. 
\textbf{b}, Zoom-in of the topological cavity corner (green box in \textbf{a}). \textbf{c}, Experimental setup. Mechanical side-bands are imprinted on a laser beam transmitted through an optical cavity, detecting the NPSD of the mechanical waves. Acronyms: optical wave meter ($\lambda$ meter), variable optical attenuator (VOA),  fiber polarization controller (FPC), optical switch (SW), erbium-doped fiber amplifier (EDFA), photodetector (PD). 
%A more detailed schematic and description of the measurement setup is provided in Supplementary Note~2.
\textbf{d}, \textbf{f}, and \textbf{g} Measured (top) and numerically estimated (bottom) NPSD, respectively, in the bulk of domain 1, on a slanted edge, and on a horizontal edge. 
Insets in \textbf{f}, and \textbf{g}: Sketches showing read-out positions and the expected local density of states.
\textbf{e} Measured NPSD as a function of frequency and read-out position on a slanted edge (highlighted in black in the sketch). Red dashed line corresponds to the spectrum in \textbf{f}. The low-frequency region (dark grey in \textbf{f} and \textbf{g}) harbors modes only inside the slanted edges (cf.~Fig.~\ref{fig1}i). Data calibration is required to compare measurements from different read-out cavities (see App.~\ref{sec:normlization}).
\textbf{h}, Optomechanical coupling for  edge states in slanted (top)  and  horizontal (bottom) domain walls (see App.~\ref{sec:scattering_matrix}). 
}
\label{fig2}
\end{figure*}

In the Valley Hall effect, the topological transport takes place through counter-propagating valley-polarized edge states which exist at the domain walls separating two topologically distinct domains of opposite so-called valley Chern number. By applying the mirror operation $M_y$, we construct from the deliberately mirror-symmetry-broken design described above, a second domain with opposite valley Chern numbers (see Fig.~\ref{fig1}g,h). The key feature leading to robust transport is that edge excitations can navigate a path with arbitrarily sharp angles while still remaining confined within the same valley region of quasi-momentum space. On the other hand, backscattering would require large quasi-momentum transfer to reach a different valley, and is thus strongly suppressed. Our fit to the Dirac Hamiltonian describing our anisotropic structure (see App.~\ref{sec:walley_hall}) shows both a dependence of the band structure on the domain wall orientation and some deviations from the idealized theoretical limiting case. For a horizontal domain wall, this leads to in-gap edge states that extend only through part of the full band gap (see Fig.~\ref{fig1}i). Below we show that the transmission around sharp corners remains robust nevertheless, with this imperfection only reducing the relevant bandwidth.

We have fabricated several devices where an internal domain of type $2$ is surrounded by an external domain of type $1$. The ensuing closed domain wall produces a topological mechanical cavity.  In a topological cavity, counter-propagating running waves remain decoupled in spite of sharp turns and/or disorder. This give rise to a characteristic spectrum formed by a series of doublets. These doublets are degenerate, with any slight lifting of the degeneracy due to residual inter-valley scattering.

%%% SECTION Describing msmts in Fig. 2 (bulk & spatially resolved) 547
 
The first topological cavity structure that we study is shown in Fig.~\ref{fig2}a and Fig.~\ref{fig2}b, consisting of an equilateral triangle of $28$ snowflake unit cells along each side. A schematic of our optical setup used to measure the phononic properties of the topological cavity structure is shown in Fig.~\ref{fig2}c. A tunable external cavity diode laser coupled to an optical fiber taper is used to optically excite individual optical nanocavities within the multiscale OMC array. The out-coupled laser light, which contains the local mechanical motion of the structure imprinted as intensity modulations, is detected on a photodiode and analyzed on an electronic spectrum analyzer. Owing to the thermal nature of the measured mechanical motion in this work, the measured electronic spectrum analyzer signal represents a local mechanical noise power spectral density (NPSD). By moving the taper position we are able to address any unit cell of the larger-scale phononic lattice, obtaining a site-resolved spectrum of the thermally populated phonon modes (see App.~\ref{sec:measuring} for further details). As an example, we show in the top plot of Fig.~\ref{fig2}d the resulting optically-transduced local mechanical spectrum for an optical fiber taper position at site (d) in Fig.~\ref{fig2}a, which is in the bulk region of domain $1$. The measured spectrum is seen to be in close agreement with our theoretical predictions based on FEM simulations (bottom plot of Fig.~\ref{fig2}d), both of which show a bulk band gap which covers an interval from $316$~MHz to $338$~MHz. We note that the amplitude of the thermal motion transduced in these measurements is only on order $10$~fm, highlighting the sensitivity of our optical read-out scheme.

We now focus on the domain wall region. Exploiting our single-site resolution capability, we have measured the mechanical NPSD as a function of read-out position, as shown in Fig.~\ref{fig2}e. This reveals two dramatically different transport regimes. For the mechanical cavity modes at lower frequencies ($321$~MHz - $327$~MHz), we observe a strong modulation versus site position in each of the mechanical mode peaks. These fringe-like features indicate that thermal phonon excitations are reflected and form standing waves. This is due to the absence of topological edge modes inside the horizontal domain wall at these frequencies, resulting in standing waves inside the slanted domain wall portions of the mechanical cavity path. By contrast, we observe no such fringes in the higher-frequency regime ($327$~MHz - $337$~MHz). This indicates backscattering-immune running waves, providing a direct visual signature of the formation of a topological mechanical cavity. Below, we refer to this frequency range as the topological bandwidth.  In between these regimes, there is a crossover region (light grey in Figs.~\ref{fig2}f-g), where the horizontal edge already supports edge states but backscattering is still possible because very small quasi-momentum transfers are sufficient to flip right-moving into left-moving horizontal edge states due to their proximity to the Brillouin zone boundary of the horizontal edge structure (see bandstructure plot in Fig.~\ref{fig1}i).
%I think this explanation is right, but one needs to read this very carefully.  The necessary quasi-momentum transfer for backscattering isn't much different here for the slanted portions of the cavity, but rather the horizontal edge state dispersion flattens out here and the horizontal edge states are very close to the edge of the 1D waveguide Brillouin zone boundary (pi in Fig. 1i).  in this case, yes, for the horizontal edge states the quasi-momentum for backscattering is tiny. 

We further substantiate the absence of backscattering in the topological bandwidth by comparing  the frequency dependence of the measured NPSD with theory predictions that assume perfect transmission at the corners. They are based on scattering matrix calculations that take FEM simulations as input (see App.~\ref{sec:scattering_matrix}). The theoretical spectra are in good agreement with measurement results both on the slanted and the horizontal edges, as shown in Figs.~\ref{fig2}f and \ref{fig2}g, respectively. Even the behaviour of the peak heights, distinctly different for both types of edges, is captured very well by including both the group velocity dispersion and the frequency-dependent vacuum optomechanical coupling $\gzero^{(\rm 1D)}$ (see Fig.~\ref{fig2}h) in our analysis. 
%This confirms that our measurement results are compatible with negligible backscattering in the topological bandwidth.

\begin{figure}
\begin{center}
\includegraphics[width=\columnwidth]{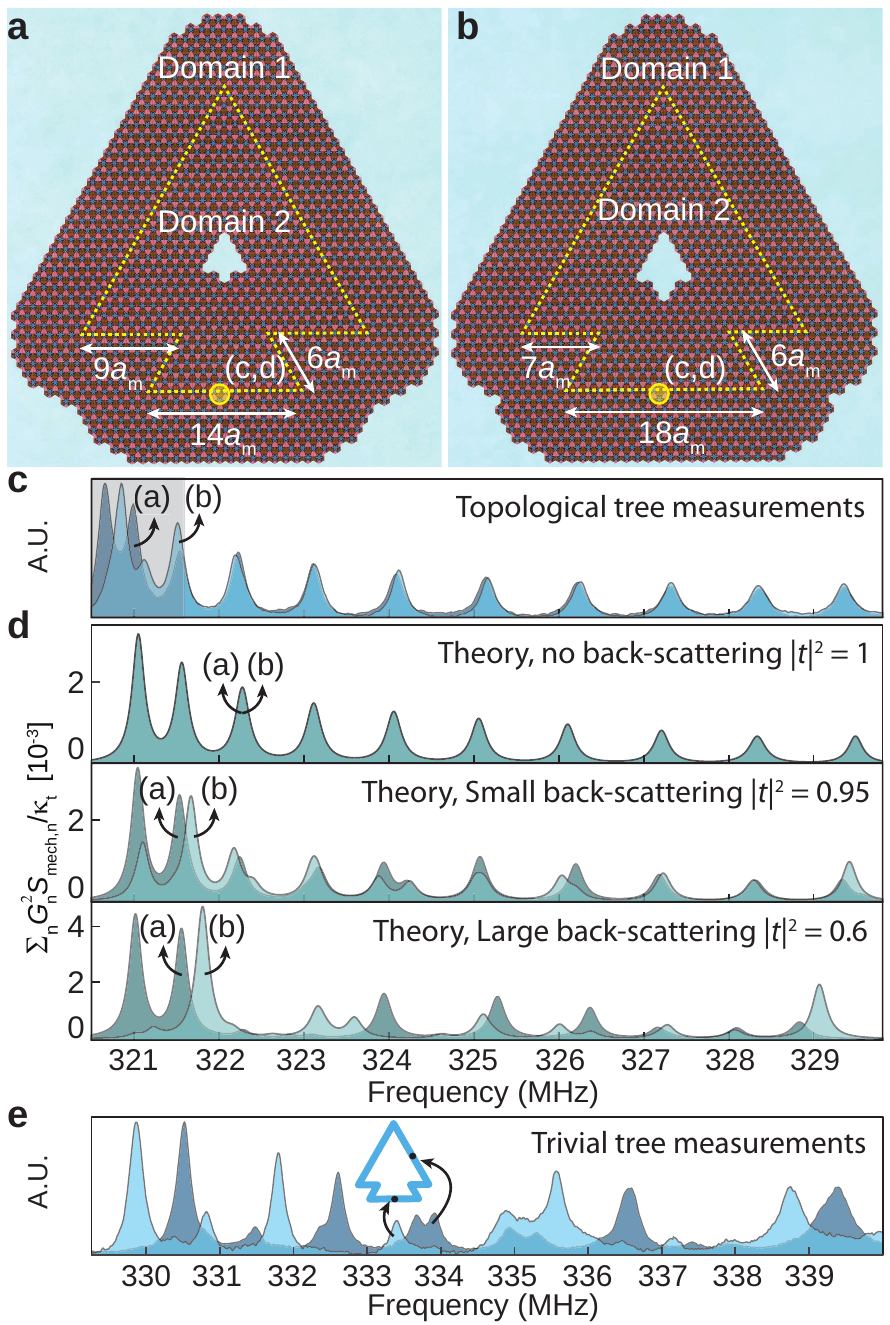}
\caption{\textbf{Robustness against backscattering.}
\textbf{a} and \textbf{b}, Optical microscope image of two different tree-shaped topological mechanical cavities. 
\textbf{c}, Comparison of measurement results for the two tree-shaped topological cavities. \textbf{d}, Theoretical prediction for three backscattering strengths. Darker (lighter) spectra  correspond to the tree-shaped topological cavities in \textbf{a} (\textbf{b}). In \textbf{c}, the cross-over region where backscattering can occur without requiring large quasi-momentum transfer is highlighted in light grey.
\textbf{e}, Measurements for a trivial waveguide mechanical cavity (see App.~\ref{sec:triv_waveguide}). Transduced NPSD measured at a slanted (dark blue) and a horizontal (light blue) edge, revealing strong backscattering.
}
\label{fig3}
\end{center}
\end{figure}

%%% SECTION Need for more complex geometries 134
While the triangle geometry is the simplest closed-loop geometry, already producing a topological mechanical cavity, we also sought to test the robustness and immunity to waveguide imperfections in more complex cavity structure where we could independently vary the length of waveguide segments in between sharp corners. The effects of such variations should be most pronounced in a geometry with appreciable backscattering at the corners, eventually producing separate standing wave patterns in the segments whose free spectral range would depend on the segment length. By contrast, the ideal case of robust topological transport should only be sensitive to the overall length of the domain wall circumference. Producing samples with different local geometrical details, but the same circumference, allows us to test these ideas by comparing their spectra.

%%% SECTION Tree samples (layout,spectra,backscattering,trivial case) 399
To this end, we designed and fabricated two tree-shaped topological cavity structures. Each of these has a total domain wall circumference of $96$ unit cells and includes seven $60^{\circ}$ corners, but individual segment lengths differ. Figure.~\ref{fig3}c shows the mechanical spectra measured near the horizontal edge of both tree geometries, superimposed onto each other. The most important observation is that, outside of the grey region, the two spectra agree almost perfectly, despite the different geometries. This is a clear and direct experimental signature of the near-perfect absence of backscattering, as predicted for the topological edge states. The grey region is close to the band gap for the horizontal edge, where no suppression of backscattering is expected (see above). 

In order to estimate the sensitivity of the spectra to backscattering, we performed calculations assuming varying levels of backscattering for both tree-shaped topological cavity geometries (Fig.~\ref{fig3}d), where $|t|^2$ ($1-|t|^2$) is the transmission (reflection) probability at each corner. These results show that even a small reflection probability of the order of $5\%$ is enough to produce clearly visible differences between the spectra, including a splitting of the peaks. Both the direct measurements and the comparison with theory confirm that the phononic topological edge states robustly transmit through sharp corners. 

%may be attributable to differences in device fabrication: these two devices are separated by $\sim 3$~mm on the SOI wafer, which results in non-identical geometries during electron-beam lithography and inductively coupled plasma reactive ion etching (see Methods).

For further comparison, we also designed and fabricated a tree-shaped trivial cavity. It is created by pulling a bulk band into the bulk band gap along a line defect embedded into an otherwise uniform domain 1 (see App.~\ref{sec:triv_waveguide} for details). In this case, the mechanical spectra measured at two different locations (on a slanted and a horizontal edge) show signatures of backscattering from the sharp corners (see Fig.~\ref{fig3}e), with irregular peak spacing and different peak locations for the two spectra.

%%%%
%%% SECTION Conclusions 203

In conclusion, we have demonstrated a multiscale optomechanical crystal and observed topological transport of thermal phonons in the $0.3$~GHz band over a bandwidth of $15$~MHz. This novel design opens the door to implementing on-chip phononic circuits \cite{Habraken2012,peano2015topological,brendel2018snowflake,sanavio2020} with robust topological waveguides that have access to the full toolbox of optomechanics.  Beyond cooling, mechanical lasing, sensitive read-out, and optical generation of nonclassical quantum states, this would also include the active optical control of topological circuits via local manipulation of mechanical modes (e.g., switching links between edge states). Another very promising avenue for applications consists in pushing towards even higher frequencies in the hypersonic regime – up to $100$~GHz should be possible with advanced lithographic methods – inverting the scale hierarchy between photonics and phononics. This would allow one to manipulate thermal phonons in myriad of new ways, including broad-band cooling of entire microscale objects, not just individual mechanical modes. Unidirectional edge channels like those found in a Chern insulator would allow one to implement thermal diodes, and, when supplemented by an energy pump, topologically protected phonon amplification and lasing~\cite{peano2016topological,mittal2018topological,bandres2018topological,zeng_electrically_2020}. An exciting long-term perspective is to use topological phononic circuits as the basis of a new platform to explore quantum acoustodynamics for quantum information processing and storage, with coupling to dopants or superconducting qubits.
%, and rely on optomechanical read-out being much more sensitive than electronic approaches at high frequencies

% \textbf{Data Availability} - The data that support the findings of this study are available from the corresponding author upon reasonable request.

% Main text references
%\bibliographystyle{natbib}
%\bibliographystyle{IEEEtran}
% \bibliographystyle{naturemag}
% \bibliography{Biblio_v1}
%\bibliography{ChiralSound}
%{\color{red} less than 70 references.}

% \putbib[Biblio_v1]
% \end{bibunit}
%\bibliography{Biblio_v1}


\vspace{2mm}
\noindent\textbf{Acknowledgements}\\ 
The authors would like to thank Sameer Sonar and Utku Hatipoglu for the help with nanofabrication and measurement. This work was supported by the Gordon and Betty Moore Foundation (award \#7435) and the Kavli Nanoscience Institute at Caltech. H.R. was supported by the National Science Scholarship from A*STAR, Singapore. T.S. and F.M. acknowledge support from the European Union’s Horizon 2020 research and innovation programme under the Marie Sklodowska-Curie grant agreement No. 722923 (OMT). V.P. acknowledges support by the Julian Schwinger Foundation (Grant No. JSF-16-03-0000).  F.M. acknowledges support from the European Union’s Horizon 2020 Research and Innovation program under Grant No. 732894, Future and Emerging Technologies (FET)-Proactive Hybrid Optomechanical Technologies (HOT).

% \vspace{2mm}

% \noindent\textbf{Author contributions.}\\ 
% HR, TS, CB, FM, VP, and OP came up with the concept and planned the experiment. HR, TS, HP, and CB performed the device design and fabrication. HR performed the measurements. HR, TS, FM, VP, and OP analyzed the data. All authors contributed to the writing of the manuscript.

% \vspace{2mm}

% \noindent\textbf{Additional information}\\
% \textbf{Supplementary information} is available in the online version of the paper.  \\
% \textbf{Competing interests.} The authors declare no competing interests.\\
% \textbf{Materials \& Correspondence.} Correspondence and requests for materials should be sent to OP (opainter@caltech.edu).

%\begin{bibunit} % References for the methods section
%\section*{Methods}

% Methods Section references
%\putbib[Biblio_v1]
%\end{bibunit}

\clearpage
\onecolumngrid

\appendix

% \subsection*{Valley Hall Effect: Theoretical Model with Anisotropy}
\section{Valley Hall Effect: Theoretical Model with Anisotropy}
\label{sec:walley_hall}

In the Valley Hall effect, the relevant topological invariant is the so-called valley Chern number $C_v$ \cite{Martin_2008_PRL,zhang_valley_2013}. The valley Chern number is defined within one valley in the framework of an effective two-band description and assumes two possible half-integer values, $C_v= \pm1/2$. Interfaces between regions with opposite valley Chern numbers support in-gap valley-polarized edge states. Since the two valleys are mapped into each other by time-reversal, their edge states are counter-propagating. 

We note  that due to both our elongated cavity design and the anisotropic silicon crystal (see App.~\ref{sec:silicon_anisotropy}), our system is not invariant under ${\cal C}_3$-rotations. This is a notable difference compared to previous larger-scale implementations of the valley Hall effect \cite{ju_topological_2015,gao_topologically_2018,zeng_electrically_2020,lu2017observation, Miniaci_PRX_2018,schaibley_valleytronics_2016}.
Taking into account the residual bulk symmetry ${\cal T}M_x$, 
we find that our system is approximated by the effective two-band Dirac Hamiltonian (see App.~\ref{sec:dirac_equation_derivation})
\begin{equation}
{\hat H}_{D}=\bar{\Omega}+(v_0+v_{x}{\hat \sigma}_{x}){\hat p}_{x}+v_{y}{\hat \sigma}_{y}{\hat p}_{y}+\{\Theta(\hat{\mathbf{r}}),\left(m+m'{\hat p}_{x}\right)\}{\hat \sigma}_{z}.\label{Dirac_equation}
\end{equation}
Here, we set $\hbar=1$, ${\hat \sigma}_{x,y,z}$ are the Pauli matrices, $\{,\!\}$ denotes the anti-commutator, and $\Theta(\mathbf{r})=1/2$ ($\Theta(\mathbf{r})=-1/2$) inside domain $1$ (domain $2$). Moreover,  $\mathbf{p}=(p_x,p_y)$ is the quasi-momentum counted from a point on the $k_x$-axis where the Bloch waves are mapped into each other via $M_y$, see Fig. 1e,f. 
The most obvious difference to ${\cal C}_3$-symmetric systems is that the speed of the edge state now depends on the domain wall orientation. The solutions for slanted and horizontal domain walls and other surprising features are discussed in the App.~\ref{sec:dirac_equation_derivation}.

We now focus on the valley close to the $\vec{K}$ point.  Fixing the gauge by choosing $\sigma_z=1$ for the Bloch wave (f) (as shown in Fig.~\ref{fig1}), a fit yields $m= 2\pi\times 10.8$MHz, $m'/\amech=-2\pi\times 5.4$MHz, 
$v_x / \amech=2\pi\times  12.5$MHz, and $v_y / \amech=2\pi\times 14.9$MHz. 
The valley Chern number for the lowest band is  $C_v=-{\rm sign}(\Theta(\mathbf{r})mv_xv_y)/2$, see App.~\ref{sec:dirac_equation_derivation}. Thus, we find $C_v=-1/2$ ($C_v=1/2$) for domain 1 (domain 2). According to the bulk-boundary correspondence, the edge state will be a right-mover if one crosses the domain wall from domain 1 to domain 2 \cite{asboth_short_2016}. This is consistent with our strip FEM simulations, see Fig.~\ref{fig1}i.
The expansion leading to Eq.~(\ref{Dirac_equation}) is valid if $m/\amech\ll v_y,(v^2_x+m'^2)^{1/2}$ (see App.~\ref{sec:dirac_equation_derivation}). This condition is not strictly fulfilled in our experiment, which leads to the deviations from the ideal case remarked upon in the Main Text. 

% \subsection*{Measuring the mechanical thermal fluctuations}
\section{Measuring the mechanical thermal fluctuations}
\label{sec:measuring}

The thermal mechanical motion of phonons within the multiscale OMCs of this work are measured by driving the system with the laser locked to a blue detuning of $340$~MHz from the optical nanocavity resonance. This frequency offset is chosen to align with the center frequency of the mechanical Dirac cones, increasing the sensitivity of the optical read-out for phonons propagating in the topological edge states. An optical fiber taper with a localized dimple region couples light evanescently into and out of an individual optical cavity with high efficiency. By moving the taper, we can address any unit cell of the larger-scale phononic lattice. Mechanical motion is imprinted on the phase of the laser light inside the optical nanocavity, which when extracted via the optical fiber taper maps the mechanical motion into intensity modulations in the transmitted laser light. The transmitted laser signal in the optical fiber is sent through an erbium-doped fiber amplifier (EDFA) to amplify the optical intensity modulations, and then onto a high-speed photoreceiver. The RF voltage from the photoreceiver is sent into a spectrum analyzer to determine the noise power spectral density (NPSD). The NPSD of the photocurrent contains a component proportional to the sum of the mechanical NPSD $S_{{\rm mech},n}$ of the mechanical normal modes of the structure, weighted by the square of the local optomechanical coupling $G_n(\mathbf{j})$, where $n$ labels the mechanical mode and  $\mathbf{j}$ labels the (unit cell of the) read-out cavity (see App.~\ref{sec:power_spectrum}). Since only the vibrations within a single unit cell contribute to the optomechanical coupling $G_n(\mathbf{j})$, the transduced mechanical NPSD can be viewed as a (coarse-grained) mechanical local density of states.

%A tunable external cavity diode laser is used to optically excite the embedded optical nanocavities within the different multiscale OMC arrays. An optical fiber taper with a localized dimple region couples light evanescently into and out of an individual optical cavity with high efficiency. By moving the taper, we can address any unit cell of the larger-scale phononic lattice, obtaining site-resolved mechanical spectra. The thermal Brownian motion of the mechanical modes of the structure is measured by driving a given optical cavity with laser light of frequency $340$~MHz above the optical cavity resonance (blue detuned pumping). This frequency offset is chosen to align with the center frequency of the mechanical Dirac cones, increasing the sensitivity of the optical read-out for phonons propagating in the topological edge states. Mechanical motion is imprinted on the phase of the laser light inside the optical nanocavity, which when extracted via the optical fiber taper and detected on a photodiode, maps the mechanical motion to modulations in the photocurrent. An electronic spectrum analyzer is used to extract the mechanical noise power spectral density (NPSD), from which a spectrum of the local thermally populated phonon modes is determined.

\section{Finite Element Simulations}
\label{sec:FEM_sim}
In Fig.1 of Main Text, we show the phononic band structures and the photonic crystal cavity modes. The mechanical normal modes are obtained by numerically solving the eigenvalue equation,
\begin{equation}
    {\rm div} \left[\mathbf{C}:\left[{\rm grad}\, \mathbf{Q}_n(\mathbf{r}) + \left({\rm grad}\, \mathbf{Q}_n(\mathbf{r}) \right)^T \right]\right]=-2\Omega_n^2\rho(\mathbf{r})\mathbf{Q}_n(\mathbf{r}).
\end{equation}
Here and throughout the Appendix, $\mathbf{Q}_n(\mathbf{r})$ $(\mathbf{Q}_{{\mathbf k},n}(\mathbf{r}))$ denotes the three-dimensional mechanical displacement $\mathbf{\mathbf{Q}_n(\mathbf{r},t)}=\Re \left[\mathbf{\mathbf{Q}_n(\mathbf{r})} \cdot e^{-i\Omega_n t} \right]$ for a normal mode (Bloch wave) with eigenfrequency $\Omega_n$. Moreover, $\mathbf{C}$ is the elasticity tensor, $\rho$ the mass density, and $:$ is a symbol for the tensor product, where $[\mathbf{C} : {\rm grad}\, {\bf \boldsymbol{\psi}}]_{ij}=C_{ijkl}\partial _l \psi _k$.

Optically, our structure is described by the Maxwell's equations, which in the absence of a source, takes the form of the following eigenvalue equation
\begin{equation}
    c^2{\rm curl}\left[\frac{1}{\varepsilon(\mathbf{r})} {\rm curl}\, \mathbf{H}_n(\mathbf{r})\right]=\omega_n^2\mathbf{H}_n(\mathbf{r}).
\end{equation}
where $\mathbf{H}_n(\mathbf{r})$ denotes the magnetic field $\mathbf{H_n(\mathbf{r},t)}=\Re \left[\mathbf{\mathbf{H}_n(\mathbf{r})} \cdot e^{-i\omega_n t} \right]$ for the photonic crystal cavity mode with eigenfrequency $\omega_n$. $c$ is the speed of light in vacuum, and $\varepsilon(\mathbf{r})$ is the relative permittivity of the medium.
Both of these equations are solved with the finite-element method (FEM) solver \cite{COMSOL}.

\section{Device fabrication}
\label{sec:fab}
% Fab section, will futher update later
The devices were fabricated from a silicon-on-insulator (SOI) wafer (SEH, $220$~nm silicon device layer, 3~$\mu$m buried-oxide layer) using electron-beam lithography followed by inductively coupled plasma reactive ion etching (ICP/RIE). The devices were then cleaned by an oxygen plasma treatment before a final released in vapor-HF to remove buried-oxide layer.
%later and diluted-HF etch to remove the chemically-grown oxide.
Note that the mechanical frequencies are different in the tree versus triangle cavity geometries, because parameters of the snowflake structures ($\amech$, $r$ and $w$) in the tree geometries have been scaled by an overall factor of $1.02$ with respect to the triangle cavity samples (the photonic crystal properties were kept identical).

\section{Optical cavity design and characterization}
\label{sec:opt_cav_design}
\begin{figure}[h]
\begin{center}
\includegraphics[width=1\columnwidth]{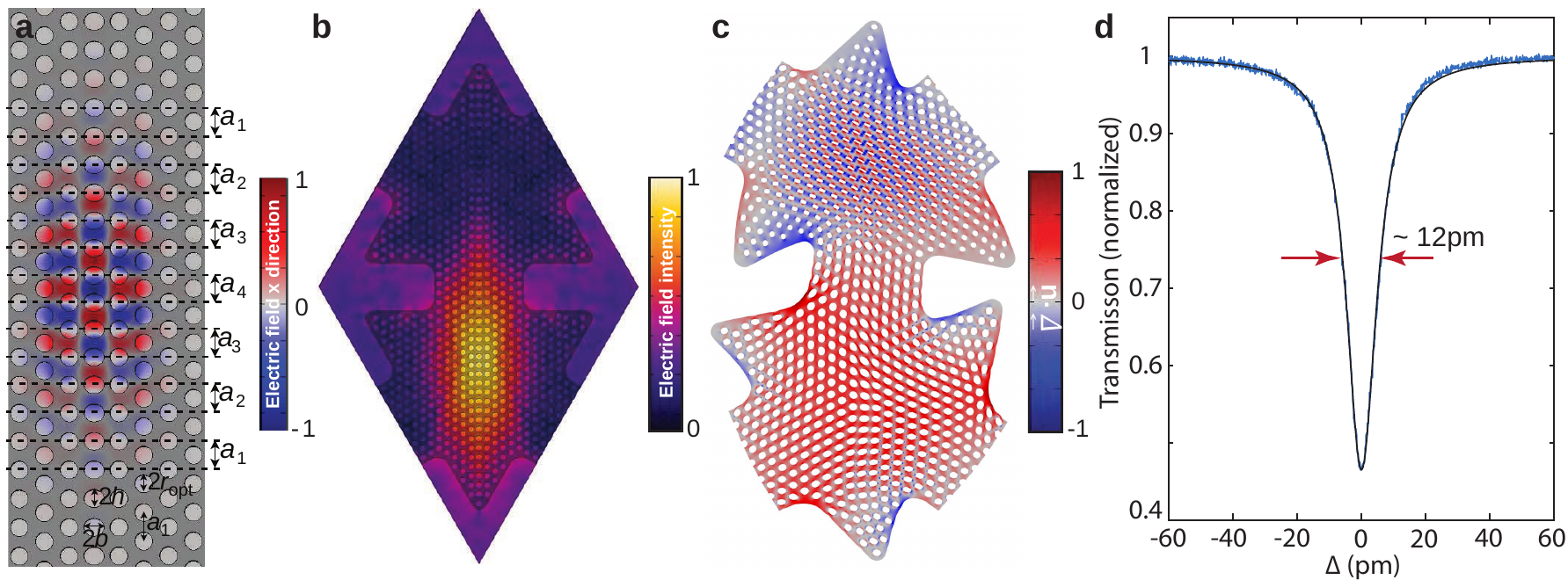}
\caption{
\textbf{Optical cavity design and characterization.} 
\textbf{a}, Design of optical cavity, local lattice constant is increased smoothly from $a_1 = 450nm$ to $a_4 = 470nm$.
\textbf{b}, Optimum position of optical cavity in order to simultaneously achieve a large $\gzero$ and optical quality factor. 
\textbf{c}, Snapshots of the mechanical deformation at the degeneracy breaking point for upper band. 
The colours represent the change of volume ($\vec{\nabla} \cdot \vec{u}$), with red (blue) corresponding to expansion (compression).
\textbf{d}, Optical spectrum of a optical cavity measured using a swept laser scan.
}
\label{SI_opt_cav}
\end{center}
\end{figure}

%\textit{some materials in METHOD.}
%\textit{add a spectrum to SI figure.}

The device in this work is designed around the silicon-on-insulator (SOI) materials platform, which naturally provides a thin Si device layer with  typical widths of a few hundred nanometers.
%In these devices, both microwave-frequency acoustic modes and near-infrared optical modes can be guided in the vertical direction.
% VP: what does this sentence mean?
In our multiscale optomechanical crystal design, photonic crystal optical cavities are embedded inside the triangular membranes forming the larger scale snowflake phononic crystal. The role of the optical cavities is to amplify the radiation pressure force of the laser light. The radiation pressure force is given by $\hbar G \vert a \vert ^2$, where $G$ is the optomechanical frequency shift per displacement and $\vert a \vert ^2 = \ncav$ is the number of intra-cavity photons. %Our photonic crystal design choice should maximise this force, while taking into account a large number of design constraints - (i) wavelength of cavity should be in the infrared range of 1550 nm, (ii) it should have high optical quality factor of $10^5$ for an efficient light-vibration energy conversion, (iii) photonic crystal must have a triangular lattice in order to fit inside the snowflake triangular lattice phononic crystal, (iv) cavity mode must have the right polarisation pattern such that it can be easily excited with a fibre-taper setup.

%A common approach to forming photonic crystal optical circuits on the thin Silicon device layer in SOI materials platform is to etch a pattern of holes into the thin dielectric film. VP Do we really need this?
Our optical cavity has been engineered starting from an existing design 
%used a modified optical cavity design adopted from 
\cite{Sekoguchi14}. 
A cavity based on this design has displayed the highest observed optical quality factor reported in the literature ($Q\sim 10^7$). Here, we have modified the original design to achieve a stronger coupling of the cavity resonance to the Dirac cone mechanical normal modes while maintaining the high optical quality factor.
The underlying basic design is a triangular lattice of cylindrical holes. An effective means of forming resonant cavities in such 2D slab photonic crystal structures is to weakly modulate the properties of a line-defect waveguide (W1 waveguide) \cite{johnson2000linear,chutinan2000waveguides}. Leaky optical resonances are localized inside the slab and yet have wave vector components which radiate energy into the surrounding cladding, which is a major source of optical loss in real fabricated structures, causing light scattering out of the plane of the slab. A line-defect waveguide in a triangular lattice of cylindrical holes can be designed to have no leaky mode bands crossing the localized cavity mode frequency.
In order to form a localized cavity resonance, the local lattice constant is increased smoothly from a nominal value of $a_1 = 450$~nm to a value of $a_4 = 470$~nm in the center of the cavity (see Fig.~\ref{SI_opt_cav}a), such that a localized resonance is created from shifting the higher frequency waveguide band into the bandgap. 
%FEM simulations of the optical and mechanical properties of the resulting optical cavity and mechanical unit cell structure were performed. The simulated change of volume ($\nabla \cdot \overrightarrow{u}$) of mechanical breathing mode and electric field intensity of the fundamental confined optical mode is shown in Fig. \ref{SI_opt_cav} (a) and (b), respectively. This optical mode has a resonance wavelength of $\lambdaopt = 1550$~nm, a theoretical radiation-limited $\Qopt$ of $2.1 \times 10^5$. 

%in order to generate high single-photon optomechanical coupling rate, $\gzero$.
The Dirac cone mechanical modes are in-plane modes with a breathing motion about the centre of the triangle. 
The  change of volume ($\nabla \cdot \overrightarrow{u}$) of such a mechanical breathing mode, simulated using FEM calculations, is shown in Fig. \ref{SI_opt_cav}c.
The mechanical breathing modes have naturally large optomechanical couplings, since breathing modes can efficiently interact with the optical cavity by moving the dielectric boundaries near the optical cavity as well as produce strain which overlaps with the electrical field of the optical resonance. 
In order to boost both the ``moving-boundary" and the ``photo-elastic" type of optomechanical coupling even further, we have added elliptical holes in the center of the W1 waveguide defect.
%So, the Noda cavity is placed at the centre of the triangles of snowflake phononic crystal as shown in Fig. \ref{SI_opt_cav} (b).
%We have also shifted the center of the optical cavities in the \textit{y}-axis in order to increase the optical quality factor ($\Qo$). 
Moreover, we have optimized  the position of the center of the optical cavity along the \textit{y}-axis in order to increase the optical quality factor ($\Qo$) within the limited triangular membrane region.
This modification contributes as well to an increase  of the optomechanical coupling. %TALK ABOUT TYPICAL SIMULATED OPTICAL Q HERE.

The optical properties of the resonances of the optical cavities are determined by scanning the tunable laser across the $\lambda =  1520 - 1570$~nm wavelength band, and measuring the transmitted optical power on a photodetector (PD1, cf.~Main Text Fig.~2a). 
From the normalized transmission spectrum, the wavelength of the optical resonance, the total optical cavity decay rate, and the external coupling rate to the fiber taper waveguide for a device being tested can be determined. An example of the transmission spectrum is shown in Fig.~\ref{SI_opt_cav}d, where the
optical cavity decay rate and external coupling rate are fitted to be $\kappa/{2\pi} \approx 1.5$~GHz and $\kappaE/{2\pi} \approx 0.47$~GHz, respectively, corresponding to a loaded (extrinsic) optical $\Qo$ factor of $\approx 129k$ ($\QoE$ of $\approx 410k$). 

%The optical properties of the resonances of the optical cavities are determined by scanning the tunable laser across the $\lambda =  1520 - 1570$~nm wavelength band, and measuring the transmitted optical power on a photodetector (PD1). 
%From the normalized transmission spectrum we can determine the wavelength of the optical resonance, the total optical cavity decay rate, and the external coupling rate to the fiber taper. As already stated in the Main Text, our measurements show that the typical optical cavity decay rate and external coupling rate are $\kappa \approx 2$~GHz and $\kappaE \approx 0.67$~GHz, respectively, corresponding to a loaded (intrinsic) optical quality factor $\Qo\approx 10^5$ ($\QoE$  $\approx 3\times10^5$). 

%An example optical transmission spectrum corresponding to Fig.~\ref{fig2}(d) is shown in SI Note NUMBER HERE.
Finally, the unit-cell single-photon optomechanical coupling strengths between the localized optical resonance and the two gapped Dirac cone modes in the unit cell geometry 
%as shown in Fig. \ref{fig2}b 
are calculated to be $\gzero/2\pi = 5.5$kHz (lower band)
%, Fig. \ref{fig2}f) 
and $33.7$kHz (upper band).

\section{Theoretical description of the edge states using the Dirac equation}
\label{sec:dirac_equation_derivation}
Here, we derive the Dirac Hamiltonian of the Main Text and solve for its eigenstates and eigenvalues. 

% Here and throughout the Supplemental Information, we denote as $\mathbf{Q}_n(\mathbf{r})$ $(\mathbf{Q}_{{\mathbf k},n}(\mathbf{r}))$ the displacement field for a normal mode (Bloch wave). We note that  $\mathbf{Q}_n(\mathbf{r})$  might indicate a formally complex running-wave solution of the time-independent elasticity equations with eigenfrequency $\Omega_n$. In this case, the displacement for the corresponding time-dependent solution of the elasticity equation  is ${\rm Re}[\exp(-i\Omega_n t)\mathbf{Q}_n(\mathbf{r})]$ 

\subsection*{Derivation of the massless Dirac Hamiltonian}
We first consider the special case  in which Dirac cones are gapless. For our discussion only the in-plane symmetries are important.  The general framework is a system with Wallpaper group  cmm  (point group ${\cal C}_{2\nu}$) with a pair of Dirac cones on the $k_x$-axis. Since the two valleys are mapped one onto the other by the time-reversal symmetry it is entirely sufficient to discuss the dynamics for just one valley. Most of the discussion will be general but, when concreteness  requires it, we  focus on the valley whose degeneracy point lies on the positive $k_x$-axis in the first Brillouin zone (BZ).

For the cmm scenario, the point group  includes the mirror symmetry $M_y$ (with the mirror plane $zx$, flipping $y\mapsto -y$). Since $M_y$ does not change the quasi-momentum on the $k_x$ axis, the  Bloch waves can be chosen to be its eigenstates there. In other words, the bands on the $k_x$-axis can be labelled by their parity (odd or even) under $M_y$. Two bands with different parity are immune to level repulsion and they, thus, can cross leading  to Dirac cones. The cones are robust because a perturbation that does not break $M_y$ will slightly displace the crossing point but can not eliminate it. In the following we denote as $\mathbf{Q}_+(\mathbf{r})$ (even) and $\mathbf{Q}_-(\mathbf{r})$ (odd) the Bloch waves at the degeneracy point. By definition, we have
\begin{equation}\label{eq:MQpm}
M_{y}\boldsymbol{Q}_{\pm}(\mathbf{r})=\pm\boldsymbol{Q}_{\pm}(\mathbf{r}).
\end{equation}
We note that  $M_y$ is the only  unitary symmetry that maps one valley onto itself. The remaining unitary symmetries $M_x$ and $M_xM_y$ (a $\pi$-rotation about the $z$-axis)  map the two partner valleys one onto the other and, thus, do not directly constrain the form of the Dirac Hamiltonian for a fixed valley. However, they do it indirectly when combined with the time-reversal symmetry ${\cal T}$ to form the corresponding anti-unitary   valley-preserving  symmetries ${\cal T}M_x$ and ${\cal T}M_xM_y$. In the following, we can choose to focus on ${\cal T}M_x$ because a Hamiltonian that is invariant under $M_y$ and ${\cal T}M_x$ will automatically be invariant under ${\cal T}M_xM_y$. 
With the goal of constraining the form of the Dirac Hamiltonian as much as possible, we fix (at least partially) the complex phase  of $\boldsymbol{Q}_{\pm}(\mathbf{x})$
by requiring that 
\begin{equation}\label{eq:partial_gauge_fix}
{\cal T}M_x\boldsymbol{Q}_{\pm}(\mathbf{x})=\boldsymbol{Q}_{\pm}(\mathbf{x}).\end{equation}
We note that the above definition  still allows to change the sign of $\boldsymbol{Q}_{+}(\mathbf{x})$ and/or $\boldsymbol{Q}_{-}(\mathbf{x})$.
In view of performing a two band approximation we define the basis,
\begin{equation}\label{eq:defQ_A/B}
\boldsymbol{Q}_{A/B,\mathbf{p}}(\mathbf{r})=e^{i\mathbf{p}\cdot\mathbf{r}}\boldsymbol{Q}_{A/B}(\mathbf{r}),\quad\boldsymbol{Q}_{A/B}(\mathbf{r})=\frac{1}{\sqrt{2}}(\boldsymbol{Q}_{+}(\mathbf{r})\pm\boldsymbol{Q}_{-}(\mathbf{r})).
\end{equation}
where $\mathbf{p}$ is the quasi-momentum counted off from the degeneracy point. Note that by changing the sign of $Q_{+}$ or $Q_{-}$ (but not both signs at the same time) will swap the labels $A$ and $B$.   We mention in passing that the Bloch waves $Q_{A,\mathbf{p}}$ and $Q_{B,\mathbf{p}}$ transform  under the cmm group symmetries in the same way as plane waves on each of the two sublattices of  graphene. This should facilitate the reading to readers familiar with the Dirac equation in this more well known context.

In view of finding the form of the local Hamiltonian, one needs preliminarily  to derive  how the basis $\boldsymbol{Q}_{A/B,\mathbf{p}}(\mathbf{r})$ transforms under the symmetry transformations $M_y$ and ${\cal T}M_x$. From Eqs. (\ref{eq:MQpm},\ref{eq:partial_gauge_fix},\ref{eq:defQ_A/B}), we find
\begin{align}\label{eq:MY_of_A/B}
&M_{y}\boldsymbol{Q}_{A,(p_x,p_y)}(\mathbf{r})=\boldsymbol{Q}_{B,(p_x,-p_y)}(\mathbf{r}).\nonumber\\
&{\cal T}M_{x}\boldsymbol{Q}_{A/B,(p_x,p_y)}(\mathbf{r})=\boldsymbol{Q}_{A/B,(p_x,-p_y)}(\mathbf{r}).
\end{align}
Next, we introduce a set of Pauli matrices  $\hat{\sigma}_{i=x,y,z}$ such that $\hat{\sigma}_z$ is diagonal on the  $A/B$ basis and $\sigma_z=1$ ($\sigma_z=-1$) for $\boldsymbol{Q}_{A,\mathbf{p}}(\mathbf{r})$  ($\boldsymbol{Q}_{B,\mathbf{p}}(\mathbf{r})$). From Eqs. (\ref{eq:MY_of_A/B}) one, thus, see
\begin{equation}\label{eq:transform_sigma}
    M_y\hat{\sigma}_{x}M_y=\hat{\sigma}_{x},\quad  M_y\hat{\sigma}_{y/z}M_y=-\hat{\sigma}_{y/z},\quad {\cal T}M_x\hat{\sigma}_{x/z}{\cal T}M_x=\hat{\sigma}_{x/z},\quad {\cal T}M_x\hat{\sigma}_{y}{\cal T}M_x=-\hat{\sigma}_{y}.
\end{equation}
while at the same time changing the quasi-momentum: under both $M_y$ and ${\cal T}M_x$ we have
\begin{equation}\label{eq:transform_quasimomentum}
    (p_x,p_y)\to (p_x,-p_y).
\end{equation}
Thus, for example the interaction $p_x\hat{\sigma}_y$ will transform to $-p_x\hat{\sigma}_y$ under the mirror symmetry $M_y$ because $\sigma_y$ changes sign, cf.~\ref{eq:transform_sigma}, while $p_x$ remains invariant, cf.~Eqs.  \ref{eq:transform_quasimomentum}.
Using Eqs. (\ref{eq:transform_sigma}) and (\ref{eq:transform_quasimomentum}) we can easily  determine the form of the Dirac equation. We are interested in a (small) region about the degeneracy point and, thus, we can restrict ourselves to linear terms in the quasi-momentum. Considering all possible linear terms and keeping only those that are invariant under both $M_y$ and ${\cal T}M_x$, we arrive at the massless Dirac Hamiltonian
\begin{equation}\label{eq:massless_dirac}
H_{D}=\bar{\Omega}+(v_{0}+v_{x}\hat{\sigma}_{x})p_{x}+v_{y}\hat{\sigma}_{y}p_{y}.
\end{equation}
This gives rise to gapless cones described by the band structure
\begin{equation}
    \Omega_{e/g}=\bar{\Omega}+v_{0}p_x\pm\sqrt{(v_xp_x)^2+(v_yp_y)^2}.
\end{equation}
Compared to the standard setting  with ${\cal C}_3$ symmetry, here, the speed depends on the direction (because $v_x\neq v_y$). Moreover, in the same direction the speed is different  for the ground and excited band (because $v_0\neq 0$). Above the critical value of $|v_0|$, $|v_0|=|v_x|$, the band structure become qualitatively different leading to so-called type II Dirac cones \cite{Huang_2016_type_II}. For type II cones there is a direction where the speed is zero for one of the two bands. In the following, we do not discuss further this scenario as our experiment is in the regime where $|v_0|<|v_x|$.

\subsection*{Derivation of the full Dirac Hamiltonian within one domain}

Next, we consider the case where the mirror symmetry  ${M}_{y}$ is broken
  but  $M_{x}$ is still a symmetry. Thus, the relevant Wallpaper group is now cm (point group ${\cal C}_\nu$). Using Eqs.  (\ref{eq:transform_sigma}) and (\ref{eq:transform_quasimomentum}) to find all possible terms that preserves the symmetry ${\cal T}M_x$, we arrive at the Dirac Hamiltonian
\begin{equation}\label{eq:full_Dirac}
H_{D}=\bar{\Omega}+(v_{0}+v_{x}\hat{\sigma}_{x})\hat{p}_{x}+v_{y}\hat{\sigma}_{y}\hat{p}_{y}+(m+m'\hat{p}_{x})\sigma_{z}.
\end{equation}
This give rise to the band structure
\begin{equation}\label{eq:bands_Dirac_2}
    \Omega_{e/g}=\bar{\Omega}+v_0 p_x\pm \sqrt{(m+m'p_x)^2+(v_xp_x)^2+(v_yp_y)^2}.
\end{equation}
We note that within the expansion in $\mathbf{p}$ that underlies the Dirac Hamiltonian, we must assume $|m'p_x|\ll |m| $. Moreover, when the design with broken ${M}_{y}$-symmetry is obtained by modifying a design with conserved ${M}_{y}$ symmetry (as in our case) all the parameters in Eq. (\ref{eq:massless_dirac}) are renormalized, including the point on the $k_x$-axis from which the quasi-momentum is counted. We also note that for $p_x=p_y=0$  in Eq. (\ref{eq:full_Dirac})  the Bloch waves are eigenstates of $\sigma_z$. Since $\sigma_z$ is by definition diagonal in the A/B basis from Eq. (\ref{eq:MY_of_A/B}) it follows that for this particular quasi-momentum the Bloch waves are mirror-symmetric partners of each other. This feature can serve as a definition of the origin $p_x=0$. This implies that  the Bloch waves shown in Fig. 1(e-f) of the Main Text, which look very much like mirror-symmetric partners in the $xz$ plane, have displaced quasi-momentum $\mathbf{p}$ very close to the origin. Once the point where $p_x=0$ is known, we use Eq. (\ref{eq:bands_Dirac_2}) to fit the parameters in the Dirac Hamiltonian (the fitted values are given in the Main Text). We note that this way of fitting does not allow to fix the sign of $v_x$, $v_y$, and $m$. In particular, the sign of $m v_x v_y$ is still unknown. As we discuss below this sign fixes the value of the valley Chern number and to be able to determine it, it is not enough to examine the band structure alone but one rather  needs to have information regarding the Bloch waves, see below.

\subsection*{Berry curvature and Valley Chern numbers }
It is convenient to introduce the set of rotated Pauli matrices
\begin{equation}
    \tau_x=\frac{v_x}{\sqrt{v^2_x+m'^2}}\sigma_x+\frac{m'}{\sqrt{v^2_x+m'^2}}\sigma_z,\quad\tau_y=  \sigma_y,\quad   \tau_z=\frac{v_x}{\sqrt{v^2_x+m'^2}}\sigma_z-\frac{m'}{\sqrt{v^2_x+m'^2}}\sigma_x.
\end{equation}
In terms of the new set of Pauli matrices, the Dirac Hamiltonian Eq. (\ref{eq:full_Dirac}) takes the simpler form
\begin{align}\label{eq:full_Dirac_2}
&H_{D}=\bar{\Omega}+(v_{0}+V_{x}\hat{\tau}_{x})(\hat{p}_{x}-p_x^{(0)})+V_{y}\hat{\tau}_{y}\hat{p}_{y}+M\tau_{z},\\
V_x&=(v^2_x+m'^2)^{1/2},\quad V_y=v_y,\quad M=\frac{mv_x}{V_x},\quad p_x^{(0)}=-\frac{mm'}{V^2_x}.
\end{align}
 The band structure in terms of the rescaled parameters reads  
\begin{equation}\label{eq:bands_Dirac}
    \Omega_{e/g}=\bar{\Omega}+v_0 p_x\pm \sqrt{M^2+V^2_x(p_x-p^{(0)}_x)^2+(V_yp_y)^2}.
\end{equation}
 Thus,  $p_x^{(0)}$ is the bottom of the valley  where the band splitting is minimum (when counted off from the quasi-momentum where the two Bloch waves are one the mirror symmetry of the other in the $zx$-plane) and $M$ is the bulk band gap.

The Berry curvature for the lowest band of the Dirac Hamiltonian is (see \cite{asboth_short_2016} for the first equality)
\begin{align}\label{eq:Berry_curvature}
 & {\cal B}(\mathbf{p})=-2{\rm Im}\frac{\langle g,\mathbf{p}|\nabla_{p_{x}}H_D(\mathbf{p})|e,\mathbf{p}\rangle\langle e,\mathbf{p}|\nabla_{p_{y}}H_D(\mathbf{p})|g,\mathbf{p}\rangle}{(\Omega_{e}(\mathbf{p})-\Omega_{g}(\mathbf{p}))^{2}}=\frac{MV_xV_y}{2\left(M^2+V^2_x(p_x-p^{(0)}_x)^2+(V_yp_y)^2\right)^{3/2}},
% & E_{e}-E_{g}=2\left(M^2+V_x^2\left(p_x-p^{(0)}_x\right)^2+(V_yp_y)^2\right)^{1/2},
\end{align}
%where $E_g(\mathbf{p})$ and $E_e(\mathbf{p})$ are the ground and excited bulk bands with 
where$|g,\mathbf{p}\rangle$ and  $|e,\mathbf{p}\rangle$ are the ground and excited Bloch waves, respectively. As usual, the  Chern number is  defined as an integral of the Berry curvature \cite{asboth_short_2016}. For the valley Chern number, the integral over the BZ is replaced by an  integral over the 2D plane,
\begin{equation}\label{eq:valley_chern_number}
C_v=    -\frac{1}{2\pi}\int d^2\mathbf{p}{\cal B}(\mathbf{p})=-\frac{1}{2}{\rm sgn}(mv_xv_y).
\end{equation}

\subsection*{Limits of validity of the Dirac approach}

The valley Chern numbers are well defined if the Berry curvature of the real bands is strongly peaked  in an isolated region surrounded by  a small Berry curvature region. In fact, it makes sense to identify each valley exactly with such isolated large Berry curvature region. 
%The valleys are well separated if the width of the Berry curvature peak, as predicted by the Dirac equation, is much smaller that the typical dimension of the unit cell and the distance between the two valley bottoms. Both of this quantities are of the order $1/a_m$. 
The Berry curvature as calculated using the Dirac Hamiltonian is  peaked in an ellipse-shaped region with axes $M/V_x$ and $M/V_y$ about $\mathbf{p}=(p^{(0)}_x,0)$, cf.~Eq. (\ref{eq:Berry_curvature}). 
We have to require that this region
remains within the quasi-momentum region where the linear expansion leading to the Dirac equation is valid. The typical size of this region is $\sim 1/a_m$. We, thus, arrive to the condition,
\begin{equation}
 M/a_m\ll V_x,V_y .  
\end{equation}
%We note that when this condition is fulfilled the Dirac cones are automatically well separated because the two valleys are centered close to the high symmetry points $\vec{K}$ and $\vec{K'}$ whose distance is also $\sim 1/a_m$.

\subsection*{Identifying the valley Chern numbers from FEM simulations}

Next, we want to determine the valley Chern number for our particular structure. We aim to use the formula $C_v=-\frac{1}{2}{\rm sgn}(mv_xv_y)$ and use  some limited input from our FEM simulations to identify the sign of $mv_xv_y$. As discussed above our definitions  Eqs. (\ref{eq:MQpm},\ref{eq:partial_gauge_fix},\ref{eq:defQ_A/B}) have the disadvantage of not completely fixing the gauge in the Dirac equation: we have the freedom to identify either of the Bloch waves in Fig.~1(e,f) with $\boldsymbol{Q}_A$. Swapping the two states will have the effect of changing the sign of both $m$ and $v_y$ but will not change the gauge invariant quantity, $C_v$. Ideally we would like an alternative definition that completely fix the gauge. Such definition would, thus, determine which state should be identified with $\boldsymbol{Q}_A$. In this scenario, the sign of $v_xv_y$ will be fixed by the gauge choice while the sign of $m$ could be read out directly from the band structure and the symmetry of the Bloch waves for $\mathbf{p}=0$. 

We can achieve exactly this if we view our system with space group cm (for the gapped cones) and cmm (for the gapless cones) as derived by the ${\cal C}_{3}$ symmetry breaking in a system
%A remarkable feature of the Bloch waves shown in Fig. 1 of the Main Text  is their breathing motion which is instrumental to obtain  an efficient optomechanical coupling. This feature can be explained if
with space group p3m1 (gapped cones) and p6m (gapless cones).  In our particular case, the symmetry breaking comes from the silicon anisotropy and the elongated shape of the optical cavities. For the ${\cal C}_{3}$ symmetric case, we will use the ${\cal C}_{3}$ symmetry to fix the gauge and identify a robust feature to identify the normal mode $\boldsymbol{Q}_A$. Afterwards, in the symmetry broken case, we will appeal to a continuity argument to find the valley Chern number, see below.
%In the ${\cal C}_3$ symmetric scenario, it is possible to  fix the gauge using this additional unitary symmetry. 

The scenario with  ${\cal C}_{3}$ symmetry is discussed in detail in \cite{brendel2017pseudomagnetic}. 
In this case, we recover Eq. (\ref{eq:full_Dirac}) with $v=v_x=v_y$ and $m',v_0=0$ if: (i) The quasi-momentum is counted off from the high-symmetry point $\mathbf{K}$. (ii) We identify the Bloch waves $\boldsymbol{Q}_{A/B}(\mathbf{r})$ with eigenstates of the ${\cal C}_3$ rotations with quasi-angular momentum $m_c$ about the ${\cal C}_6$ rotocenter of the original ${\cal C}_{6\nu}$ symmetric design (the center of the snowflakes), $m_c=-\sigma_z$. (iii) We fix their  phases  to fulfill Eq. (\ref{eq:MY_of_A/B}) \cite{brendel2017pseudomagnetic}. By requiring $m_c=-\sigma_z$ we have fixed the sign of $v_x v_y$ to be positive, thereby,  there is no further ambiguity in the sign of $m$ which now determines the  valley Chern number,  $C_v=-{\rm sign }(m)/2$. 
%Thus, one can easily infer $C_v$  from the quasi-momentum and energy of the Bloch waves at the high-symmetry point $\mathbf{K}$. 
For our particular geometry it is useful to keep in mind that  the p3m1 has three rotocenters. In our geometry, the two additional  rotocenters lie at the centers of the upward and downward-pointing triangles. The Bloch waves $\boldsymbol{Q}_{A/B}(\mathbf{r})$ are simultaneous   eigenstates of all rotations  about any of the three rotocenters. The quasi-angular momentum $m_{d/u}$ for the rotations about the center of the downward/upward-pointing triangles is \cite{brendel2017pseudomagnetic} 
\begin{equation}
m_{d/u}=(m_{c}\pm 1+1){\rm mod}3-1\label{eq:angular-mom}.
\end{equation}
For the Bloch wave 
$\boldsymbol{Q}_{A}(\mathbf{r})$ we have
 $m_{d}=0$, which means that it
displays a breathing motion in the downward-pointing triangles.  Likewise,  the mode $\boldsymbol{Q}_{B}(\mathbf{r})$ displays a breathing motion in the upper triangle. 

Once the perturbation breaking the ${\cal C}_3$ symmetry is introduced  the Bloch waves $\boldsymbol{Q}_{A/B}(\mathbf{r})$ are not anymore exact eigenstates of the ${\cal C}_3$ rotations, nevertheless, the breathing motion is  still clearly visible, cf.~Fig. 1 (e-f).  This allows us to identify the Bloch wave in Fig 1(f)  (breathing motion in the downward-pointing triangles) with $\boldsymbol{Q}_{A}(\mathbf{r})$ with the expectation that the product  $v_xv_y$ will be positive (the velocities should remain similar as in the ${\cal C}_3$ symmetric limit,   $v_x,v_y\sim v$). Moreover, from the band structure Fig. 1(d) we see that the mass $m$ is positive. We can conclude that the valley Chern number is $C_v=-1/2$ for domain 1, cf.~Eq.  (\ref{eq:valley_chern_number}).

\subsection*{Derivation of the Dirac equation in a system combining domain 1 and domain 2}

If the parameters of Eq. (\ref{eq:full_Dirac}) for domain 1 are known one can easily find the parameters for domain 2 (which is the mirror image of domain 1 in the $zx$ plane)  by  transforming   Eq. (\ref{eq:full_Dirac}) under the mirror symmetry $M_y$. Using Eqs. (\ref{eq:transform_sigma}) and (\ref{eq:transform_quasimomentum}), we see that $v_{0}$, $v_{x}$, and $v_{y}$ remain the same in the two domains while $m$ and $m'$ change sign. Thus, the valley Chern number,  cf.~Eq. (\ref{eq:valley_chern_number}), changes sign in domain 2,  $C_v=1/2$.

In a system combining both domain 1 and domain 2 we assume that a normal mode  $\mathbf{Q}_n(\mathbf{r})$ can be obtained by multiplying the Bloch waves Eq. (\ref{eq:defQ_A/B}) for $\mathbf{p}=0$ by a smooth envelope,
\begin{equation}
\mathbf{Q}_{n}(\mathbf{r})=\psi_{n,A}(\mathbf{r})\mathbf{Q}_{A}(\mathbf{r})+ \psi_{n,B}(\mathbf{r})\mathbf{Q}_{B}(\mathbf{r}).
\end{equation}
This leads to the time-independent Schroedinger equation
\begin{equation}\label{eq:Dirac_equation}
    \Omega_{n}\boldsymbol{\psi}_{n}(\mathbf{r})={\hat H}_{D}\boldsymbol{\psi}_{n}(\mathbf{r}),
\end{equation}
where $\boldsymbol{\psi}_{n}(\mathbf{r})$ groups the smooth envelopes in a vector, $\boldsymbol{\psi}_{n}(\mathbf{r})=(\psi_{n,A}(\mathbf{r}),\psi_{n,B}(\mathbf{r}))$, and ${\hat H}_{D}$ is the Dirac Hamiltonian Eq. (1) of the Main Text 
\begin{equation}\label{eq:Dirac_two_domain}
{\hat H}_{D}=\bar{\Omega}+(v_0+v_{x}{\hat \sigma}_{x}){\hat p}_{x}+v_{y}{\hat \sigma}_{y}{\hat p}_{y}+\{\Theta(\hat{\mathbf{r}}),\left(m+m'{\hat p}_{x}\right)\}{\hat \sigma}_{z}.
\end{equation}
Here $\Theta(\mathbf{r})=1/2$ ($\Theta(\mathbf{r})=-1/2$) in domain 1 (2). We note that, here,  position and quasi-momentum are non-commuting operators, thus, requiring the introduction of the anti-commutator $\{,\}$ to make sure that the Hamiltonian is hermitian.

%This Hamiltonian describes gapped anisotropic Dirac cones with spectrum
%\begin{equation}
%E_{\pm}(\mathbf{p})=\bar{\Omega}+v_{0}p_{x}\pm\sqrt{(m+m'p_{x})^{2}+(v_{x}p_{x})^{2}+(v_{y}p_{y})^{2}}.\label{eq:bulk_spectrum}
%\end{equation}

\subsection*{Solution of the Dirac equation for a strip configuration}

Next, we look for gapless eigenstates of Hamiltonian Eq. (\ref{eq:Dirac_two_domain}) in the presence of translationally invariant domain walls. In this scenario, the quasi-momentum in the translationally invariant direction is a conserved quantity and the smooth envelope depends only on the coordinate transverse to the domain wall.

\subsubsection*{Edge states for a horizontal strip.} 
For a horizontal strip, the displacement field for smooth envelope Bloch waves takes the form
\begin{equation}
\mathbf{Q}_{{p_x}}(\mathbf{r})=\psi_{A,p_x}(y)e^{ip_x x}\mathbf{Q}_{A}(\mathbf{r})+ \psi_{B,p_x}(y)e^{ip_x x}\mathbf{Q}_{B}(\mathbf{r}).
\end{equation}
This results in a Hamiltonian of the form 
\begin{equation}\label{eq:Dirac_horizontal_domain}
{\hat H}_{D}=\bar{\Omega}+(v_0+v_{x}{\hat \sigma}_{x}) p_{x}-iv_{y}{\hat \sigma}_{y}\frac{d}{dy}+2\Theta(y)\left(m+m' p_{x}\right){\hat \sigma}_{z}.
\end{equation}
We are interested in the same configuration as in Fig 1(g) of the Main Text  where domain 1 (domain 2) is in the lower-half (upper-half) plane. This choice corresponds to  $\Theta(y)=1/2$ for $y<0$, and $\Theta(y)=-1/2$ otherwise. Until now we have assumed $m>0$. We note that if we keep $\Theta(y)$ fixed, and  changing the sign of both $m$ and $m'$  describes a scenario where the two domains are swapped. In the following, we want  to compare these two scenarios. For this purpose, we look for gapless edge eigenstates of the Hamiltonian  Eq. (\ref{eq:Dirac_horizontal_domain})   without committing on  the sign of $m$ and $m'$. We find
\begin{equation}\label{eq:edge_horizontal}
\boldsymbol{\psi}_{p_{x}}(y)=e^{-|(m+m'p_{x})y/v_{y}|}\begin{pmatrix}1\\
{\rm sign}(v_y m)
\end{pmatrix},\quad \Omega_{p_{x}}=\bar\Omega+({\rm sign}(v_y m)v_{x}+v_{0})p_{x}.
\end{equation}
This solution is valid for $|m|>|m'p_{x}|$ (within the limit of validity of the linear expansion that underlies the Dirac equation). In the following, we also assume that $|v_x|>|v_0|$ (in the scenario where  this condition is violated the Dirac cones are of type II and the mass term does not lead to a global band gap.)
From Eq. (\ref{eq:edge_horizontal})  we see that, as expected, the  propagation direction changes when the domains are swapped according to the bulk boundary correspondence \cite{asboth_short_2016}. More precisely, the edge state is a right  mover if domain 1 is in the lower-half plane (for $m>0$, $v_xv_y>0$). Vice versa,  it is a left  mover if domain 1 is in the upper-half plane (for $m<0$, $v_xv_y>0$). This is consistent with the bulk-boundary correspondence because the valley Chern number $C_v$ is $-1/2$ in domain 1  and  $1/2$ in domain 2 and, thus, the edge state is a right (left) mover if the Chern number increases (decreases) by one  across the domain wall. 
%This is exactly the behavior predicted by the bulk-boundary correspondence \cite{asboth_short_2016}.
This is also in agreement with our FEM simulations, cf.~Fig1(i, left panel) of the Main Text where the Dirac cone close to the $\vec{K}$-point which has positive quasi-momentum $k_x$ in the first Brillouin zone of the bulk has  negative quasi-momentum  in the first Brillouin zone of a horizontal strip.

Until now, we have discussed general features of the Valley Hall effect that are not unique to our setting with broken ${\cal C}_3$ symmetry. Eq. (\ref{eq:edge_horizontal}) predicts also other more surprising features that are unique to our setting.
Most remarkably, the speed  $|v_{x}+{\rm sign}(m)v_0|$ changes if the two domains are swapped (changing the sign of $m$.) 
%This is an effect of the breaking of the ${\cal C}_{3}$ symmetry. 
Moreover, the
localization length of the edge state depends on the longitudinal quasi-momentum $p_{x}$. Both these features are confirmed  in  FEM simulations of a strip with a horizontal domain wall.

%A linear fit of localization length obtained from finite-element simulations
%as a function of $p_{x}$ yields the parameter $m'$ for our structure.
%Other parameters for our experiment are obtained by fitting from the
%bulk spectrum (Eq. \ref{eq:bulk_spectrum}) as
%\begin{align*}
%m & =2\pi\times9.25{\rm MHz},\:\frac{v_{y}}{a_{m}}=2\pi\times14.90{\rm MHz},\:\frac{v_{x}}{a_{m}}=2\pi\times12.47{\rm MHz},\:\frac{v_{0}}{a_{m}}=-2\pi\times1.19{\rm MHz},\\
%\frac{m'}{a_{m}} & =-2\pi\times5.40{\rm MHz},\:\delta pa_{m}=-\pi\times0.09.
%\end{align*}

\subsubsection*{Edge states for a slanted strip.} 

Next, we calculate the solution when the domain wall is along the
line  $\tilde{y}=-y/2+\sqrt{3}x/2=0$ ($240^{\circ}$ strip). In this scenario, it is convenient to change to a rotated frame with coordinates  $\tilde{y}$ (transverse to the domain wall) and $\tilde{x}=-x/2-\sqrt{3}y/2$ (longitudinal to the domain wall). In this case, the conserved quasi-momentum is  $p_{\tilde{x}}$ (in the direction $\tilde{x}$)  and the envelope is a function of $\tilde{y}$, $\boldsymbol{\psi}_{p_{\tilde{x}}}(\tilde{y})$. The Dirac Hamiltonian in terms of the rotated quasi-momenta and Pauli matrices, 
\begin{align*}
p_{x} & =-p_{\tilde{x}}/2+\sqrt{3}p_{\tilde{y}}/2,\quad p_{y}=-p_{\tilde{y}}/2-\sqrt{3}p_{\tilde{x}}/2,\\
\sigma_{x}= & -\sigma_{\tilde{x}}/2+\sqrt{3}\sigma_{\tilde{y}}/2,\quad\sigma_{y}=-\sigma_{\tilde{y}}/2-\sqrt{3}\sigma_{\tilde{x}}/2,
\end{align*}
reads
\begin{equation}\label{eq:Dirac_Ham_slanted}
\hat{H}_D=
-i\partial_{\tilde{y}}\left(h'_{0}+h'_{\tilde{x}}\sigma_{\tilde{x}}+h'_{\tilde{y}}\sigma_{\tilde{y}}\right)+\{-i\partial_{\tilde{y}},\Theta(\tilde{y})\}h'_{z}\hat{\sigma}_z+
h_{0}+h_{\tilde{x}}\sigma_{\tilde{x}}+h_{\tilde{y}}\sigma_{\tilde{y}}+2\Theta(\tilde{y})h_{z}\sigma_{z}
\end{equation}
where
\begin{align*}
 & h'_{0}=\frac{\sqrt{3}}{2}v_{0},\quad h'_{\tilde{x}}=\frac{\sqrt{3}}{4}(v_y-v_x),\quad h'_{\tilde{y}}=\frac{1}{4}(v_y+3v_x),\quad h'_{z}=\frac{\sqrt{3}}{2}m'\\
 & h_{0}=\bar{\Omega}-\frac{v_{0}}{2}p_{\tilde{x}},\quad h_{\tilde{x}}=\frac{1}{4}\left(v_x+3v_y\right)p_{\tilde{x}},\quad h_{\tilde{y}}=\frac{\sqrt{3}}{4}(v_y-v_x)p_{\tilde{x}},\quad h_z=m-\frac{1}{2}m'p_{\tilde{x}}.
\end{align*}
In this case, to solve the Dirac equation (\ref{eq:Dirac_equation}) with the Dirac Hamiltonian in the more general form Eq. (\ref{eq:Dirac_Ham_slanted}) we have to use the  ansatz,
\begin{equation}\label{eq:ansatz_slanted}
\boldsymbol{\psi}_{p_{\tilde{x}}}(\tilde{y})=(1+i\epsilon\Theta(\tilde{y}))e^{-|\tilde{y}|/\xi+i\Xi\tilde{y}}\begin{pmatrix}1\\
b
\end{pmatrix}.
\end{equation}
Compared to Eq. (\ref{eq:edge_horizontal}) this more general ansatz allows for oscillations of the wave function with period $2\pi/\Xi$. 
By plugging the anstaz Eq. (\ref{eq:ansatz_slanted}) into the Dirac equation (\ref{eq:Dirac_equation}), we find three types of terms: (i) terms containing a delta function, (ii) terms in the form
$M_{i}\Theta(\tilde{y})\psi_{p_{\tilde{x}}}(\tilde{y})$ where $M_{i}$ are matrices independent of $\tilde{y}$, and (iii) terms that depends on $\tilde{y}$ only via the wave function $\boldsymbol{\psi}_{p_{\tilde{x}}}(\tilde{y})$. By isolating the terms proportional to $\delta(\tilde{y})$ and requiring their sum to be zero, we find the equation
\begin{equation}
\left[ih_{z}'\sigma_{z}+\epsilon\left(h'_{0}+h'_{\tilde{x}}\sigma_{\tilde{x}}+h'_{\tilde{y}}\sigma_{\tilde{y}}\right)\right]\begin{pmatrix}1\\
b
\end{pmatrix}=0.
\end{equation}
It has a solution if 
\begin{equation}
{\rm det}\left[ih_{z}'\sigma_{z}+\epsilon\left(h'_{0}+h'_{\tilde{x}}\sigma_{\tilde{x}}+h'_{\tilde{y}}\sigma_{\tilde{y}}\right)\right]=0.
\end{equation}
One immediately finds two possible solutions
\begin{equation}
\epsilon_{\pm}=\pm\frac{h'_{z}}{\sqrt{h_{\tilde{x}}'^{2}+h_{\tilde{y}}'^{2}-h_{0}'^{2}}}.
\end{equation}
The corresponding $b$ is
\begin{equation}\label{eq:bpm}
b_{\pm}=-\frac{h'_{0}h'_{\tilde{x}}\mp\sqrt{h_{\tilde{x}}'^{2}+h_{\tilde{y}}'^{2}-h_{0}'^{2}}h'_{\tilde{y}}+i(h'_{0}h'_{\tilde{y}}\pm\sqrt{h_{\tilde{x}}'^{2}+h_{\tilde{y}}'^{2}-h_{0}'^{2}}h'_{\tilde{x}})}{h_{\tilde{x}}'^{2}+h_{\tilde{y}}'^{2}}.
\end{equation}
Note that $|b_{\pm}|=1$ and, thus, the vectors $(1,b_{\pm})^{T}$ lie on the equator of the Bloch sphere. Next, we require that the sum of the terms in the form
$\propto\Theta(\tilde{y})\boldsymbol{\psi}_{p_{\tilde{x}}}(\tilde{y})$ is zero. We find the equation
\begin{equation}
\left[\xi(h_{z}+h'_{z}\Xi)\sigma_{z}+i\left(h'_{0}+h'_{\tilde{x}}\sigma_{\tilde{x}}+h'_{\tilde{y}}\sigma_{\tilde{y}}\right)\right]\begin{pmatrix}1\\
b
\end{pmatrix}.
\end{equation}
where $b$ should be equal either to $b_{+}$ or $b_{-}$. By requiring that the determinant is zero we find
\begin{equation}
\xi=\frac{1}{|h_{z}+h'_{z}\Xi|}\left(h_{\tilde{x}}'^{2}+h_{\tilde{y}}'^{2}-h_{0}'^{2}\right)^{1/2}.
\end{equation}
By solving for $b$ we find $b=b_{+}$ $(b=b_{-})$ if $h_{z}+h'_{z}\Xi<0$ $(h_{z}+h'_{z}\Xi>0)$ independent of $\Xi$. Next, we need to require that the sum of the terms in the form $\propto\boldsymbol{\psi}_{p_{\tilde{x}}}(\tilde{y})$ is zero. We find the equation
\begin{equation}
\left(i\tilde{h}_{z}\sigma_{z}+\tilde{h}_{\tilde{x}}\sigma_{\tilde{x}}+\tilde{h}_{\tilde{y}}\sigma_{\tilde{y}}+\tilde{h}_{0}\right)\begin{pmatrix}1\\
b
\end{pmatrix}=0,\end{equation}
where
\begin{equation}
\tilde{h}_{\tilde{x}}=h_{\tilde{x}}+\Xi h'_{\tilde{x}},\quad \tilde{h}_{\tilde{y}}=h_{\tilde{y}}+\Xi h'_{\tilde{y}},\quad\tilde{h}_{z}=\frac{|h_{z}+h'_{z}\Xi|}{\left(h_{\tilde{x}}'^{2}+h_{\tilde{y}}'^{2}-h_{0}'^{2}\right)^{1/2}}h'_{z},\quad\tilde{h}_{0}=h_{0}+\Xi h'_{0}-\Omega_{p_{\tilde{x}}}.
\end{equation}
To solve this it is convenient to define 
\begin{equation}
\tilde{\phi}={\rm \arg}b,\quad\text{\ensuremath{\tilde{\sigma}_{\tilde{x}}=\cos\phi}}\sigma_{\tilde{x}}+\sin\tilde{\phi}\sigma_{\tilde{y}},\quad\text{\ensuremath{\tilde{\sigma}_{\tilde{y}}=\cos\tilde{\phi}}}\sigma_{\tilde{y}}-\sin\tilde{\phi}\sigma_{\tilde{x}}.
\end{equation}
and rewrite the equation in terms of the Pauli matrices $\tilde{\sigma}_{\tilde{x}}$ and $\tilde{\sigma}_{\tilde{y}}$,
\begin{equation}\label{eq:constanttermsslantedrot}
\left(i\tilde{h}_{z}\sigma_{z}+(\cos\tilde{\phi}\tilde{h}_{\tilde{x}}+\sin\tilde{\phi}\tilde{h}_{\tilde{y}})\tilde{\sigma}_{\tilde{x}}+(\cos\tilde{\phi}\tilde{h}_{\tilde{y}}-\sin\tilde{\phi}\tilde{h}_{\tilde{x}})\tilde{\sigma}_{\tilde{y}}+\tilde{h}_{0}\right)\begin{pmatrix}1\\
b
\end{pmatrix}=0.    
\end{equation}
From  Eq. (\ref{eq:bpm}), we can read out
\begin{equation}
\cos\tilde{\phi}	=-\frac{h'_{0}h'_{\tilde{x}}\pm\sqrt{h_{\tilde{x}}'^{2}+h_{\tilde{y}}'^{2}-h_{0}'^{2}}h'_{\tilde{y}}}{h_{\tilde{x}}'^{2}+h_{\tilde{y}}'^{2}},\quad
\sin\tilde{\phi}	=-\frac{h'_{0}h'_{\tilde{y}}\mp\sqrt{h_{\tilde{x}}'^{2}+h_{\tilde{y}}'^{2}-h_{0}'^{2}}h'_{\tilde{x}}}{h_{\tilde{x}}'^{2}+h_{\tilde{y}}'^{2}}.
\end{equation}
We note that $\tilde{\phi}$ is independent of $\Xi$ and $p_{x}$ and that by construction
\begin{equation}
\tilde{\sigma}_{\tilde{x}}\begin{pmatrix}1\\
b
\end{pmatrix}=\begin{pmatrix}1\\
b
\end{pmatrix},\quad\sigma_{z}\begin{pmatrix}1\\
b
\end{pmatrix}=i\tilde{\sigma}_{\tilde{y}}\begin{pmatrix}1\\
b
\end{pmatrix}=\begin{pmatrix}1\\
-b
\end{pmatrix}.    
\end{equation}
Plugging the above relations into Eq. (\ref{eq:constanttermsslantedrot}), we immediately find
\begin{equation}
\tilde{h}_{z}-\tilde{h}_{\tilde{y}}\cos\tilde{\phi}+\tilde{h}_{\tilde{x}}\sin\tilde{\phi}=0,\quad
	\tilde{h}_{\tilde{x}}\cos\tilde{\phi}+\tilde{h}_{\tilde{y}}\sin\tilde{\phi}+\tilde{h}_{0}=0.    
\end{equation}
From the second equation we find
\begin{equation}
\Omega_{{p_{\tilde{x}}}}
=h_{0}
+\Xi h'_{0}+(h_{\tilde{x}}+\Xi h'_{\tilde{x}})\cos\tilde{\phi}+(h_{\tilde{y}}+\Xi h'_{\tilde{y}})\sin\tilde{\phi}
\end{equation}
where $\Xi$ is obtained by solving the  first equation (which is a simple linear  equation).
%\begin{equation}
%\frac{|h_{z}+h'_{z}\Xi|}{\left(h_{\tilde{x}}'^{2}+h_{\tilde{y}}'^{2}-h_{0}'^{2}\right)^{1/2}}h'_{z}-\cos\tilde{\phi}(h_{\tilde{y}}+\Xi h'_{\tilde{y}})+\sin\tilde{\phi}(h_{\tilde{x}}+\Xi h'_{\tilde{x}})=0    
%\end{equation}
%Thus, for $h_{z}+h'_{z}\Xi>0$, we find
The full expression for $\Xi$ and $\bar{\Omega}$ is very cumbersome and does not give much physical insight and, thus, we omit it here. Instead, it is interesting to comment on its leading order expansion in $m'/\bar{v},v_0/\bar{v},\delta v/\bar{v}$ $\left(\bar{v}=(v_x+v_y)/2, \delta v=v_x-v_y\right)$,
\begin{align}\label{eq:dispersion_slanted}
\Xi &\approx \frac{\sqrt{3}m}{2\bar{v}}\frac{m'}{\bar{v}}+\frac{\sqrt{3}}{2}\left(-{\rm sign}(m\bar{v})\frac{v_0}{\bar{v}}+\frac{\delta v}{\bar{v}}\right)p_x, \\
\Omega_{{p_{\tilde{x}}}} &\approx\bar{\Omega} +\bar{v}\left({\rm sign}(m\bar{v})-\frac{v_0}{2\bar{v}}-{\rm sign}(m\bar{v})\frac{\delta v}{4\bar{v}}\right)p_x.
\end{align}
From this expression we see that again the edge state (for the valley close to the $\vec{K}$ point) is a right mover if the domain 1 is in the lower $\tilde{y}$-plane (for $m>0$). Also in this case (as for the horizontal domain wall) the speed  changes if the two domains are exchanged. Compared to the horizontal domain wall, the edge state amplitude does not only decay away from the domain wall but it also displays oscillations with period $2\pi/\Xi$.
%As above, we first solve the constraint
%\begin{equation}
%{\rm det}\left(i\tilde{h}_{z}\sigma_{z}+\tilde{h}_{\tilde{x}}\sigma_{\tilde{x}}+\tilde{h}_{\tilde{y}}\sigma_{\tilde{y}}+\tilde{h}_{0}\right)=0.    
%\end{equation}
%This equation gives the possible solutions
%\begin{equation}
%\tilde{h}_{0}=\pm\sqrt{\tilde{h}_{\tilde{x}}^{2}+\tilde{h}_{\tilde{y}}^{2}-\tilde{h}_{\tilde{z}}^{2}}.    
%\end{equation}
%Thus, $E=E_{+}$ or $E=E_{-}$ where 
%\begin{equation}
%E_{\pm}=h_{0}+\kappa h'_{0}\pm\sqrt{\tilde{h}_{\tilde{x}}^{2}+\tilde{h}_{\tilde{y}}^{2}-\tilde{h}_{\tilde{z}}^{2}}.    
%\end{equation}
%We can finally find $\kappa$ by solving Eq. XXX
%\begin{equation}
%i\tilde{h}_{z}\pm\sqrt{\tilde{h}_{\tilde{x}}^{2}+\tilde{h}_{\tilde{y}}^{2}-\tilde{h}_{\tilde{z}}^{2}}+(\tilde{h}_{\tilde{x}}-i\tilde{h}_{y})b=0\frac{i\tilde{h}_{z}\pm\sqrt{\tilde{h}_{\tilde{x}}^{2}+\tilde{h}_{\tilde{y}}^{2}-\tilde{h}_{\tilde{z}}^{2}}}{-\tilde{h}_{\tilde{x}}+i\tilde{h}_{\tilde{x}}}    
%\end{equation}

%
\section{Effect of Anisotropic Material Properties of Silicon}
\label{sec:silicon_anisotropy}

\begin{figure}
\begin{center}
\includegraphics[width=1\columnwidth]{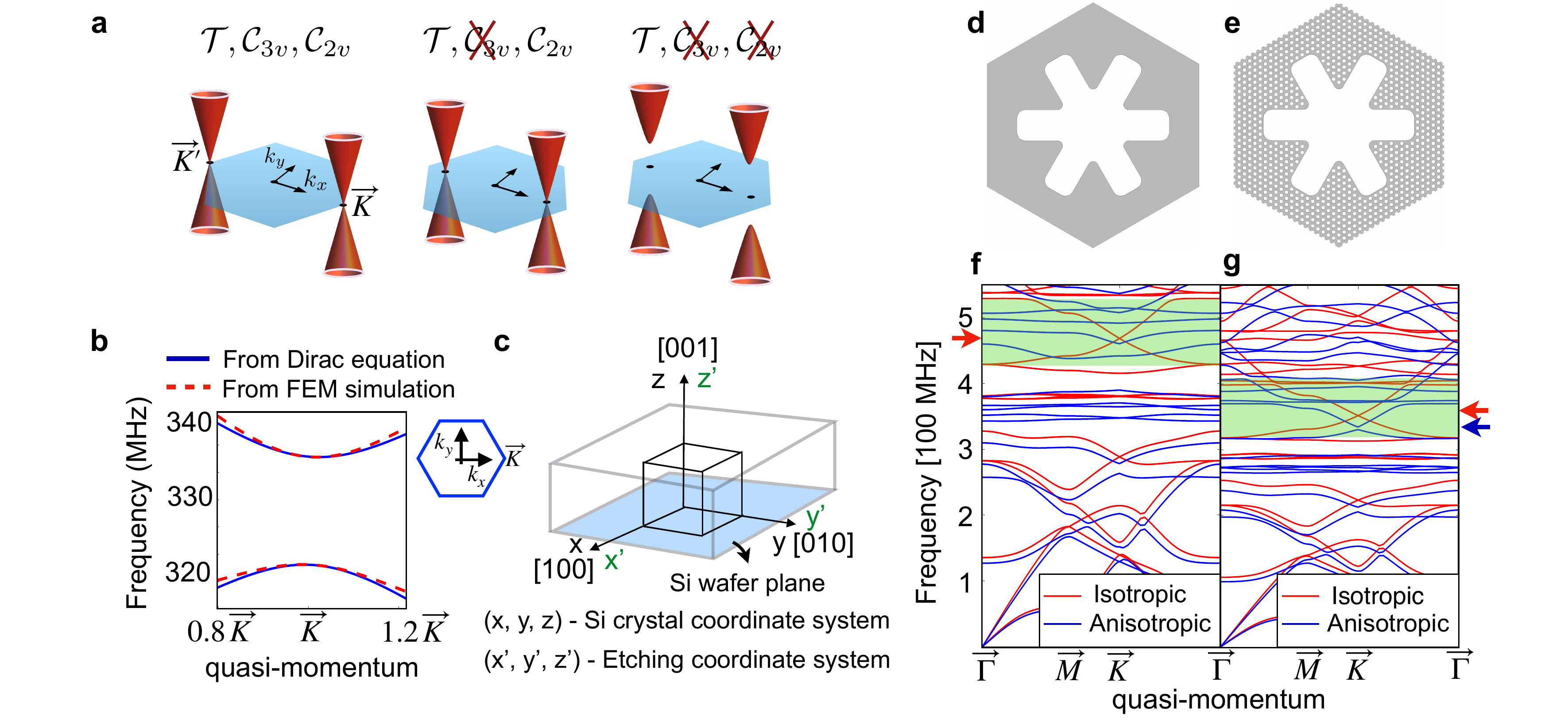}
\caption{
\textbf{Effect of symmetries and anisotropic material properties of Silicon on the Dirac band dispersion.} 
\textbf{a}, 
Effects of different symmetries on the Dirac cones. Dirac cones are displaced from the $K$ point when the $\mathcal{D}_{6v}$ symmetry (60-degree rotations and mirror symmetry) is broken. The degenerate cones split up when the $\mathcal{D}_{2v}$ symmetry is also broken.
\textbf{b}, 
Comparison of the dispersion for the gapped Dirac cones, calculated from FEM simulations and the Dirac equation.
\textbf{c}, 
The (x',y',z') axes of the fabricated devices are oriented with the [100], [010] and [001] planes of the Silicon cubic crystal, respectively.
\textbf{d} and \textbf{e}, 
Unit cell geometry of the snowflake crystal in the presence or absence of the circular photonic crystal holes. \textbf{f} and \textbf{g}, the simulated phononic band structure with isotropic Silicon material properties and anisotropic Silicon material properties,respectively. The arrows indicate the Dirac cone frequencies. The Dirac cones are significantly distorted for the snowflake crystals without photonic crystal holes (d and f), compared to the one with photonic crystal holes (e and g).
}
\label{SI_aniso}
\end{center}
\end{figure}
%In this section maybe we discuss about 2D unitcell first, then refer back to maintext how we made the strips (also assuming topo bands are already duscussed in maintext), then discuss anisotropic for horizontal and slanted strips again. -Jared
The Silicon crystal structure has a cubic primitive lattice, which leads to anisotropic material properties, and thus, the phononic band structures of our phononic crystal devices depend on the orientation of alignment during fabrication with respect to the Silicon crystal. 
The surface plane of the silicon wafer we used is parallel to the (001) crystal plane of Silicon. The alignment co-ordinate axes (x',y',z') during fabrication are oriented to the axes ([100], [010], [001]) of the Silicon wafer, cf.~Fig \ref{SI_aniso}c. 
%In contrast to the optical simulations, we have treated the Silicon material used in acoustic simulations with anisotropic elastic tensor \cite{Si_young_mod}. 
Our FEM simulations take fully into account the anisotropy using an appropriate anisotropic elastic tensor  \cite{hopcroft_2010}. 
In the presence of discrete translational symmetry, we may express the solutions to the acoustic wave equation as Bloch modes and numerically solve for the phononic band structure with an FEM solver \cite{COMSOL}. Periodic boundary conditions are used to solve for the desired number of bands at any desired point within the Brillouin zone (BZ). 
The simulated phononic band structures for the two-dimensional snowflake unit cell are shown in Fig.~\ref{SI_aniso}f and g. 
%Note the bandgap here is full bandgap for all 3 dimensions.
%In contrast to optics, phonons cannot propagate in vacuum and thus radiation out of the device plane is prohibited. This in principle allows for the formation of complete phononic bandgaps.
% The relevant pseudo-gap for the even symmetry modes is shaded in blue. In a perfectly symmetric structure this is of no consequence, as different symmetries are uncoupled from each other. In a real, fabricated structure, on the other hand, imperfections will always exist which break the perfect mirror symmetries of the structure and couple modes of different symmetry. While this coupling is typically not strong enough to prevent localization of a mechanical mode, it has important ramifications for the quality of the mechanical cavities formed from such a gap.
We investigate the role of anisotropic material properties on the Dirac band dispersion. This is done by comparing the two-dimensional snowflake unit cell phononic band structure without (see Fig~\ref{SI_aniso}d) and with (see Fig~\ref{SI_aniso}e) circular photonic crystal holes for the two cases of isotropic (red solid lines) and anisotropic (blue solid lines)  Silicon elasticity tensors. We notice that the Dirac bands for snowflakes without optical cavity holes (see Fig~\ref{SI_aniso}f) are significantly distorted for the anisotropic silicon elasticity tensor. Therefore, the small circular photonic crystal holes counter the distortion of Dirac bands by the anisotropic silicon (see Fig~\ref{SI_aniso}g).
\section{Optical readout of the thermal mechanical power spectrum}
\label{sec:power_spectrum}
We measure the local thermal power spectrum of the mechanical modes in the bulk and domain wall regions. 
Readout of the mechanics is performed by observing the transduced sidebands in the optical cavity spectrum.
We strongly drive an optical cavity at a blue-detuning
of $\bar{\Omega}$, corresponding to the middle of the mechanical bulk band gap. In the frame rotating with the
laser frequency $\omega_{L}$, the optical cavity (annihilation operator $a$) interacting with the
mechanical normal modes (annihilation operator $b_n$ for the n-th mode) is described by the  set of coupled Langevin equations,
\begin{align}
\dot{a} & =\left(i\Delta-\frac{\kappa_{t}}{2}\right)a+i\sum_{n}g_{0n}a\left(b_{n}+b_{n}^{\dagger}\right)+\sqrt{\frac{\kappa_{e}}{2}}a_{in}\label{eq:OMequations}\\
\dot{b}_{n} & =\left(-i\Omega_{n}-\frac{\Gamma_{n}}{2}\right)b_{n}+ig_{0n}a^{\dagger}a+\sqrt{\Gamma_{n}}b_{in,n}\nonumber 
\end{align}
Here,   $\Delta=\omega_{L}-\omega_{cav}=\bar{\Omega}$ is the detuning and $\kappa_{t}=\kappa_{i}+\kappa_{e}$ is the total optical decay rate. (We have a double-sided fiber taper coupling,
because of which we observe only 50\% of the output photons from
the cavity.) The input phonon noise annihilation operator $b_{in,n}$ represents the interaction
of the mechanical system with the thermal bath. The optical noise is negligible. The mechanical decay rates $\Gamma_{n}$ are almost constant, $\Gamma_{n}\approx 200$kHz, and we do not seek to model them. On the other hand, we determine  the mechanical eigenfrequencies $\Omega_n$
%Here, we ignore the
%input optical noise. We have a double-sided fiber taper coupling,
%because of which, we observe only 50\% of the output photons from
%the cavity $\kappa_{t}=\kappa_{i}+\kappa_{e}$. 
and the single-photon optomechanical coupling rates $g_{0n}$   combining FEM simulation of a strip configuration to scattering matrix calculations, see App.~\ref{sec:scattering_matrix}.
%In the bulk region, these modes are the eigenvalues for each discrete quasi-momentum $k$. While, in the domain wall region, these are the standing wave modes. 
%The input phonon
%noise annihilation operator $b_{in,j}$ represents the interaction
%of the mechanical system with the thermal bath. Here, we ignore the
%input optical noise. We have a double-sided fiber taper coupling,
%because of which, we observe only 50\% of the output photons from
%the cavity $\kappa_{t}=\kappa_{i}+\kappa_{e}$. In the domain wall region, the optomechanical
%coupling $g_{0j}(x)$ between an optical cavity mode at the strip unit cell $x$ and the standing wave mode is given by
%\begin{equation}
%g_{0j}(x)=g_{0j}^{(strip)}(x)\psi_{j}(x)
%\end{equation}
%Here, $g_{0j}^{(strip)}(x)$ is the optomechanical coupling rate between
%the optical mode and the strip eigenfunction at $k(\Omega_{j})$.
%We get $g_{0j}^{(strip)}(x)$ for both horizontal and slanted strip
%unit cell from the finite-element simulations. $\psi_{n}(x)$ is the
%amplitude of the displacement field at $x$ (Eq. (\ref{eq:amp_displacement})). In the bulk region, the optomechanical coupling is given by $g_{0j}(x)=g_{0j}^{(strip)}(x)/\sqrt{N}$.

We linearize the equations of motion about the static equilibrium \cite{Aspelmeyer2014}, and set 
$a=\alpha+\delta ae^{i\theta}$, where 
$\alpha=\vert\alpha\vert e^{i\theta}$ ($\vert\alpha\vert^{2}$ is the number
of intra-cavity photons),
and
$b_{n}=\beta_{n}+\delta b_{n}$, with
\begin{align}
\beta_{n} & =\frac{ig_{0n}\vert\alpha\vert^{2}}{i\Omega_{n}+\Gamma_{n}/2}\thickapprox\frac{g_{0n}\vert\alpha\vert^{2}}{\Omega_{n}}\\
\alpha & =\frac{-\sqrt{\kappa_{e}/2}a_{in}}{i\Delta-\kappa_{t}/2+i\sum_{n}g_{0n}(\beta_{n}+\beta_{n}^{*})}\thickapprox\frac{-\sqrt{\kappa_{e}/2}a_{in}}{i\Delta-\kappa_{t}/2+2i\vert\alpha\vert^{2}\sum_{n}g_{0n}^{2}/\Omega_{n}}\thickapprox\frac{-\sqrt{\kappa_{e}/2}a_{in}}{i\Delta-\kappa_{t}/2}
\end{align}
Here, we have used $\Omega_{j}\gg\Gamma_{j}$ and $\kappa_{t}\gg\vert\alpha\vert^{2}\sum_{n}g_{0n}^{2}/\Omega_{n}$.
Ignoring the nonlinear interaction, the resulting equation of motion
is
\begin{align}
\delta\dot{a} & =\left(i\bar{\Delta}-\frac{\kappa_{t}}{2}\right)\delta a+i\sum_{n}g_{0n}\vert\alpha\vert\left(\delta b_{n}+\delta b_{n}^{\dagger}\right)\\
\delta\dot{b}_{n} & =\left(-i\Omega_{n}-\frac{\Gamma_{n}}{2}\right)\delta b_{n}+ig_{0n}\vert\alpha\vert\left(\delta a+\delta a^{\dagger}\right)+\sqrt{\Gamma_{n}}b_{in,n}
\end{align}
Here, $\bar{\Delta}=\Delta+\sum_{n}g_{0n}\left(\beta_{n}+\beta_{n}^{*}\right)$. For the frequency domain operators defined by $O[\omega]=\frac{1}{2\pi}\intop_{-\infty}^{+\infty}d\omega e^{i\omega t}O(t)$
and $O^{\dagger}[\omega]=\frac{1}{2\pi}\intop_{-\infty}^{+\infty}d\omega e^{i\omega t}O^{\dagger}(t)=\left[O[-\omega]\right]^{\dagger}$,
the above equation can be recasted to the following linear system
of algebraic equations
\begin{align}
\delta a[\omega] & =i\vert\alpha\vert\chi_{opt}[\omega]\sum_{n}g_{0n}\left(\delta b_{n}[\omega]+\delta b_{n}^{\dagger}[\omega]\right)\\
\chi_{n}^{-1}[\omega]\delta b_{n}[\omega]= & \sqrt{\Gamma_{n}}b_{in,n}[\omega]-i\vert\alpha\vert^{2}g_{0n}\left(\chi_{opt}[\omega]-\chi_{opt}^{*}[-\omega]\right)\sum_{n'}g_{0n'}\left(\delta b_{n'}[\omega]+\delta b_{n'}^{\dagger}[\omega]\right)\\
\left(\chi_{n}^{-1}[-\omega]\right)^{*}\delta b_{n}^{\dagger}[\omega]= & \sqrt{\Gamma_{n}}b_{in,n}^{\dagger}[\omega]-i\vert\alpha\vert^{2}g_{0n}\left(\chi_{opt}[\omega]-\chi_{opt}^{*}[-\omega]\right)\sum_{n'}g_{0n'}\left(\delta b_{n'}[\omega]+\delta b_{n'}^{\dagger}[\omega]\right)
\end{align}
where $\chi_{n}[\omega]=\left[\Gamma_{n}/2-i\left(\omega-\Omega_{n}\right)\right]^{-1}$
and $\chi_{opt}[\omega]=\left[\kappa_{t}/2-i\left(\omega+\bar{\Delta}\right)\right]^{-1}$
are the mechanical and optical susceptibilities, respectively, in the
absence of optomechanical coupling. We can ignore the optical backaction on the mechanics because our experiment is in a regime of small optomechanical cooperativity. Thus,
\begin{align}
\delta a[\omega] & =i\vert\alpha\vert\chi_{opt}[\omega]\sum_{n}g_{0n}\sqrt{\Gamma_{n}}\left(\chi_{n}[\omega]b_{in,n}[\omega]+\left(\chi_{n}[-\omega]\right)^{*}b_{in,n}^{\dagger}[\omega]\right)
\end{align}
The output from the cavity is obtained via the input-output relations as
\begin{align}
a_{out}[\omega] & =a_{in}-\sqrt{\frac{\kappa_{e}}{2}}\left(\alpha+\delta a[\omega]e^{i\theta}\right)\\
 & =a_{in}\frac{i\Delta-\kappa_{i}/2}{i\Delta-\kappa_{t}/2}+a_{in}\frac{\kappa_{e}/2}{i\Delta-\kappa_{t}/2}i\chi_{opt}[\omega]\sum_{n}g_{0n}\sqrt{\Gamma_{n}}\left(\chi_{n}[\omega]b_{in,n}[\omega]+\left(\chi_{n}[-\omega]\right)^{*}b_{in,n}^{\dagger}[\omega]\right)=\bar{a}_{out}+\delta a_{out}[\omega]
\end{align}
In our detection scheme, the strong laser beats with the optical sidebands. This generates photo-current proportional to the cavity output amplitude
quadrature $I(t)=\delta a_{out}(t)+\delta a_{out}^{\dagger}(t)$. The cavity output power spectrum density is given by
\begin{equation}
S_\text{II}[\omega]=\int_{-\infty}^{\infty}dte^{i\omega t}\left\langle I(t)I(0)\right\rangle =2\pi\int_{-\infty}^{\infty}d\omega'\left\langle I[\omega]I[\omega']\right\rangle 
\label{eq:SIIdefination}
\end{equation}
For a thermal bath of average phonon occupancy $n_{b}$, the correlation
of the noise operators are $\left\langle b_{in,n}^{\dagger}[\omega]b_{in,k}[\omega']\right\rangle =n_{b}/(2\pi)\delta(\omega+\omega')\delta_{n,k}$
and $\left\langle b_{in,n}[\omega]b_{in,k}^{\dagger}[\omega']\right\rangle =(n_{b}+1)/(2\pi)\delta(\omega+\omega')\delta_{n,k}$.
At room temperature $T=300K$ and $\Omega_{n}= 330$MHz, $n_{b}\thickapprox k_{B}T/\hbar\Omega_{n}\thickapprox18940$
phonons. The boson occupancy is practically identical for all the
standing wave mechanical normal modes, and we also assume $n_{b}+1\thickapprox n_{b}$
for simplicity. For $\omega\thickapprox \bar{\Omega}$ and $\Omega_{n}\gg\Gamma_{n}$,
we find
\begin{align}\label{eq:noise_spectrum}
S_\text{II}[\omega] & =\frac{\kappa_{e}}{2}\left|\alpha\chi_{opt}[\omega]-\alpha^{*}\chi_{opt}^{*}[-\omega]\right|^{2}\sum_{n}\frac{g_{0n}^{2}n_{b}\Gamma_{n}}{\left(\omega-\Omega_{n}\right)^{2}+\left(\Gamma_{n}/2\right)^{2}}=
\frac{\kappa_{e}}{2}\left|\alpha\chi_{opt}[\omega]-\alpha^{*}\chi_{opt}^{*}[-\omega]\right|^{2}\sum_{n}G_{n}^{2}S_{{\rm mech},n}[\omega]\\
G_{n} & =\frac{g_{0n}}{x_{zpf,n}},\:S_{{\rm mech},n}[\omega]=\frac{x_{zpf,n}^{2}n_{b}\Gamma_{n}}{\left(\omega-\Omega_{n}\right)^{2}+\left(\Gamma_{n}/2\right)^{2}},
\end{align}
where $x_{zpf,n}$ are the zero point fluctuations of the $n$-th mode.
%VP I thought it was strange here to multiply and divide by \kappa_t

So far the discussion has been generic and could refer to any cavity coupled to multiple mechanical modes. Next, we discuss the particular features that arise in the position resolved noise spectrum of our topological mechanical cavity because of  the underlying topology.
As discussed in the Main Text, the spectrum $\Omega_j$ in the topological region is formed by a series of quasidegenerate doublets because of the suppression of backscattering (see the derivation below based on the scattering matrix approach). Each of these modes is a standing wave mode leading to a dependence of the optomechanical coupling on  the optical cavity via its position (alternatively, as long as they are quasi-degenerate, they can be treated as running waves that are counterpropagating, leading to the same final result displayed below). Away from a corner, the domain wall hosting the topological mechanical cavity can be approximated as a 1D strip, and we expect the coupling to display a sinusoidal dependence on the cavity position, $G_{n}(j)\propto \cos(k_l(\Omega_j)ja_m+\phi_j)$ where $a_m$ is the length of strip unit cell, $ja_m$ is the position of the $j$-th cavity and $k_l(\Omega_n)$ is the quasi-momentum (which depends on the direction of the domain wall $l$), see below for derivation. For quasidegenerate levels ($\Omega_n\approx \Omega_{n+1}$), the couplings $G_{n}(j)$ and $G_{n+1}(j)$ will then be sinusoidal waves with the same period but a phase delay $\phi_{n+1}-\phi_{n}\approx\pi/2$. From  Eq. (\ref{eq:noise_spectrum}) we see that  two quasidegenerate levels with a splitting $\Omega_n-\Omega_{n+1}$, much  smaller than the mechanical decay rate $\Gamma_n$, give rise to a single mechanical noise spectrum peak with a height proportional to the sum of the squares of their respective optomechanical couplings,
\begin{equation}\label{eq:no_backscattering}
S_\text{II}[\Omega_n]\propto G^2_n+G^2_{n+1}\approx \cos^2(k_l(\Omega_n)ja_m+\phi_n)+\sin^2(k_l(\Omega_n)ja_m+\phi_n)\approx {\rm const}.
\end{equation}
Thus, we find that the peak height is the same for all cavities that are localized along the same edge of the polygonal-shaped domain wall and away from the corners. This feature is visible in the measured position resolved spectrum shown in Fig. 2(d) of the Main Text.

\section{Semi-analytical calculation of the spectra and optomechanical coupling for the mechanical topological cavity}
\label{sec:scattering_matrix}
Here, we discuss how we calculate  the mechanical eigenfrequencies $\Omega_j$ and optomechanical couplings $g_{0j}$ used for the theoretical estimation of the power spectra of the  mechanical topological cavity. A full FEM simulation of our device is not feasible due to the multiscale nature of  our optomechanical crystal and the large system size. Instead, we adopt a hybrid approach where the spectra and optomechanical couplings are obtained from a scattering matrix calculation that uses the band structure and optomechanical couplings of strip configurations (with the two relevant orientations of the domain wall) obtained using FEM simulation as input. 

\subsection*{Scattering matrix calculation of the  spectrum $\Omega_n$}

\label{subsec:scat_matrix}
\begin{figure}
\begin{center}
\includegraphics[width=1\columnwidth]{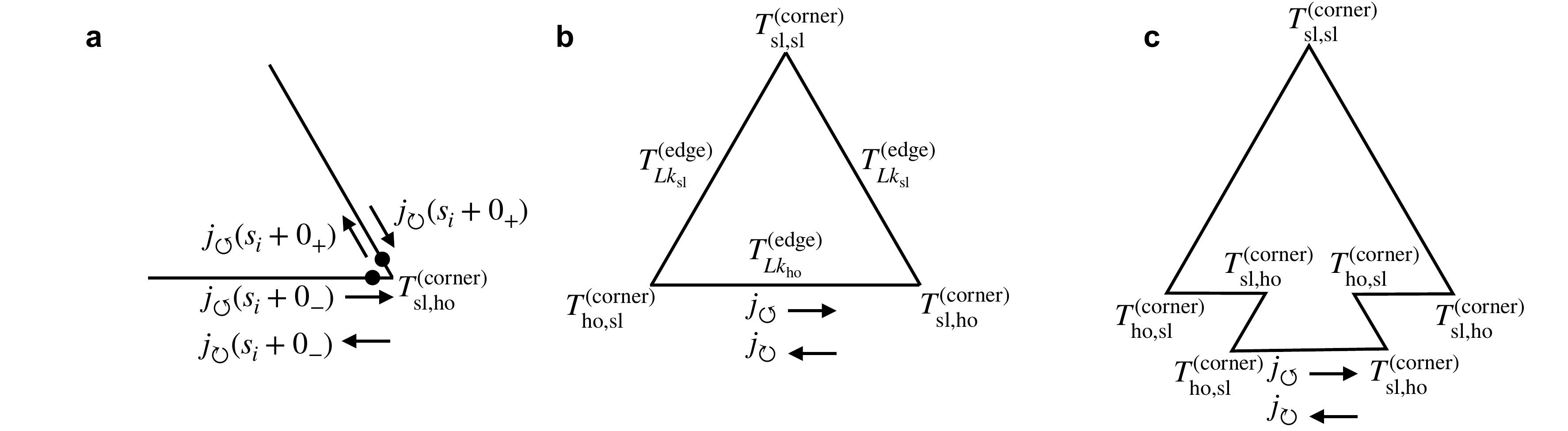}
\caption{
\textbf{Schematic for the one-dimensional scattering matrix model.} 
\textbf{a}, 
Wave transport from the horizontal to the slanted waveguide through the corner is modelled with the transfer matrix $T^{(\rm corner)}_{{\rm sl,ho}}$. 
\textbf{b} (\textbf{c}), 
Modelling wave transport on the triangular (tree) topological mechanical cavities with different transfer matrices.
}
\label{SI_scat_mat_setup}
\end{center}
\end{figure}

We model the topological mechanical cavity as a closed sequence of  edge channels  connected by scattering centers. We assume that away from the corners the mechanical waves propagate unimpeded as if they were flowing in an infinitely long edge channel. On the other hand, we describe the transmission across the corners phenomenologically using the standard scattering-matrix approach. In the following, we assume that the domain 1 region  has a polygonal form with sides in the horizontal direction (along the $x$-axis) or two different slanted directions ($120^\circ$ or $240^\circ$ from the $x$-axis). This general scenario applies to all the devices we have built.

We introduce the coordinate $s$ that follows the domain wall length and denote as $ j_\circlearrowleft (s)$ and $ j_\circlearrowright (s)$ the mechanical edge current amplitudes circulating clockwise and anti-clockwise about domain 2, respectively (thus, $ |j_\circlearrowleft (s)|^2$ and $ |j_\circlearrowright (s)|^2$ are the corresponding  mechanical energy fluxes).

Away from a corner, we assume that  the mechanical wave propagates as in an infinitely long strip, 
\begin{equation}\label{T-matrix_waveguide}
\begin{pmatrix}j_\circlearrowleft (s+na_m)\\
j_\circlearrowright (s+na_m)
\end{pmatrix}=T^{(\rm edge)}_{n a_m k_l(\Omega)}\begin{pmatrix}j_\circlearrowleft (s)\\
j_\circlearrowright (s)
\end{pmatrix}, 
\quad
T^{(\rm edge)}_{\theta}=\begin{pmatrix}e^{-i\theta} & 0\\
0 & e^{i\theta}.
\end{pmatrix}
\end{equation}
Here, $n$ is an integer, $a_m$ is the length of the strip unit cell, $k_l(\Omega)$ is the  quasi-momentum at frequency $\Omega$ in the positive branch, $k_l(\Omega)>0$,  for an  infinite strip. We note that the dispersion relation $k_l(\Omega)$ depends on the domain wall orientation at  $s$,  horizontal ($l={\rm ho}$) or slanted ($l={\rm sl}$). The two slanted directions are equivalent because of the mirror symmetry about the $yz$ plane. We note that $j_\circlearrowleft$ describes a wave propagating anticlockwise because $dk_l(\Omega)/d\Omega<0$ in the positive quasi-momentum branch, cf.~Main Text Fig. 1.

Next, we denote as $s_i$ the coordinate of a corner. The domain wall has different orientations $l$ and $r$ for $s=s_i+0_-$ (immediately before the corner) and $s=s_i+0_+$ (immediately after the corner), respectively.  We connect the  solutions at the two opposite sides of the corner using the  appropriate transfer matrix,
\begin{equation}\label{T-matrix}
\begin{pmatrix}j_\circlearrowleft (s_i+0_+)\\
j_\circlearrowright (s_i+0_+)
\end{pmatrix}=T^{(\rm corner)}_{r,l}\begin{pmatrix}j_\circlearrowleft (s_i+0_-)\\
j_\circlearrowright (s_i+0_-)
\end{pmatrix}, 
\quad
T^{(\rm corner)}_{r,l}\left(r,t\right)=\frac{1}{t_{rl}}\begin{pmatrix}\frac{t_{rl}}{t_{rl}^{*}} & -r_{rl}^{*}\frac{t_{rl}}{t_{rl}^{*}}\\
-r_{rl} & 1
\end{pmatrix}.
\end{equation}
Here, $r_{rl}$ and $t_{rl}$ are the complex reflection and transmission coefficients,
respectively,  and, thus, $\vert r_{lr}\vert^{2}+\vert t_{lr}\vert^{2}=1$. The form of the T-matrix above reflects the symmetry under time-reversal and the conservation of the energy flux.
%We denote  as $ j_{l}^{(in)}$
%and $j_{r}^{(in)}$ ($j_{l}^{(out)}$ and $j_{r}^{(out)}$) the mechanical edge current amplitudes incident on (traveling away from)
% a corner at its  two opposite sides  (labeled as $l$ and $r$), see sketch in Fig.~\ref{SI_scat_mat_setup} (a).
%As usual, the outgoing currents  can be  expressed in terms of the incoming current using the S-matrix
%\begin{equation}
%\begin{pmatrix}j_{l}^{(out)}\\
%j_{r}^{(out)}
%\end{pmatrix}=S_{rl}\left(r,t\right)\begin{pmatrix}j_{l}^{(in)}\\
%j_{r}^{(in)}
%\end{pmatrix}, 
%\quad
%S_{lr}\left(r_{rl},t_{rl}\right)=\begin{pmatrix}r_{rl} & t_{rl}\\
%t_{rl} & -r_{rl}^{*}\frac{t_{rl}}{t_{rl}^{*}}
%\end{pmatrix}
%\end{equation}
%
By applying the mirror symmetry about the $yz$ plane to Eq.  (\ref{T-matrix}), one finds 
\begin{equation}\label{T_matrix_parity}
T^{(\rm corner)}_{l,r}=\frac{1}{t_{rl}}\begin{pmatrix}\frac{t_{rl}}{t_{rl}^{*}} & r_{rl}\\
r_{rl}^{*}\frac{t_{rl}}{t_{rl}^{*}} & 1
\end{pmatrix}.
\end{equation}
For the special case $r=l={\rm sl}$, combining the above equation with Eq.  (\ref{T-matrix})  gives the constraint $r_{{\rm sl\, sl}}^*t_{{\rm sl\, sl}}+t_{{\rm sl\, sl}}^*r_{{\rm sl\, sl}}=0$. Thus, we are left with five independent transfer matrix parameters: $t_{\rm ho, sl}$, $r_{\rm ho, sl}$ (two independent phases and one independent  amplitude) and $t_{\rm sl, sl}$ (independent  amplitude and phase.)

Using the building blocks Eqs.  (\ref{T-matrix_waveguide},\ref{T-matrix}) and the symmetry constraint Eq. (\ref{T_matrix_parity}) we can build a transfer matrix $T_{\rm loop}$ that evolves the current amplitude from a point $s$ on a closed loop around the whole domain wall length. For example, for the triangle of side $L$ we find the loop T-matrix
\begin{equation}
T_{\rm loop}=T^{(\rm corner)}_{{\rm ho,sl}}T^{(\rm edge)}_{L k_{\rm sl}}T^{(\rm corner)}_{{\rm sl,sl}}T^{(\rm edge)}_{L k_{\rm sl}}T^{(\rm corner)}_{{\rm sl,ho}}T^{(\rm edge)}_{L k_{\rm ho}}
\end{equation}
The loop T-matrix allows to impose the appropriate periodic boundary conditions, requiring
\begin{equation}\label{eq:T_loop_pb}
  {\rm det}\left(T_{\rm loop}-1\!\!1\right)=0.  
\end{equation}
This equations can then be  solved to give the spectrum  $\Omega_n$ in terms of the transfer matrix parameters. Vice versa, when the spectrum is known, one can use the same equation to extract information about  the transfer matrix parameters.

\subsubsection*{Analytical calculation of the spectrum $\Omega_j$ in the absence of backscattering}
We note that in the limit of zero backscattering (corresponding to the special case $r_{\rm sl, sl}=r_{\rm ho, sl}=0$) the matrix $T_{\rm loop}$ is diagonal and as a consequence the spectrum is formed by degenerate doublets that solve the simple equation
\begin{equation}\label{eq:topological_spectrum}
L_{\rm sl}k_{\rm sl}(\Omega)+L_{\rm ho}k_{\rm ho}(\Omega)+N_{\rm sl,sl}\arg (t_{\rm sl,sl})+N_{\rm sl,ho}\arg( t_{\rm sl,ho}) =n2\pi.
\end{equation}
Here, $L_{\rm sl}$ ($L_{\rm ho}$) is the total length along slanted sides, and $N_{\rm sl,sl}$  ($N_{\rm sl,ho}$) is the number of corners connecting two slanted sides (a slanted and a horizontal side). Importantly, according to this formula (that assumes no backscattering) we expect the same spectrum for the two different tree-shaped cavity geometries. This is the tell-tale signature for the absence of backscattering (as a consequence of the topological nature of the transport) that we will be looking for in the experiment.

By deriving Eq. (\ref{eq:topological_spectrum}) with respect to the frequency we find a simple expression for the free spectral range between two  doublets
\begin{equation}
\Omega_{j+2}- \Omega_{j}\approx \frac{2\pi v_{\rm sl}v_{\rm ho}}{L_{\rm sl}v_{\rm ho}+L_{\rm ho}v_{\rm sl}}.
\end{equation}
where $v_{\rm sl}=d\Omega/dk_{\rm sl}$ ($v_{\rm ho}=d\Omega/dk_{\rm ho}$) is the group velocity on a slanted (horizontal) side. Here, we have assumed that the transmission amplitudes $t_{\rm sl,sl}$ and $t_{\rm sl,ho}$ are frequency independent.   We note that the assumption of constant scattering parameters clearly breaks down at the crossover region between the topological and trivial region (about the edge of the horizontal strip bandwidth) where the transmission across the slanted-horizontal corners $t_{\rm sl,ho}$ must go to zero. This is the reason for the mismatch between the measured and theoretical noise spectra in that region. We do not seek to model the crossover region.

\subsection*{Calculation of the  vacuum OM couplings $g_{0n}$}
% \label{sec:g0}
In this section we show how to estimate the vacuum optomechanical couplings $\{g_{0n}\}$ for the normal modes $\{\mathbf{Q}_n(\mathbf{r})\}$ of the topological mechanical cavity. By definition, this is the cavity shift by a displacement field with amplitude equal to the zero-point fluctuations. By requiring that the energy stored in the vibrations $\Omega_j^2\int_V|\mathbf{Q}_n(\mathbf{r})|^2\rho(\mathbf{r})d^3\mathbf{r}$ is equal to the zero point energy $\hbar\Omega_j/2$ we find the normalization condition
\begin{equation}
\int_{V}|\mathbf{Q}_{n}(\mathbf{r})|^{2}\rho(\mathbf{r}) d^{3}\mathbf{r}=\frac{\hbar}{2\Omega_{n}}.\label{eq:normmech}
\end{equation}
The cavity shift $g_{0j}$ is then given by the standard perturbative formulas $g_{0n}=g_{0n}^{(\rm PE)}+g_{0n}^{(\rm Bnd)}$ with the   moving boundary contributions and photoelastic contributions
\begin{align}\label{eq:g_vacuum}
 g_{0n}^{(\rm Bnd)}&=-\frac{\omega_{0}}{2}\frac{\int_{A}\left(\mathbf{Q}_{n}(\mathbf{r})\cdot\mathbf{n}\right)\left((\varepsilon(\mathbf{r})-\varepsilon_0)|\mathbf{e}^{\parallel}(\mathbf{r})|^{2}-(\varepsilon(\mathbf{r})-\varepsilon_0)^{-1}|\mathbf{d}^{\perp}(\mathbf{r})|^{2}\right)dA}{ \int_{V}|\mathbf{e}(\mathbf{r})|^{2}\varepsilon(\mathbf{r}) d^{3}\mathbf{r}}  \\
 g_{0n}^{(\rm PE)}&=-\frac{\omega_{0}}{2}\frac{\int_{V}\mathbf{e}(\mathbf{r})\cdot \delta\varepsilon(\mathbf{r})\cdot \mathbf{e}(\mathbf{r})}{ \int_{V}|\mathbf{e}(\mathbf{r})|^{2}\varepsilon(\mathbf{r}) d^{3}\mathbf{r}}
\end{align}
where  $\mathbf{e}(\mathbf{r})$ and $\mathbf{d}(\mathbf{r})$ are the electric and the electric displacement fields in  the cavity mode, respectively. Moreover, $\varepsilon(\mathbf{r})$ is the permittivity,
and the tensor $\delta\varepsilon(\mathbf{r})$ is the local change of permittivity due to the strain.  For the purpose of our discussion it is only important that the mapping between the displacement field $\mathbf{Q}_{n}(\mathbf{r})$ and the tensor $\delta\varepsilon(\mathbf{r})$  (via the strain tensor)  is linear, see Ref. \cite{Safavi-Naeini2014} for more details.  Since the electric and the displacement fields are exponentially localized  within a  single triangular membrane, we can approximate the displacement field as
\begin{equation}\label{eq:Q_nofu}
\mathbf{Q}_{n}(\mathbf{r})\approx A_{n,i}e^{ik_l(\Omega)s}\mathbf{u}_{n,l,\Omega_n}(\mathbf{r})+A_{n,i}^*(s)e^{-ik_l(\Omega)s}\mathbf{u}^*_{n,l,\Omega_n}(\mathbf{r}),
\end{equation}
where $\exp[{ik_l(\Omega)s}]\mathbf{u}_{n,l,\Omega_n}(\mathbf{r})$ are the edge state Bloch waves for an infinite strip with the appropriate orientation. Thus, $\mathbf{u}_{n,l,\Omega_n}(\mathbf{r})$ is a periodic function in the domain wall direction, e.g. it is periodic in $x$ for $l={\rm ho}$, and  is exponentially localized about the domain wall in the transverse direction. The  complex amplitude $A_{n,i}$ is assumed to be constant in the region of the cavity but its value,  to be calculated using the scattering matrix approach and imposing the normalization condition Eq. (\ref{eq:normmech}),  may depend on the side of the polygon-shaped domain wall labeled by the second index $i$, see below. 

Next, we define  the vacuum 1D optomechanical coupling (per unit cell) $g_{0,l}^{(1D)}(\Omega)$ (the quantity plotted in Fig. 2 of the Main Text) as the cavity shift in the presence of a 1D Bloch wave $\mathbf{u}_{n,l,\Omega}(\mathbf{r})$ of amplitude set by the normalization condition
\begin{equation}\label{eq:norm_Bloch}
\int_{U}|\mathbf{u}_{n,l,\Omega_n}(\mathbf{r})|^{2}\rho(\mathbf{r}) d^{3}\mathbf{r}=\frac{\hbar}{2\Omega_{n}},
\end{equation}
where U indicates the unit cell of the strip. This definition has the merit to be independent of the size of the system (length of the strip) and is, thus, suitable to be computed using finite element simulations.  We note that we can fix the phase of $\mathbf{u}_{n,l,\Omega}(\mathbf{r})$ by requiring that $g_{0,l}^{(1D)}(\Omega)$ is real. Using this definition and substituting Eq. (\ref{eq:Q_nofu}) into Eq. (\ref{eq:g_vacuum}), we find
\begin{equation}\label{eq:g0n}
g_{0n}(s)=g_{0,l}^{(1D)}(\Omega_n) 2|A_{n,i}|\cos(k_l(\Omega)s+\arg(A_{n,i}))
\end{equation}
where $s$ is the position of the cavity.
Thus, the task of calculating the optomechanical coupling reduces to the task of calculating the amplitude $A_{n,i}$. 
%(Below we show that the amplitude $A_{n,i}$ actually depends only on the side of the polygon-shaped domain wall.) 
This is done by noting that using Eq. (\ref{eq:Q_nofu}), we can identify the amplitude of the clockwise and anti-clockwise  mechanical energy fluxes with
\begin{equation}\label{eq:current_of_A}
 j_\circlearrowright (s)=\left(v_l\frac{\hbar\Omega_n}{2a_m}\right)^{1/2}A_{n,i}e^{ik_l(\Omega)s} ,\quad  j_\circlearrowleft (s)=\left(v_l\frac{\hbar\Omega_n}{2a_m}\right)^{1/2}A_{n,i}e^{-ik_l(\Omega)s},
\end{equation}
respectively. Thus, we can calculate $A_{n,i}$ from Eq. (\ref{eq:T_loop_pb}) modulus a normalization factor (the global phase of the solution of Eq. (\ref{eq:T_loop_pb}) does not have a physical meaning and is fixed by requiring that $j_\circlearrowright (s)=j^*_\circlearrowleft (s)$. This is always possible because of the time-reversal symmetry).  The normalization factor is fixed by Eq. (\ref{eq:normmech}). Using Eq. (\ref{eq:Q_nofu}) and Eq. (\ref{eq:norm_Bloch}), the latter constraint can be rewritten  as 
\begin{equation}\label{eq:norm_of_A}
1=\sum_i 2|A_{n,i}|^2N_i,
%=  \sum_j 4\frac{a_m}{\hbar\Omega}\frac{|j_i|^2}{v_{l_i}}N_i  .
\end{equation}
where  $N_i$ is the length of the side $i$ (in number of unit cells). 

For the case of zero backscattering the flux $|j_\circlearrowright(s)|$ is constant, cf.~Eq. (\ref{eq:T_loop_pb}).  Thus, from Eq. (\ref{eq:current_of_A}) and Eq. (\ref{eq:norm_of_A}) we find 
\begin{equation}
%1= 2|A_{n,{\rm ho}|^2(N_{\rm ho}+N_{\rm sl}\frac{v_{\rm ho}}{v_{\rm sl}},
%=  \sum_j 4\frac{a_m}{\hbar\Omega}\frac{|j_i|^2}{v_{l_i}}N_i  .
|A_{n,{\rm ho}}|=\left(\frac{a_mv_{\rm sl}}{2(L_{\rm ho}v_{\rm sl}+L_{\rm sl}v_{\rm ho})}\right)^{1/2},\quad
|A_{n,{\rm sl}}|=\left(\frac{v_{\rm ho}}{v_{\rm sl}}\right)^{1/2}|A_{n,{\rm ho}}|.
\end{equation} 
where $|A_{n,{\rm ho}}|$  ($|A_{n,{\rm sl}}|$) is the  amplitude on all  horizontal (slanted) sides.  We note that for a pair of degenerate solutions have  the same amplitudes, $|A_{n,{\rm l}}|=|A_{n+1,{\rm l}}|$, while $|\arg(A_{n,{\rm l}}/A_{n+1,{\rm l}})|=\pi/2$ to ensure that the two solutions are orthogonal.

It is interesting to estimate the typical amplitude of the zero-point and the thermal fluctuations in our topological cavity. For the zero point fluctuations we have \cite{Safavi-Naeini2014}
\begin{equation}
    x_{\rm zpf,n}={\rm Max}[ \mathbf{Q}_n(\mathbf{r})]_V\approx |A_{n,{\rm ho}}|{\rm Max} [\mathbf{u}_{n,l,\Omega_n}(\mathbf{r})]_U\approx N^{-1/2} {\rm Max} [\mathbf{u}_{n,l,\Omega_n}(\mathbf{r})]_U
\end{equation}
We calculate the quantity ${\rm Max} [\mathbf{u}_{n,l,\Omega_n}(\mathbf{r})]_U$  using FEM simulations. It turns out that ${\rm Max} [\mathbf{u}_{n,l,\Omega_n}(\mathbf{r})]_U\sim 1$fm. Taking into account that for our devices  $N\sim 100$ we find 
\begin{equation}
x_{\rm zpf,n}\sim 0.1{\rm fm}.
\end{equation} 
We can then readily find the typical  amplitude of the thermal vibrations in our experiment to be 
\begin{equation}
x_{\rm th}\sim \sqrt{\frac{k_BT}{\hbar\bar{\Omega}}}x_{\rm zpf,n}\sim 10 {\rm fm}.  
\end{equation}

\subsection*{Fitting parameters}
In calculating the spectrum $\Omega_n$ and the corresponding optomechanical couplings $g_{0n}$, we have made the simplifying assumption that the propagation along the domain wall is similar as the propagation in an infinitely long domain wall and abruptly switch to the propagation in a domain wall with a different orientation  after turning a sharp angle. This allows for a simple theoretical description but is not entirely realistic  in the region close to the corners. A more realistic point of view is that Eq. (\ref{T-matrix_waveguide}) is valid only away from the corners and Eq.(\ref{T-matrix}) describes the propagation across a finite region about the corners. In view of this physical interpretation, we replace in our  calculation of the noise spectrum the lengths of the sides of the polygon-shaped domain wall  with  effective lengths  that are determined using the total length of the domain wall  as a fitting parameter and rescaling accordingly the single side lengths. 

For the plots where we have assumed perfect transmission we take also the overall phase acquired by crossing the three corners (for the triangle $\arg(t_{\rm sl, sl}+2t_{\rm sl, ho})$) as a fitting parameter. 

In addition we have used a uniform shift of the band structure as a fitting parameter. This is justified because,  within the topological  bandwidth of our devices, the main effect  of a rescaling of the whole band structure by as little as $\sim 1\%$  would be a uniform shift  of the order of the free spectral range. This could be  caused by any residual  mismatch between the fabrication and the nominal parameters used for our FEM simulations. Indeed, we have observed a uniform shift of the band structure  even  post fabrication comparing measurement taken on the same device in different days. We attribute this drift, that tends to saturate after some time, to oxidation of the device surface.  
%We emphasize that this fitting parameter does not change qualitatively the spectrum but only allows a uniform rescaling.
%
\begin{figure}
\begin{center}
\includegraphics[width=1\columnwidth]{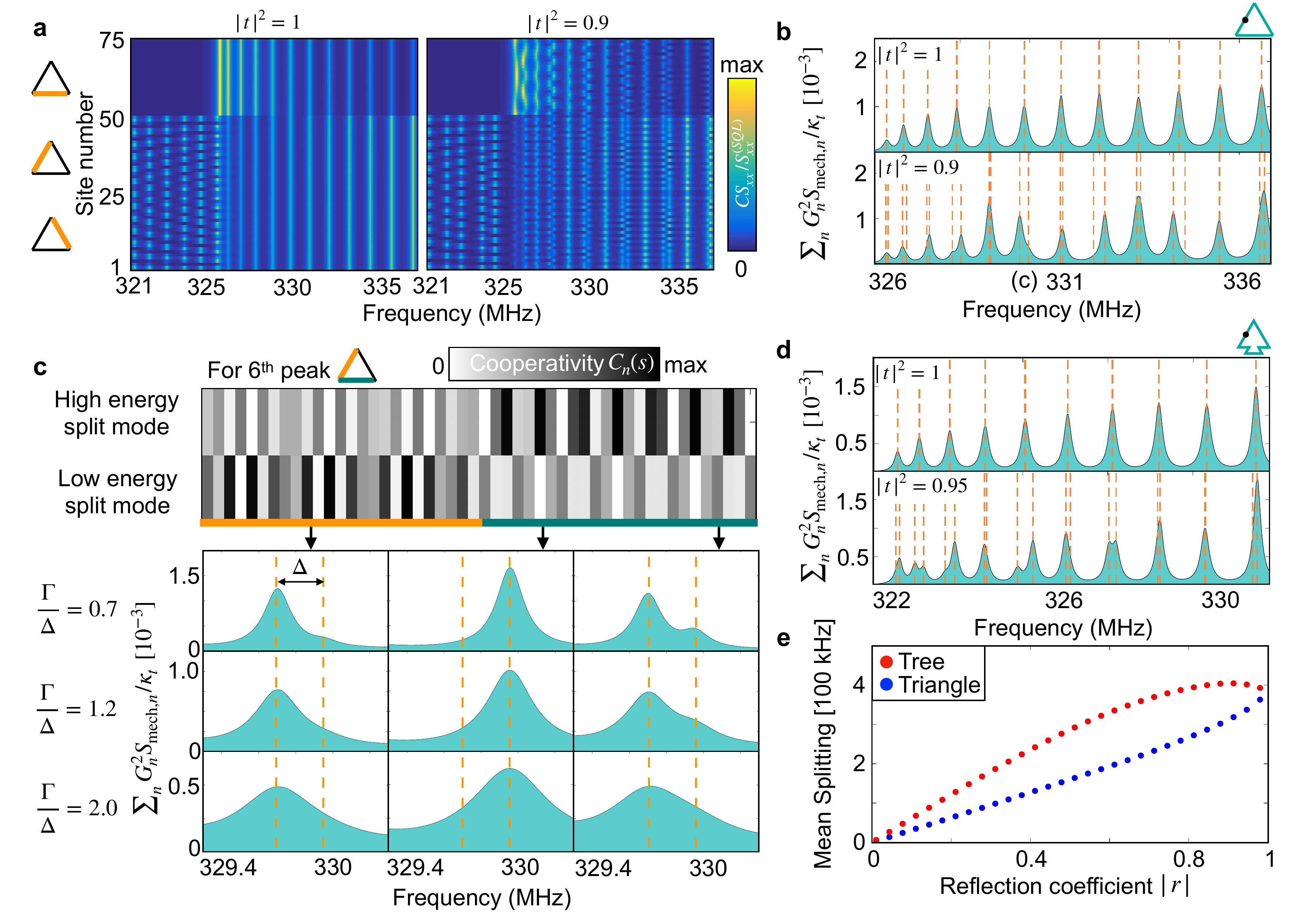}
\caption{\textbf{Signatures of backscattering on the spectrum.} 
\textbf{a}, 
Theoretically predicted spectra on all the sites of the triangular topological mechanical cavity (shown in the form of a density plot). For non-vanishing backscattering $\vert t\vert ^2=0.9$, the spectra become position-dependent, i.e. they are no longer identical for all the sites on the slanted or the horizontal waveguide.
\textbf{b} (\textbf{d}), 
Local spectrum for triangular (tree) topological mechanical cavity at the middle of the long slanted waveguide, in the absence and presence of small backscattering. The orange dashed lines indicate the eigenvalues. In the presence of small backscattering, the degenerate peaks are split and the amplitudes of the peaks are modulated.
\textbf{c}, 
Intensity mode profile (displayed in terms of optomechanical cooperativity) of the two split modes of the 6\textsuperscript{th} doublet peak for the triangular topological mechanical cavity. Resulting spectrum around this peak at different sites for small, intermediate and large values of mechanical damping $\Gamma$.
\textbf{e},
Comparison of the mean splitting for the triangular and tree topological mechanical cavities. The splitting is larger for the tree cavity because it has more corners than the triangular cavity.
}
\label{SI_backscattering}
\end{center}
\end{figure}

\subsection*{Effect of finite backscattering on the spectrum}

Here, we study the effect of backscattering on the NPSD. As discussed above in order to calculate the NSPD we first have to calculate the eigenfrequencies $\Omega_n$ and the optomechanical couplings $g_{0n}$ in the transfer matrix approach using Eqs. (\ref{eq:T_loop_pb},\ref{eq:norm_of_A},\ref{eq:g0n}). Then, we can plug these quantities into the general expression Eq. \ref{eq:noise_spectrum}. For simplicity we always assume $t_{\rm sl, sl}=t_{\rm sl, ho}=t$.

Fig. \ref{SI_backscattering}(a) shows the triangle NPSD for $|t|^2=0.9$ (right).
For comparison, we also show  the results calculated assuming perfect transmission,  $|t|^2=1$ (left). 
%As discussed above, in the backscattering free scenario  the spectrum is formed by a series of doublets that give rise to a single NPSD peak, cf.~the top panels of  Fig. \ref{SI_backscattering} (b) and (d), whose height is position independent, cf.~the high frequency region of Fig \ref{SI_backscattering} (a). Conversely, when a significant backscattering is intro

In the presence of backscattering $\left(\vert r\vert ^2\neq 0\right)$, the underlying doublets $\Omega_n$ and $\Omega_{n+1}$  (indicated by orange lines in Fig \ref{SI_backscattering} (b) and (d)) are split. The splitting averaged over all doublets as a function of  $|r|$ is plot in Fig. \ref{SI_backscattering} (e). We note that the average splitting is always larger for the tree-shaped topological cavity than for the triangular  cavity. This should be expected because of the larger number of corners acting as scattering centers for the tree geometry. The splitting of the underlying doublets changes the lineshape of the NPSD. The position dependence of the lineshape is governed by the ratio of the splitting and the decay rate and is studied  for the $6$-th peak in \ref{SI_backscattering} (c).
%We notice in Figs \ref{SI_backscattering} (b) and (d) that the eigenvalues (indicated by orange lines) does not have a one-one correspondence with the peak positions in the spectrum. This is explained in Fig \ref{SI_backscattering} (c), where we illustrate the position dependence of the spectrum near the 6\textsuperscript{th} degenerate peak of the triangular topological cavity. The intensity mode profiles of the two split eigenvalues are calculated using Eq.~\ref{eq:amp_displacement}.
 We note that for splittings $\Delta$ similar to the decay rate  $\Gamma$ (top and center panel), we see either one or two peaks in the spectrum, depending on the position dependence of the intensity profiles of the two modes (via the position dependent  optomechanical couplings, cf.~Eq. (\ref{eq:g0n})) at the measurement location. On the other hand, in the regime of our experiment where the splitting $\Delta$ is smaller than the decay rate $\Gamma$, there is always a single peak. In this regime  the residual small splitting $\Delta$  is revealed by what looks like a position dependent drift of the peak location.

%-----------------------------------------------------

\section{Calibration of frequency drift and normalization of NPSD spectra of the mechanical topological cavity}
\label{sec:normlization}

\begin{figure}[h]
\begin{center}
\includegraphics[width=\columnwidth]{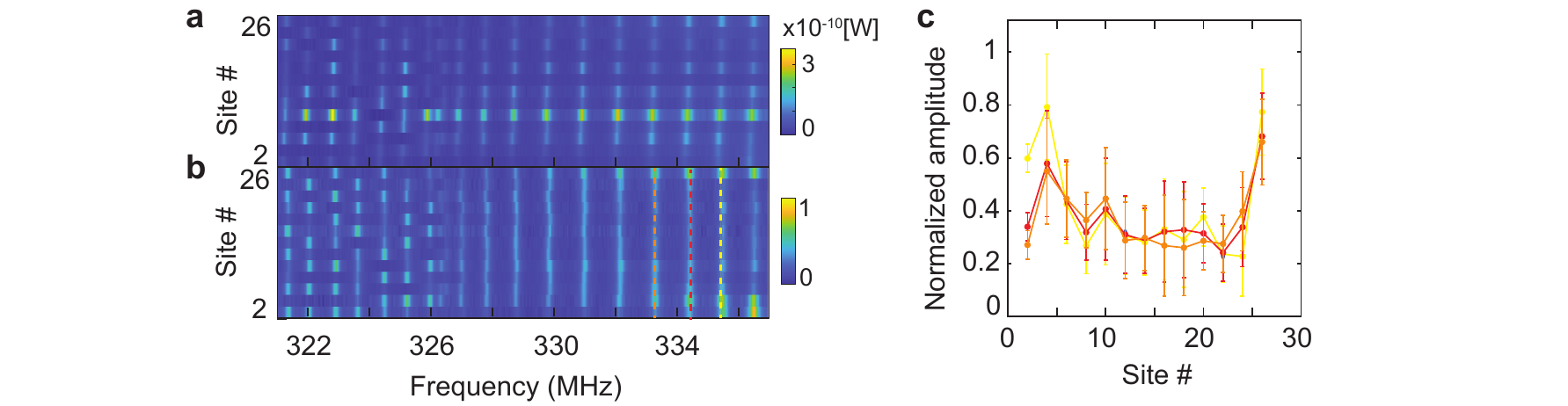}
\caption{\textbf{Calibration of frequency drift and normalization of NPSD spectra of the mechanical topological cavity}
\textbf{a}, Density plot of the measured NPSD before frequency drift calibration amplitude normalization, as a function of frequency and read-out position along a slanted edge.
\textbf{b}, Density plot of the same measured NPSD after frequency drift calibration and amplitude normalization. 
\textbf{c}, Normalized peak amplitudes verses measurement locations for 3 peaks, corresponding frequencies are indicated with dashed lines in \textbf{b}. Error bars correspond to the standard deviation of the 5 peak amplitudes used for the  amplitudes normalization.
}
\label{figNoormalization}
\end{center}
\end{figure}

We have observed a uniform shift of the band structure during measurements comparing spectra taken on the same device on different days. We attribute this drift, that tends to saturate after some time, to surface oxidation of the silicon device. In addition, the total optical loss $\kappa$, external coupling efficiency $\kappaE/\kappa$, as well as the total optical optical power delivered to each optical cavity during measurements are different, hence the amplitude of 
the optically transduced mechanical NPSD received on spectrum analyzer are different for each site. Therefore, we have applied spectra normalization, as discussed below, for the Main Text figures Fig.2 and Fig.3.

The frequency drift is calibrated using the two peaks between $323$~MHz and $325$~MHz. The frequency spectra of all the sites are shifted such that these two peaks align as well as possible with the corresponding two peaks at the (arbitrarily selected) reference site $\#26$. To illustrate this procedure, the raw spectra are plotted in Fig.~\ref{figNoormalization}a; it can be seen that there is a drift in mechanical frequency to lower frequencies from site $\#26$ to site $\#2$. In Fig.~\ref{figNoormalization}b, the frequency drifts have been removed by the calibration.

The amplitudes of the measured spectra are also normalized. Essentially, there are differences in the intensities between measurements at different sites due to these sites being measured individually in succession, with different $\kappa$,  $\kappaE/\kappa$, and total optical optical power delivered to the optical cavity during measurements. These spurious differences (that are not connected to the intrinsic physics of the device) are removed by calibration in the following manner.
The five highest peaks in the grey low-frequency region (cf.~Main Text Fig. 1i) are averaged, and the mean value is used to normalize each spectrum. The theoretical expectations for the NPSD as a function of frequency and read-out position on a slanted edge are plotted in Fig.~\ref{SI_backscattering}a. 
Figure~\ref{figNoormalization} shows the normalized peak amplitudes versus measurement location for 3 peaks. It can be seen that the normalized NPSD near the center of the slanted side of the triangle mechanical topological cavity are constant within error.

We remark briefly on why we have chosen our calibration procedure in this way. Our aim in this site-resolved measurement was to test experimentally whether the high-frequency region indeed harbors running wave modes that have little to no backscattering, as expected from the topological nature of the edge channels. Such modes would then show up with an intensity that is independent of position (in the idealized absence of changes of coupling strength between the measurements at different sites). In order to test this hypothesis without bias, we decided against normalizing the overall intensity based on peaks in this high-frequency region (because this could be viewed as enforcing at least an average tendency towards location-independent intensities). We rather took peaks in the low-frequency region as a reference for the normalization. Since in that region we expect standing waves (due to backscattering at the ends of the edge), each individual frequency peak is already strongly location-dependent in its intensity. To avoid hampering the overall normalization by this fact, we averaged over the five highest peaks.

\section{Calibration of the optomechanical coupling rate}
\label{sec:g0}

% The method similar to the calibration of g0 we used, I will update later

The optomechanical coupling of the between localized optical resonance and mechanical resonance can be estimated using two different methods. 
The first method, the increase in the mechanical linewidth of each mechanical mode can be fitted as a function of optical power to find the optomechanical coupling for each mode. However, the increased mechanical linewidth of our devices is on the order of few hundred Hz ($\gammaOM < 1$~kHz), even with large number of cavity photons ($\ncav \approx 10000$), which is small compared to the intrinsic mechanical linewidth ($\gammai \approx 200$~kHz).
The second method involves calibration of the optical powers and electronic detection system, and uses the fact that the transduced thermal Brownian motion of the mechanical resonator is proportional to $g_\text{0n}^2$.

To calibrate the detection efficiency of the setup we first measure the efficiency of transmission from laser to the input power of the dimpled fiber taper. These values are measured once when the optical components are connected and do not change.
%in the circulator from port 1 to port 2 ($\eta_{\text{12}} = 88 \%$), and from port 2 to port 3 ($\eta_{\text{23}} = 84 \%$). These values are measured once when the optical components are connected and do not change. These calibrations are used to determine the reflection efficiency of the device and the overall detection efficiency of the heterodyne setup. 
To measure the efficiency of dimpled fiber taper, the laser is tuned off-resonance from the optical mode (where the device optical transmission should be flat) and a continuous-wave signal of input power $P_{\text{in}}$ is sent into input port of the dimpled fiber taper. 
%leading to a power $\eta_{\text{taper}}^2 P_{\text{in}}$ exiting output of the taper.
The optical losses incurred in the path input port of taper to the device-under-test are accumulated into an efficiency factor $\eta_{\text{taper}}$. These losses are incurred twice in transmission to the output of the taper (loss in the input side of taper is assumed to be the same as output side), so a power of $\eta_{\text{taper}}^2 P_{\text{in}}$ propagates out of output port of the taper. This signal is sent to a power meter (PM), and thus the coupling efficiency of taper is determined as $\eta_{\text{taper}} = \sqrt{P_{\text{PM}}/P_{\text{in}}} = 51 \%$ .

To calibrate the overall detection efficiency ($\eta_{\text{det}}$), we must also determine the efficiency of the rest of the detection path and detector.
%the heterodyne detector itself, which includes the intrinsic quantum efficiency of the BPD, the alignment of the polarization between the LO and the signal, and the degree to which the LO power overcomes the electronic noise of the detector. 
This is accomplished by using the amplitude modulator to create optical sidebands detuned from the signal by the mechanical frequency while the laser is tuned off-resonance from the optical mode. 
%The optical switches SW1 and SW2 are used to route the signal through a tunable filter to select a single sideband which is sent through the device and onto the BPD. 
The power $P_{\text{cal}}$ in this sideband is calculated using $V_{\pi}$ of EOM and $\eta_{\text{taper}}$.
%directly measured on the PM at SW3, 
The photocurrent NPSD ($S_{\text{pc}}[\omega]$) as transduced on the spectrum analyser is given by 

\begin{equation}
S_{\text{pc}}[\omega] = S_{\text{dark}} + \frac{G_{\text{e}}^2 }{R_{\text{L}}}S_{\text{SN}}^2 \left( 1 + \frac{\eta_{\text{det}}S_{\text{cal}}[\omega]}{\hbar\omega_{\text{o}}}  \right) ,
\end{equation}

\noindent where $S_{\text{dark}}[\omega]$ is the electronic NPSD of the detector, $S_{\text{SN}} = \sqrt{2\hbar\omega_{\text{o}}P_{\text{drive}}}$ is the optical shot-noise NPSD arising from driving optical power at optical frequency $\omega_{\text{l}}$, which lies an order of magnitude above the electronic noise, and $S_{\text{cal}}$ is the NPSD of the signal, where $\int_{-\infty}^{\infty}  S_{\text{cal}}[\omega] \frac{d\omega}{2\pi} = \eta_{\text{taper}}^2 P_{\text{cal}}$. 
The gain factor $G_{\text{e}}$ represents the conversion from optical power to voltage while $R_{\text{L}}$ is the input impedance of the spectrum analyser. 
The total noise floor $S_{\text{noise}} = \frac{G_{\text{e}}^2 }{R_{\text{L}}}S_{\text{SN}}^2+ S_{\text{dark}}$ is measured with the EOM drive turned off (no optical sidebands), while $S_{\text{dark}}$ is measured independently with both signal and drive laser beams blocked (laser is blocked at BOA). The calibration tone (with NPSD $S_{\text{cal}}[\omega]$) picks up losses in the optical setup (fibers, fiber unions), fast optical detector, and microwave cable, which are parametrized into $\eta_{\text{det}}$. The efficiency of the thermal vibration NPSD detection path is extracted as 

\begin{equation}
\eta_{\text{det}}  = \frac{\hbar\omega_{\text{o}}}{\eta_{\text{taper}}^2P_{\text{cal}}} \int_{-\infty}^{\infty} \frac{S_{\text{pc}}[\omega] - S_{\text{noise}}}{S_{\text{noise}}- S_{\text{dark}}} \frac{d\omega}{2\pi} = 9.7 \% .
\end{equation}

%Overall measurement efficiency $\eta$ used for calibrating optomechanical coupling rate based on mechanical thermometry can be written as

%\begin{equation}
%\eta = \eta_{\text{taper}} \eta_{\text{det}} = 4.95 \% %\text{.}
%\end{equation}

In order to calibrate $g_\text{0n}$, the fiber taper is parked on the device-under-test, which is one of the center sites of slanted edge of a tree-shaped cavity geometry in this calibration. The optical drive laser is locked to a blue detuning (from the optical cavity resonance) of $340$~MHz, and the optical drive power is tuned such that the same amount of power is received on the photodetector. With

\begin{equation}
S_{\text{pc}}[\omega] = S_{\text{dark}} + \frac{G_{\text{e}}^2 }{R_{\text{L}}}S_{\text{SN}}^2 \left( 1 + \frac{\eta_{\text{det}}S_{\text{II}}[\omega]}{\hbar\omega_{\text{o}}}  \right) ,
\end{equation}

\noindent where $S_{\text{II}}$ is the NPSD of optomechanically generated photons, given by Eq.~\ref{eq:noise_spectrum}. For the tree-shaped topological cavity, we get calibrated $g_\text{0n}/2\pi = 2.56$~kHz for one peak ($328.21$~MHz) within topological bandwidth, which is in good agreement with the theoretically predicted value (Eq.~\ref{eq:g0n}) for the same peak, $g_\text{0n}/2\pi = 2.25$~kHz. We note that this agreement implies that we are indeed observing thermal motion ($S_{\text{II}}$ in Eq.~\ref{eq:noise_spectrum} is calculated for the room temperature $300$~K).

\section{Trivial waveguide design}
\label{sec:triv_waveguide}
\begin{figure}[h]
\begin{center}
\includegraphics[width=\columnwidth]{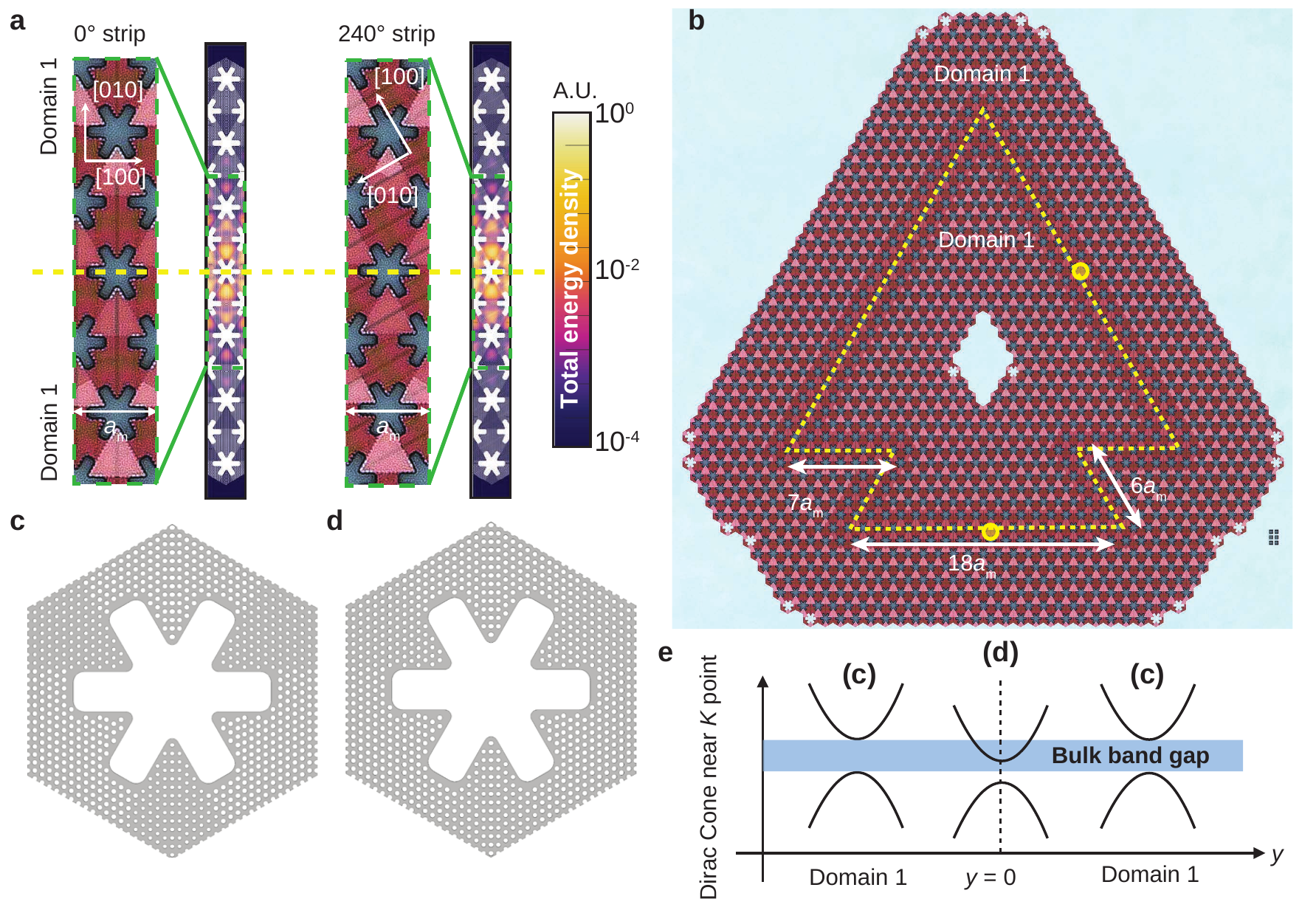}
\caption{\textbf{Trivial waveguide}
\textbf{a}, Optical microscope images and simulated mechanical mode profiles of the strip unit cell for $0^{\circ}$ (horizontal) or $240^{\circ}$ (slanted) vs. the silicon [100] crystal orientation. A defect trivial waveguide is built by modifying the radius of the holes in the upward-pointing triangles near the center three snowflakes.
%produced at the central defect interface of the two similar domains. 
%The two cases have the domain wall oriented by
%1-D edge channel CHANGE NAME produced at the boundary (red dotted line) of two topologically inequivalent domains that are related to each other by mirror symmetry. We show a transverse cross-section of the geometry (optical microscope image) and of the resulting phononic edge channel mode profile. Triangles with an optical cavity are dark red. The two cases have the domain wall oriented by $0^{\circ}$ (horizontal, \textbf{g}) or $240^{\circ}$ (slanted, \textbf{h}) vs. the silicon [100] crystal orientation.
\textbf{b}, Optical microscope image of the tree-shaped trivial mechanical cavity. The dimensions are similar to that of topological mechanical cavity in Fig. 3(b) of Main Text.
% Each side of the trivial mechanical, with the each slanted edge consists of 32 unit cells, and horizontal edge consist of $7 + 18 + 7$ unit cells. 
%The upward-pointing triangles are always light-red on the two sides of the defect region, thus implying that there is no domain inversion.
The center of trivial waveguide is indicated by dashed lines. Mechanical NPSD is measured at the horizontal and slanted edges of tree geometry as indicated by the yellow dots.
\textbf{c}, Unit cell in domain 1.
%, regions away from the mode localization region. 
\textbf{d}, Unit cell of the center three snowflakes. 
%By tuning the upper band of the gapped Dirac cone inside the bulk band gap. 
\textbf{e}, The Dirac cone bands of center three snowflakes are tuned into the bulk band gap by changing the size of the circular holes in the upward-pointing triangles.
%scaling down the radius of the optical cavity holes in the upward-pointing triangles in comparison to the downward-pointing triangles.
}
\label{figTrival}
\end{center}
\end{figure}

In this section, we describe the design of the trivial waveguide, which is used in Fig. 3 of Main Text to show the effects of backscattering at the sharp corners. 
The strip unit cell is built by locally tuning the upper band of the gapped Dirac cone of the central region to be inside of the bulk bandgap of the two-dimensional snowflake phononic crystal (see Fig.~\ref{figTrival} (a) and (e)). Note that there is no band inversion in this strip unit cell at its center.
Trivial waveguide is formed in the center three snowflakes by the upper band of the gapped Dirac cone, corresponding optical microscope images and simulated mechanical mode profiles of $0^\circ$ and $240^\circ$ of the strip unit cells vs. Silicon [100] are shown in Fig.~\ref{figTrival}(a).
%an be used to create the phononic waveguide. 
%We create this defect ,  cf.~Fig \ref{figTrival}(d). 
%Note that there is no band inversion, and hence, it is a trivial waveguide. 
%The geometry of the waveguide unit cell for the two different orientations are shown in Fig \ref{figTrival}(a). 
In the center three snowflake unit cells, the circular holes in the upward-pointing triangles (light red) 
%at the localization region 
are scaled smaller by a factor of 0.92 in comparison to the downward-pointing triangles (see Fig~\ref{figTrival}(d)). While in domain 1, circular holes in the upward-pointing triangles are scaled smaller by a factor of 0.78 (see Fig.~\ref{figTrival}(c)).
%(0.78), cf.~Fig \ref{figTrival}(c) and (d). 
Larger circular holes shifts the gapped Dirac cones lower in energy because of the decreasing stiffness (see Fig~\ref{figTrival}(e)). An optical microscope image of the tree geometry is also shown in Fig \ref{figTrival}(b). The modes of the tree shaped trivial mechanical cavity are optically readout at the horizontal and slanted edges of the tree geometry as indicated by yellow dots.
%The transverse localization length of the waveguide mode is approximately 3 unit cells away from the center, cf.~Fig \ref{figTrival}(a). 

%\bibliographystyle{ieeetr}
%\bibliographystyle{naturemag}
%\bibliography{Biblio_v1}

\begin{thebibliography}{53}%
\makeatletter
\providecommand \@ifxundefined [1]{%
 \@ifx{#1\undefined}
}%
\providecommand \@ifnum [1]{%
 \ifnum #1\expandafter \@firstoftwo
 \else \expandafter \@secondoftwo
 \fi
}%
\providecommand \@ifx [1]{%
 \ifx #1\expandafter \@firstoftwo
 \else \expandafter \@secondoftwo
 \fi
}%
\providecommand \natexlab [1]{#1}%
\providecommand \enquote  [1]{``#1''}%
\providecommand \bibnamefont  [1]{#1}%
\providecommand \bibfnamefont [1]{#1}%
\providecommand \citenamefont [1]{#1}%
\providecommand \href@noop [0]{\@secondoftwo}%
\providecommand \href [0]{\begingroup \@sanitize@url \@href}%
\providecommand \@href[1]{\@@startlink{#1}\@@href}%
\providecommand \@@href[1]{\endgroup#1\@@endlink}%
\providecommand \@sanitize@url [0]{\catcode `\\12\catcode `\$12\catcode
  `\&12\catcode `\#12\catcode `\^12\catcode `\_12\catcode `\%12\relax}%
\providecommand \@@startlink[1]{}%
\providecommand \@@endlink[0]{}%
\providecommand \url  [0]{\begingroup\@sanitize@url \@url }%
\providecommand \@url [1]{\endgroup\@href {#1}{\urlprefix }}%
\providecommand \urlprefix  [0]{URL }%
\providecommand \Eprint [0]{\href }%
\providecommand \doibase [0]{http://dx.doi.org/}%
\providecommand \selectlanguage [0]{\@gobble}%
\providecommand \bibinfo  [0]{\@secondoftwo}%
\providecommand \bibfield  [0]{\@secondoftwo}%
\providecommand \translation [1]{[#1]}%
\providecommand \BibitemOpen [0]{}%
\providecommand \bibitemStop [0]{}%
\providecommand \bibitemNoStop [0]{.\EOS\space}%
\providecommand \EOS [0]{\spacefactor3000\relax}%
\providecommand \BibitemShut  [1]{\csname bibitem#1\endcsname}%
\let\auto@bib@innerbib\@empty
%</preamble>
\bibitem [{\citenamefont {Aspelmeyer}\ \emph {et~al.}(2014)\citenamefont
  {Aspelmeyer}, \citenamefont {Kippenberg},\ and\ \citenamefont
  {Marquardt}}]{Aspelmeyer2014}%
  \BibitemOpen
  \bibfield  {author} {\bibinfo {author} {\bibfnamefont {M.}~\bibnamefont
  {Aspelmeyer}}, \bibinfo {author} {\bibfnamefont {T.~J.}\ \bibnamefont
  {Kippenberg}}, \ and\ \bibinfo {author} {\bibfnamefont {F.}~\bibnamefont
  {Marquardt}},\ }\href@noop {} {\bibfield  {journal} {\bibinfo  {journal}
  {Rev. Mod. Phys.}\ }\textbf {\bibinfo {volume} {86}},\ \bibinfo {pages}
  {1391} (\bibinfo {year} {2014})}\BibitemShut {NoStop}%
\bibitem [{\citenamefont {de~Groot}(2019)}]{deGroot2019}%
  \BibitemOpen
  \bibfield  {author} {\bibinfo {author} {\bibfnamefont {P.~J.}\ \bibnamefont
  {de~Groot}},\ }\href {\doibase 10.1088/1361-6633/ab092d} {\bibfield
  {journal} {\bibinfo  {journal} {Rep. Prog. Phys.}\ }\textbf {\bibinfo
  {volume} {82}},\ \bibinfo {pages} {056101} (\bibinfo {year}
  {2019})}\BibitemShut {NoStop}%
\bibitem [{\citenamefont {Massel}\ \emph {et~al.}(2012)\citenamefont {Massel},
  \citenamefont {Cho}, \citenamefont {Pirkkalainen}, \citenamefont {Hakonen},
  \citenamefont {Heikkil{\"a}},\ and\ \citenamefont
  {Sillanp{\"a}{\"a}}}]{massel2012multimode}%
  \BibitemOpen
  \bibfield  {author} {\bibinfo {author} {\bibfnamefont {F.}~\bibnamefont
  {Massel}}, \bibinfo {author} {\bibfnamefont {S.~U.}\ \bibnamefont {Cho}},
  \bibinfo {author} {\bibfnamefont {J.-M.}\ \bibnamefont {Pirkkalainen}},
  \bibinfo {author} {\bibfnamefont {P.~J.}\ \bibnamefont {Hakonen}}, \bibinfo
  {author} {\bibfnamefont {T.~T.}\ \bibnamefont {Heikkil{\"a}}}, \ and\
  \bibinfo {author} {\bibfnamefont {M.~A.}\ \bibnamefont {Sillanp{\"a}{\"a}}},\
  }\href@noop {} {\bibfield  {journal} {\bibinfo  {journal} {Nature
  communications}\ }\textbf {\bibinfo {volume} {3}},\ \bibinfo {pages} {1}
  (\bibinfo {year} {2012})}\BibitemShut {NoStop}%
\bibitem [{\citenamefont {Zhang}\ \emph {et~al.}(2015)\citenamefont {Zhang},
  \citenamefont {Shah}, \citenamefont {Cardenas},\ and\ \citenamefont
  {Lipson}}]{mian2015synchronization}%
  \BibitemOpen
  \bibfield  {author} {\bibinfo {author} {\bibfnamefont {M.}~\bibnamefont
  {Zhang}}, \bibinfo {author} {\bibfnamefont {S.}~\bibnamefont {Shah}},
  \bibinfo {author} {\bibfnamefont {J.}~\bibnamefont {Cardenas}}, \ and\
  \bibinfo {author} {\bibfnamefont {M.}~\bibnamefont {Lipson}},\ }\href
  {\doibase 10.1103/PhysRevLett.115.163902} {\bibfield  {journal} {\bibinfo
  {journal} {Phys. Rev. Lett.}\ }\textbf {\bibinfo {volume} {115}},\ \bibinfo
  {pages} {163902} (\bibinfo {year} {2015})}\BibitemShut {NoStop}%
\bibitem [{\citenamefont {Xu}\ \emph {et~al.}(2016)\citenamefont {Xu},
  \citenamefont {Mason}, \citenamefont {Jiang},\ and\ \citenamefont
  {Harris}}]{xu2016topological}%
  \BibitemOpen
  \bibfield  {author} {\bibinfo {author} {\bibfnamefont {H.}~\bibnamefont
  {Xu}}, \bibinfo {author} {\bibfnamefont {D.}~\bibnamefont {Mason}}, \bibinfo
  {author} {\bibfnamefont {L.}~\bibnamefont {Jiang}}, \ and\ \bibinfo {author}
  {\bibfnamefont {J.}~\bibnamefont {Harris}},\ }\href@noop {} {\bibfield
  {journal} {\bibinfo  {journal} {Nature}\ }\textbf {\bibinfo {volume} {537}},\
  \bibinfo {pages} {80} (\bibinfo {year} {2016})}\BibitemShut {NoStop}%
\bibitem [{\citenamefont {Kharel}\ \emph {et~al.}(2019)\citenamefont {Kharel},
  \citenamefont {Harris}, \citenamefont {Kittlaus}, \citenamefont {Renninger},
  \citenamefont {Otterstrom}, \citenamefont {Harris},\ and\ \citenamefont
  {Rakich}}]{kharel2019high_frequency}%
  \BibitemOpen
  \bibfield  {author} {\bibinfo {author} {\bibfnamefont {P.}~\bibnamefont
  {Kharel}}, \bibinfo {author} {\bibfnamefont {G.~I.}\ \bibnamefont {Harris}},
  \bibinfo {author} {\bibfnamefont {E.~A.}\ \bibnamefont {Kittlaus}}, \bibinfo
  {author} {\bibfnamefont {W.~H.}\ \bibnamefont {Renninger}}, \bibinfo {author}
  {\bibfnamefont {N.~T.}\ \bibnamefont {Otterstrom}}, \bibinfo {author}
  {\bibfnamefont {J.~G.~E.}\ \bibnamefont {Harris}}, \ and\ \bibinfo {author}
  {\bibfnamefont {P.~T.}\ \bibnamefont {Rakich}},\ }\href {\doibase
  10.1126/sciadv.aav0582} {\bibfield  {journal} {\bibinfo  {journal} {Science
  Advances}\ }\textbf {\bibinfo {volume} {5}} (\bibinfo {year} {2019}),\
  10.1126/sciadv.aav0582}\BibitemShut {NoStop}%
\bibitem [{\citenamefont {Ruesink}\ \emph {et~al.}(2016)\citenamefont
  {Ruesink}, \citenamefont {Miri}, \citenamefont {Alu},\ and\ \citenamefont
  {Verhagen}}]{ruesink2016nonreciprocity}%
  \BibitemOpen
  \bibfield  {author} {\bibinfo {author} {\bibfnamefont {F.}~\bibnamefont
  {Ruesink}}, \bibinfo {author} {\bibfnamefont {M.-A.}\ \bibnamefont {Miri}},
  \bibinfo {author} {\bibfnamefont {A.}~\bibnamefont {Alu}}, \ and\ \bibinfo
  {author} {\bibfnamefont {E.}~\bibnamefont {Verhagen}},\ }\href@noop {}
  {\bibfield  {journal} {\bibinfo  {journal} {Nature communications}\ }\textbf
  {\bibinfo {volume} {7}},\ \bibinfo {pages} {1} (\bibinfo {year}
  {2016})}\BibitemShut {NoStop}%
\bibitem [{\citenamefont {Peterson}\ \emph {et~al.}(2017)\citenamefont
  {Peterson}, \citenamefont {Lecocq}, \citenamefont {Cicak}, \citenamefont
  {Simmonds}, \citenamefont {Aumentado},\ and\ \citenamefont
  {Teufel}}]{peterson2017demonstration}%
  \BibitemOpen
  \bibfield  {author} {\bibinfo {author} {\bibfnamefont {G.~A.}\ \bibnamefont
  {Peterson}}, \bibinfo {author} {\bibfnamefont {F.}~\bibnamefont {Lecocq}},
  \bibinfo {author} {\bibfnamefont {K.}~\bibnamefont {Cicak}}, \bibinfo
  {author} {\bibfnamefont {R.~W.}\ \bibnamefont {Simmonds}}, \bibinfo {author}
  {\bibfnamefont {J.}~\bibnamefont {Aumentado}}, \ and\ \bibinfo {author}
  {\bibfnamefont {J.~D.}\ \bibnamefont {Teufel}},\ }\href@noop {} {\bibfield
  {journal} {\bibinfo  {journal} {Physical Review X}\ }\textbf {\bibinfo
  {volume} {7}},\ \bibinfo {pages} {031001} (\bibinfo {year}
  {2017})}\BibitemShut {NoStop}%
\bibitem [{\citenamefont {Bernier}\ \emph {et~al.}(2017)\citenamefont
  {Bernier}, \citenamefont {Toth}, \citenamefont {Koottandavida}, \citenamefont
  {Ioannou}, \citenamefont {Malz}, \citenamefont {Nunnenkamp}, \citenamefont
  {Feofanov},\ and\ \citenamefont {Kippenberg}}]{bernier2017nonreciprocal}%
  \BibitemOpen
  \bibfield  {author} {\bibinfo {author} {\bibfnamefont {N.~R.}\ \bibnamefont
  {Bernier}}, \bibinfo {author} {\bibfnamefont {L.~D.}\ \bibnamefont {Toth}},
  \bibinfo {author} {\bibfnamefont {A.}~\bibnamefont {Koottandavida}}, \bibinfo
  {author} {\bibfnamefont {M.~A.}\ \bibnamefont {Ioannou}}, \bibinfo {author}
  {\bibfnamefont {D.}~\bibnamefont {Malz}}, \bibinfo {author} {\bibfnamefont
  {A.}~\bibnamefont {Nunnenkamp}}, \bibinfo {author} {\bibfnamefont
  {A.}~\bibnamefont {Feofanov}}, \ and\ \bibinfo {author} {\bibfnamefont
  {T.}~\bibnamefont {Kippenberg}},\ }\href@noop {} {\bibfield  {journal}
  {\bibinfo  {journal} {Nature communications}\ }\textbf {\bibinfo {volume}
  {8}},\ \bibinfo {pages} {604} (\bibinfo {year} {2017})}\BibitemShut {NoStop}%
\bibitem [{\citenamefont {Fang}\ \emph {et~al.}(2017)\citenamefont {Fang},
  \citenamefont {Luo}, \citenamefont {Metelmann}, \citenamefont {Matheny},
  \citenamefont {Marquardt}, \citenamefont {Clerk},\ and\ \citenamefont
  {Painter}}]{fang2017generalized}%
  \BibitemOpen
  \bibfield  {author} {\bibinfo {author} {\bibfnamefont {K.}~\bibnamefont
  {Fang}}, \bibinfo {author} {\bibfnamefont {J.}~\bibnamefont {Luo}}, \bibinfo
  {author} {\bibfnamefont {A.}~\bibnamefont {Metelmann}}, \bibinfo {author}
  {\bibfnamefont {M.~H.}\ \bibnamefont {Matheny}}, \bibinfo {author}
  {\bibfnamefont {F.}~\bibnamefont {Marquardt}}, \bibinfo {author}
  {\bibfnamefont {A.~A.}\ \bibnamefont {Clerk}}, \ and\ \bibinfo {author}
  {\bibfnamefont {O.}~\bibnamefont {Painter}},\ }\href@noop {} {\bibfield
  {journal} {\bibinfo  {journal} {Nature Physics}\ }\textbf {\bibinfo {volume}
  {13}},\ \bibinfo {pages} {465} (\bibinfo {year} {2017})}\BibitemShut
  {NoStop}%
\bibitem [{\citenamefont {Xu}\ \emph {et~al.}(2019)\citenamefont {Xu},
  \citenamefont {Jiang}, \citenamefont {Clerk},\ and\ \citenamefont
  {Harris}}]{xu2019nonreciprocal}%
  \BibitemOpen
  \bibfield  {author} {\bibinfo {author} {\bibfnamefont {H.}~\bibnamefont
  {Xu}}, \bibinfo {author} {\bibfnamefont {L.}~\bibnamefont {Jiang}}, \bibinfo
  {author} {\bibfnamefont {A.}~\bibnamefont {Clerk}}, \ and\ \bibinfo {author}
  {\bibfnamefont {J.}~\bibnamefont {Harris}},\ }\href@noop {} {\bibfield
  {journal} {\bibinfo  {journal} {Nature}\ }\textbf {\bibinfo {volume} {568}},\
  \bibinfo {pages} {65} (\bibinfo {year} {2019})}\BibitemShut {NoStop}%
\bibitem [{\citenamefont {Mathew}\ \emph {et~al.}(2020)\citenamefont {Mathew},
  \citenamefont {Pino},\ and\ \citenamefont
  {Verhagen}}]{mathew_synthetic_2020}%
  \BibitemOpen
  \bibfield  {author} {\bibinfo {author} {\bibfnamefont {J.~P.}\ \bibnamefont
  {Mathew}}, \bibinfo {author} {\bibfnamefont {J.~d.}\ \bibnamefont {Pino}}, \
  and\ \bibinfo {author} {\bibfnamefont {E.}~\bibnamefont {Verhagen}},\ }\href
  {\doibase 10.1038/s41565-019-0630-8} {\bibfield  {journal} {\bibinfo
  {journal} {Nature Nanotechnology}\ }\textbf {\bibinfo {volume} {15}},\
  \bibinfo {pages} {198} (\bibinfo {year} {2020})}\BibitemShut {NoStop}%
\bibitem [{\citenamefont {Peano}\ \emph {et~al.}(2015)\citenamefont {Peano},
  \citenamefont {Brendel}, \citenamefont {Schmidt},\ and\ \citenamefont
  {Marquardt}}]{peano2015topological}%
  \BibitemOpen
  \bibfield  {author} {\bibinfo {author} {\bibfnamefont {V.}~\bibnamefont
  {Peano}}, \bibinfo {author} {\bibfnamefont {C.}~\bibnamefont {Brendel}},
  \bibinfo {author} {\bibfnamefont {M.}~\bibnamefont {Schmidt}}, \ and\
  \bibinfo {author} {\bibfnamefont {F.}~\bibnamefont {Marquardt}},\ }\href@noop
  {} {\bibfield  {journal} {\bibinfo  {journal} {Physical Review X}\ }\textbf
  {\bibinfo {volume} {5}},\ \bibinfo {pages} {031011} (\bibinfo {year}
  {2015})}\BibitemShut {NoStop}%
\bibitem [{\citenamefont {Brendel}\ \emph {et~al.}(2017)\citenamefont
  {Brendel}, \citenamefont {Peano}, \citenamefont {Painter},\ and\
  \citenamefont {Marquardt}}]{brendel2017pseudomagnetic}%
  \BibitemOpen
  \bibfield  {author} {\bibinfo {author} {\bibfnamefont {C.}~\bibnamefont
  {Brendel}}, \bibinfo {author} {\bibfnamefont {V.}~\bibnamefont {Peano}},
  \bibinfo {author} {\bibfnamefont {O.~J.}\ \bibnamefont {Painter}}, \ and\
  \bibinfo {author} {\bibfnamefont {F.}~\bibnamefont {Marquardt}},\ }\href@noop
  {} {\bibfield  {journal} {\bibinfo  {journal} {Proceedings of the National
  Academy of Sciences}\ }\textbf {\bibinfo {volume} {114}},\ \bibinfo {pages}
  {E3390} (\bibinfo {year} {2017})}\BibitemShut {NoStop}%
\bibitem [{\citenamefont {Brendel}\ \emph {et~al.}(2018)\citenamefont
  {Brendel}, \citenamefont {Peano}, \citenamefont {Painter},\ and\
  \citenamefont {Marquardt}}]{brendel2018snowflake}%
  \BibitemOpen
  \bibfield  {author} {\bibinfo {author} {\bibfnamefont {C.}~\bibnamefont
  {Brendel}}, \bibinfo {author} {\bibfnamefont {V.}~\bibnamefont {Peano}},
  \bibinfo {author} {\bibfnamefont {O.}~\bibnamefont {Painter}}, \ and\
  \bibinfo {author} {\bibfnamefont {F.}~\bibnamefont {Marquardt}},\ }\href@noop
  {} {\bibfield  {journal} {\bibinfo  {journal} {Physical Review B}\ }\textbf
  {\bibinfo {volume} {97}},\ \bibinfo {pages} {020102} (\bibinfo {year}
  {2018})}\BibitemShut {NoStop}%
\bibitem [{\citenamefont {Sanavio}\ \emph {et~al.}(2020)\citenamefont
  {Sanavio}, \citenamefont {Peano},\ and\ \citenamefont
  {Xuereb}}]{sanavio2020}%
  \BibitemOpen
  \bibfield  {author} {\bibinfo {author} {\bibfnamefont {C.}~\bibnamefont
  {Sanavio}}, \bibinfo {author} {\bibfnamefont {V.}~\bibnamefont {Peano}}, \
  and\ \bibinfo {author} {\bibfnamefont {A.}~\bibnamefont {Xuereb}},\ }\href
  {\doibase 10.1103/PhysRevB.101.085108} {\bibfield  {journal} {\bibinfo
  {journal} {Phys. Rev. B}\ }\textbf {\bibinfo {volume} {101}},\ \bibinfo
  {pages} {085108} (\bibinfo {year} {2020})}\BibitemShut {NoStop}%
\bibitem [{\citenamefont {Teufel}\ \emph {et~al.}(2009)\citenamefont {Teufel},
  \citenamefont {Donner}, \citenamefont {Castellanos-Beltran}, \citenamefont
  {Harlow},\ and\ \citenamefont {Lehnert}}]{teufel2009nanomechanical}%
  \BibitemOpen
  \bibfield  {author} {\bibinfo {author} {\bibfnamefont {J.~D.}\ \bibnamefont
  {Teufel}}, \bibinfo {author} {\bibfnamefont {T.}~\bibnamefont {Donner}},
  \bibinfo {author} {\bibfnamefont {M.}~\bibnamefont {Castellanos-Beltran}},
  \bibinfo {author} {\bibfnamefont {J.~W.}\ \bibnamefont {Harlow}}, \ and\
  \bibinfo {author} {\bibfnamefont {K.~W.}\ \bibnamefont {Lehnert}},\
  }\href@noop {} {\bibfield  {journal} {\bibinfo  {journal} {Nature
  nanotechnology}\ }\textbf {\bibinfo {volume} {4}},\ \bibinfo {pages} {820}
  (\bibinfo {year} {2009})}\BibitemShut {NoStop}%
\bibitem [{\citenamefont {Wilson}\ \emph {et~al.}(2015)\citenamefont {Wilson},
  \citenamefont {Sudhir}, \citenamefont {Piro}, \citenamefont {Schilling},
  \citenamefont {Ghadimi},\ and\ \citenamefont
  {Kippenberg}}]{wilson2015measurement}%
  \BibitemOpen
  \bibfield  {author} {\bibinfo {author} {\bibfnamefont {D.}~\bibnamefont
  {Wilson}}, \bibinfo {author} {\bibfnamefont {V.}~\bibnamefont {Sudhir}},
  \bibinfo {author} {\bibfnamefont {N.}~\bibnamefont {Piro}}, \bibinfo {author}
  {\bibfnamefont {R.}~\bibnamefont {Schilling}}, \bibinfo {author}
  {\bibfnamefont {A.}~\bibnamefont {Ghadimi}}, \ and\ \bibinfo {author}
  {\bibfnamefont {T.~J.}\ \bibnamefont {Kippenberg}},\ }\href@noop {}
  {\bibfield  {journal} {\bibinfo  {journal} {Nature}\ }\textbf {\bibinfo
  {volume} {524}},\ \bibinfo {pages} {325} (\bibinfo {year}
  {2015})}\BibitemShut {NoStop}%
\bibitem [{\citenamefont {S{\"{u}}sstrunk}\ and\ \citenamefont
  {Huber}(2015)}]{Susstrunk2015}%
  \BibitemOpen
  \bibfield  {author} {\bibinfo {author} {\bibfnamefont {R.}~\bibnamefont
  {S{\"{u}}sstrunk}}\ and\ \bibinfo {author} {\bibfnamefont {S.~D.}\
  \bibnamefont {Huber}},\ }\href {\doibase 10.1126/science.aab0239} {\bibfield
  {journal} {\bibinfo  {journal} {Science}\ }\textbf {\bibinfo {volume}
  {349}},\ \bibinfo {pages} {47} (\bibinfo {year} {2015})}\BibitemShut
  {NoStop}%
\bibitem [{\citenamefont {Nash}\ \emph {et~al.}(2015)\citenamefont {Nash},
  \citenamefont {Kleckner}, \citenamefont {Read}, \citenamefont {Vitelli},
  \citenamefont {Turner},\ and\ \citenamefont {Irvine}}]{nash2015topological}%
  \BibitemOpen
  \bibfield  {author} {\bibinfo {author} {\bibfnamefont {L.~M.}\ \bibnamefont
  {Nash}}, \bibinfo {author} {\bibfnamefont {D.}~\bibnamefont {Kleckner}},
  \bibinfo {author} {\bibfnamefont {A.}~\bibnamefont {Read}}, \bibinfo {author}
  {\bibfnamefont {V.}~\bibnamefont {Vitelli}}, \bibinfo {author} {\bibfnamefont
  {A.~M.}\ \bibnamefont {Turner}}, \ and\ \bibinfo {author} {\bibfnamefont
  {W.~T.}\ \bibnamefont {Irvine}},\ }\href@noop {} {\bibfield  {journal}
  {\bibinfo  {journal} {Proceedings of the National Academy of Sciences}\
  }\textbf {\bibinfo {volume} {112}},\ \bibinfo {pages} {14495} (\bibinfo
  {year} {2015})}\BibitemShut {NoStop}%
\bibitem [{\citenamefont {Lu}\ \emph {et~al.}(2017)\citenamefont {Lu},
  \citenamefont {Qiu}, \citenamefont {Ye}, \citenamefont {Fan}, \citenamefont
  {Ke}, \citenamefont {Zhang},\ and\ \citenamefont {Liu}}]{lu2017observation}%
  \BibitemOpen
  \bibfield  {author} {\bibinfo {author} {\bibfnamefont {J.}~\bibnamefont
  {Lu}}, \bibinfo {author} {\bibfnamefont {C.}~\bibnamefont {Qiu}}, \bibinfo
  {author} {\bibfnamefont {L.}~\bibnamefont {Ye}}, \bibinfo {author}
  {\bibfnamefont {X.}~\bibnamefont {Fan}}, \bibinfo {author} {\bibfnamefont
  {M.}~\bibnamefont {Ke}}, \bibinfo {author} {\bibfnamefont {F.}~\bibnamefont
  {Zhang}}, \ and\ \bibinfo {author} {\bibfnamefont {Z.}~\bibnamefont {Liu}},\
  }\href@noop {} {\bibfield  {journal} {\bibinfo  {journal} {Nature Physics}\
  }\textbf {\bibinfo {volume} {13}},\ \bibinfo {pages} {369} (\bibinfo {year}
  {2017})}\BibitemShut {NoStop}%
\bibitem [{\citenamefont {Miniaci}\ \emph {et~al.}(2018)\citenamefont
  {Miniaci}, \citenamefont {Pal}, \citenamefont {Morvan},\ and\ \citenamefont
  {Ruzzene}}]{Miniaci_PRX_2018}%
  \BibitemOpen
  \bibfield  {author} {\bibinfo {author} {\bibfnamefont {M.}~\bibnamefont
  {Miniaci}}, \bibinfo {author} {\bibfnamefont {R.~K.}\ \bibnamefont {Pal}},
  \bibinfo {author} {\bibfnamefont {B.}~\bibnamefont {Morvan}}, \ and\ \bibinfo
  {author} {\bibfnamefont {M.}~\bibnamefont {Ruzzene}},\ }\href {\doibase
  10.1103/PhysRevX.8.031074} {\bibfield  {journal} {\bibinfo  {journal} {Phys.
  Rev. X}\ }\textbf {\bibinfo {volume} {8}},\ \bibinfo {pages} {031074}
  (\bibinfo {year} {2018})}\BibitemShut {NoStop}%
\bibitem [{\citenamefont {Yu}\ \emph {et~al.}(2018)\citenamefont {Yu},
  \citenamefont {He}, \citenamefont {Wang}, \citenamefont {Liu}, \citenamefont
  {Sun}, \citenamefont {Li}, \citenamefont {Lu}, \citenamefont {Lu},
  \citenamefont {Liu},\ and\ \citenamefont {Chen}}]{yu_elastic_2018}%
  \BibitemOpen
  \bibfield  {author} {\bibinfo {author} {\bibfnamefont {S.-Y.}\ \bibnamefont
  {Yu}}, \bibinfo {author} {\bibfnamefont {C.}~\bibnamefont {He}}, \bibinfo
  {author} {\bibfnamefont {Z.}~\bibnamefont {Wang}}, \bibinfo {author}
  {\bibfnamefont {F.-K.}\ \bibnamefont {Liu}}, \bibinfo {author} {\bibfnamefont
  {X.-C.}\ \bibnamefont {Sun}}, \bibinfo {author} {\bibfnamefont
  {Z.}~\bibnamefont {Li}}, \bibinfo {author} {\bibfnamefont {H.-Z.}\
  \bibnamefont {Lu}}, \bibinfo {author} {\bibfnamefont {M.-H.}\ \bibnamefont
  {Lu}}, \bibinfo {author} {\bibfnamefont {X.-P.}\ \bibnamefont {Liu}}, \ and\
  \bibinfo {author} {\bibfnamefont {Y.-F.}\ \bibnamefont {Chen}},\ }\href
  {\doibase 10.1038/s41467-018-05461-5} {\bibfield  {journal} {\bibinfo
  {journal} {Nature Communications}\ }\textbf {\bibinfo {volume} {9}},\
  \bibinfo {pages} {3072} (\bibinfo {year} {2018})}\BibitemShut {NoStop}%
\bibitem [{\citenamefont {Hafezi}\ \emph {et~al.}(2011)\citenamefont {Hafezi},
  \citenamefont {Demler}, \citenamefont {Lukin},\ and\ \citenamefont
  {Taylor}}]{hafezi2011robust}%
  \BibitemOpen
  \bibfield  {author} {\bibinfo {author} {\bibfnamefont {M.}~\bibnamefont
  {Hafezi}}, \bibinfo {author} {\bibfnamefont {E.~A.}\ \bibnamefont {Demler}},
  \bibinfo {author} {\bibfnamefont {M.~D.}\ \bibnamefont {Lukin}}, \ and\
  \bibinfo {author} {\bibfnamefont {J.~M.}\ \bibnamefont {Taylor}},\
  }\href@noop {} {\bibfield  {journal} {\bibinfo  {journal} {Nature Physics}\
  }\textbf {\bibinfo {volume} {7}},\ \bibinfo {pages} {907} (\bibinfo {year}
  {2011})}\BibitemShut {NoStop}%
\bibitem [{\citenamefont {Cha}\ \emph {et~al.}(2018)\citenamefont {Cha},
  \citenamefont {Kim},\ and\ \citenamefont {Daraio}}]{cha2018experimental}%
  \BibitemOpen
  \bibfield  {author} {\bibinfo {author} {\bibfnamefont {J.}~\bibnamefont
  {Cha}}, \bibinfo {author} {\bibfnamefont {K.~W.}\ \bibnamefont {Kim}}, \ and\
  \bibinfo {author} {\bibfnamefont {C.}~\bibnamefont {Daraio}},\ }\href@noop {}
  {\bibfield  {journal} {\bibinfo  {journal} {Nature}\ }\textbf {\bibinfo
  {volume} {564}},\ \bibinfo {pages} {229} (\bibinfo {year}
  {2018})}\BibitemShut {NoStop}%
\bibitem [{\citenamefont {Ma}\ \emph {et~al.}(2020)\citenamefont {Ma},
  \citenamefont {Xi}, \citenamefont {Li},\ and\ \citenamefont
  {Sun}}]{ma2020nanomechanical}%
  \BibitemOpen
  \bibfield  {author} {\bibinfo {author} {\bibfnamefont {J.}~\bibnamefont
  {Ma}}, \bibinfo {author} {\bibfnamefont {X.}~\bibnamefont {Xi}}, \bibinfo
  {author} {\bibfnamefont {Y.}~\bibnamefont {Li}}, \ and\ \bibinfo {author}
  {\bibfnamefont {X.}~\bibnamefont {Sun}},\ }\href@noop {} {\bibfield
  {journal} {\bibinfo  {journal} {arXiv preprint arXiv:2004.03067}\ } (\bibinfo
  {year} {2020})}\BibitemShut {NoStop}%
\bibitem [{\citenamefont {Nassar}\ \emph {et~al.}(2020)\citenamefont {Nassar},
  \citenamefont {Yousefzadeh}, \citenamefont {Fleury}, \citenamefont {Ruzzene},
  \citenamefont {Al{\`u}}, \citenamefont {Daraio}, \citenamefont {Norris},
  \citenamefont {Huang},\ and\ \citenamefont
  {Haberman}}]{nassar2020nonreciprocity}%
  \BibitemOpen
  \bibfield  {author} {\bibinfo {author} {\bibfnamefont {H.}~\bibnamefont
  {Nassar}}, \bibinfo {author} {\bibfnamefont {B.}~\bibnamefont {Yousefzadeh}},
  \bibinfo {author} {\bibfnamefont {R.}~\bibnamefont {Fleury}}, \bibinfo
  {author} {\bibfnamefont {M.}~\bibnamefont {Ruzzene}}, \bibinfo {author}
  {\bibfnamefont {A.}~\bibnamefont {Al{\`u}}}, \bibinfo {author} {\bibfnamefont
  {C.}~\bibnamefont {Daraio}}, \bibinfo {author} {\bibfnamefont {A.~N.}\
  \bibnamefont {Norris}}, \bibinfo {author} {\bibfnamefont {G.}~\bibnamefont
  {Huang}}, \ and\ \bibinfo {author} {\bibfnamefont {M.~R.}\ \bibnamefont
  {Haberman}},\ }\href@noop {} {\bibfield  {journal} {\bibinfo  {journal}
  {Nature Reviews Materials}\ ,\ \bibinfo {pages} {1}} (\bibinfo {year}
  {2020})}\BibitemShut {NoStop}%
\bibitem [{\citenamefont {Hasan}\ and\ \citenamefont
  {Kane}(2010)}]{hasan2010colloquium}%
  \BibitemOpen
  \bibfield  {author} {\bibinfo {author} {\bibfnamefont {M.~Z.}\ \bibnamefont
  {Hasan}}\ and\ \bibinfo {author} {\bibfnamefont {C.~L.}\ \bibnamefont
  {Kane}},\ }\href@noop {} {\bibfield  {journal} {\bibinfo  {journal} {Reviews
  of modern physics}\ }\textbf {\bibinfo {volume} {82}},\ \bibinfo {pages}
  {3045} (\bibinfo {year} {2010})}\BibitemShut {NoStop}%
\bibitem [{\citenamefont {Aidelsburger}\ \emph {et~al.}(2018)\citenamefont
  {Aidelsburger}, \citenamefont {Nascimbene},\ and\ \citenamefont
  {Goldman}}]{aidelsburger_artificial_2018}%
  \BibitemOpen
  \bibfield  {author} {\bibinfo {author} {\bibfnamefont {M.}~\bibnamefont
  {Aidelsburger}}, \bibinfo {author} {\bibfnamefont {S.}~\bibnamefont
  {Nascimbene}}, \ and\ \bibinfo {author} {\bibfnamefont {N.}~\bibnamefont
  {Goldman}},\ }\href {\doibase https://doi.org/10.1016/j.crhy.2018.03.002}
  {\bibfield  {journal} {\bibinfo  {journal} {Comptes Rendus Physique}\
  }\textbf {\bibinfo {volume} {19}},\ \bibinfo {pages} {394 } (\bibinfo {year}
  {2018})}\BibitemShut {NoStop}%
\bibitem [{\citenamefont {Mousavi}\ \emph {et~al.}(2015)\citenamefont
  {Mousavi}, \citenamefont {Khanikaev},\ and\ \citenamefont
  {Wang}}]{mousavi_topologically_2015}%
  \BibitemOpen
  \bibfield  {author} {\bibinfo {author} {\bibfnamefont {S.~H.}\ \bibnamefont
  {Mousavi}}, \bibinfo {author} {\bibfnamefont {A.~B.}\ \bibnamefont
  {Khanikaev}}, \ and\ \bibinfo {author} {\bibfnamefont {Z.}~\bibnamefont
  {Wang}},\ }\href {\doibase 10.1038/ncomms9682} {\bibfield  {journal}
  {\bibinfo  {journal} {Nature Communications}\ }\textbf {\bibinfo {volume}
  {6}},\ \bibinfo {pages} {8682} (\bibinfo {year} {2015})}\BibitemShut
  {NoStop}%
\bibitem [{\citenamefont {Deng}\ \emph {et~al.}(2020)\citenamefont {Deng},
  \citenamefont {Huang}, \citenamefont {Lu}, \citenamefont {Peri},
  \citenamefont {Li}, \citenamefont {Huber},\ and\ \citenamefont
  {Liu}}]{deng2020acoustic}%
  \BibitemOpen
  \bibfield  {author} {\bibinfo {author} {\bibfnamefont {W.}~\bibnamefont
  {Deng}}, \bibinfo {author} {\bibfnamefont {X.}~\bibnamefont {Huang}},
  \bibinfo {author} {\bibfnamefont {J.}~\bibnamefont {Lu}}, \bibinfo {author}
  {\bibfnamefont {V.}~\bibnamefont {Peri}}, \bibinfo {author} {\bibfnamefont
  {F.}~\bibnamefont {Li}}, \bibinfo {author} {\bibfnamefont {S.~D.}\
  \bibnamefont {Huber}}, \ and\ \bibinfo {author} {\bibfnamefont
  {Z.}~\bibnamefont {Liu}},\ }\href@noop {} {\bibfield  {journal} {\bibinfo
  {journal} {Nature communications}\ }\textbf {\bibinfo {volume} {11}},\
  \bibinfo {pages} {1} (\bibinfo {year} {2020})}\BibitemShut {NoStop}%
\bibitem [{\citenamefont {Eichenfield}\ \emph {et~al.}(2009)\citenamefont
  {Eichenfield}, \citenamefont {Chan}, \citenamefont {Camacho}, \citenamefont
  {Vahala},\ and\ \citenamefont {Painter}}]{Eichenfield2009}%
  \BibitemOpen
  \bibfield  {author} {\bibinfo {author} {\bibfnamefont {M.}~\bibnamefont
  {Eichenfield}}, \bibinfo {author} {\bibfnamefont {J.}~\bibnamefont {Chan}},
  \bibinfo {author} {\bibfnamefont {R.~M.}\ \bibnamefont {Camacho}}, \bibinfo
  {author} {\bibfnamefont {K.~J.}\ \bibnamefont {Vahala}}, \ and\ \bibinfo
  {author} {\bibfnamefont {O.}~\bibnamefont {Painter}},\ }\href {\doibase
  10.1038/nature08524} {\bibfield  {journal} {\bibinfo  {journal} {Nature}\
  }\textbf {\bibinfo {volume} {462}},\ \bibinfo {pages} {78} (\bibinfo {year}
  {2009})}\BibitemShut {NoStop}%
\bibitem [{\citenamefont {Safavi-Naeini}\ and\ \citenamefont
  {Painter}(2010)}]{Safavi-Naeini2010b}%
  \BibitemOpen
  \bibfield  {author} {\bibinfo {author} {\bibfnamefont {A.~H.}\ \bibnamefont
  {Safavi-Naeini}}\ and\ \bibinfo {author} {\bibfnamefont {O.}~\bibnamefont
  {Painter}},\ }\href@noop {} {\bibfield  {journal} {\bibinfo  {journal} {Opt.
  Express}\ }\textbf {\bibinfo {volume} {18}},\ \bibinfo {pages} {14926}
  (\bibinfo {year} {2010})}\BibitemShut {NoStop}%
\bibitem [{\citenamefont {Safavi-Naeini}\ \emph {et~al.}(2014)\citenamefont
  {Safavi-Naeini}, \citenamefont {Hill}, \citenamefont {Meenehan},
  \citenamefont {Chan}, \citenamefont {Gr\"oblacher},\ and\ \citenamefont
  {Painter}}]{Safavi-Naeini2014PRL}%
  \BibitemOpen
  \bibfield  {author} {\bibinfo {author} {\bibfnamefont {A.~H.}\ \bibnamefont
  {Safavi-Naeini}}, \bibinfo {author} {\bibfnamefont {J.~T.}\ \bibnamefont
  {Hill}}, \bibinfo {author} {\bibfnamefont {S.}~\bibnamefont {Meenehan}},
  \bibinfo {author} {\bibfnamefont {J.}~\bibnamefont {Chan}}, \bibinfo {author}
  {\bibfnamefont {S.}~\bibnamefont {Gr\"oblacher}}, \ and\ \bibinfo {author}
  {\bibfnamefont {O.}~\bibnamefont {Painter}},\ }\href {\doibase
  10.1103/PhysRevLett.112.153603} {\bibfield  {journal} {\bibinfo  {journal}
  {Phys. Rev. Lett.}\ }\textbf {\bibinfo {volume} {112}},\ \bibinfo {pages}
  {153603} (\bibinfo {year} {2014})}\BibitemShut {NoStop}%
\bibitem [{\citenamefont {Ren}\ \emph {et~al.}(2020)\citenamefont {Ren},
  \citenamefont {Matheny}, \citenamefont {MacCabe}, \citenamefont {Luo},
  \citenamefont {Pfeifer}, \citenamefont {Mirhosseini},\ and\ \citenamefont
  {Painter}}]{ren2020two}%
  \BibitemOpen
  \bibfield  {author} {\bibinfo {author} {\bibfnamefont {H.}~\bibnamefont
  {Ren}}, \bibinfo {author} {\bibfnamefont {M.~H.}\ \bibnamefont {Matheny}},
  \bibinfo {author} {\bibfnamefont {G.~S.}\ \bibnamefont {MacCabe}}, \bibinfo
  {author} {\bibfnamefont {J.}~\bibnamefont {Luo}}, \bibinfo {author}
  {\bibfnamefont {H.}~\bibnamefont {Pfeifer}}, \bibinfo {author} {\bibfnamefont
  {M.}~\bibnamefont {Mirhosseini}}, \ and\ \bibinfo {author} {\bibfnamefont
  {O.}~\bibnamefont {Painter}},\ }\href@noop {} {\bibfield  {journal} {\bibinfo
   {journal} {Nature communications}\ }\textbf {\bibinfo {volume} {11}},\
  \bibinfo {pages} {1} (\bibinfo {year} {2020})}\BibitemShut {NoStop}%
\bibitem [{\citenamefont {Martin}\ \emph {et~al.}(2008)\citenamefont {Martin},
  \citenamefont {Blanter},\ and\ \citenamefont {Morpurgo}}]{Martin_2008_PRL}%
  \BibitemOpen
  \bibfield  {author} {\bibinfo {author} {\bibfnamefont {I.}~\bibnamefont
  {Martin}}, \bibinfo {author} {\bibfnamefont {Y.~M.}\ \bibnamefont {Blanter}},
  \ and\ \bibinfo {author} {\bibfnamefont {A.~F.}\ \bibnamefont {Morpurgo}},\
  }\href@noop {} {\bibfield  {journal} {\bibinfo  {journal} {Phys. Rev. Lett.}\
  }\textbf {\bibinfo {volume} {100}},\ \bibinfo {pages} {036804} (\bibinfo
  {year} {2008})}\BibitemShut {NoStop}%
\bibitem [{\citenamefont {Ju}\ \emph {et~al.}(2015)\citenamefont {Ju},
  \citenamefont {Shi}, \citenamefont {Nair}, \citenamefont {Lv}, \citenamefont
  {Jin}, \citenamefont {Velasco}, \citenamefont {Ojeda-Aristizabal},
  \citenamefont {Bechtel}, \citenamefont {Martin}, \citenamefont {Zettl},
  \citenamefont {Analytis},\ and\ \citenamefont {Wang}}]{ju_topological_2015}%
  \BibitemOpen
  \bibfield  {author} {\bibinfo {author} {\bibfnamefont {L.}~\bibnamefont
  {Ju}}, \bibinfo {author} {\bibfnamefont {Z.}~\bibnamefont {Shi}}, \bibinfo
  {author} {\bibfnamefont {N.}~\bibnamefont {Nair}}, \bibinfo {author}
  {\bibfnamefont {Y.}~\bibnamefont {Lv}}, \bibinfo {author} {\bibfnamefont
  {C.}~\bibnamefont {Jin}}, \bibinfo {author} {\bibfnamefont {J.}~\bibnamefont
  {Velasco}}, \bibinfo {author} {\bibfnamefont {C.}~\bibnamefont
  {Ojeda-Aristizabal}}, \bibinfo {author} {\bibfnamefont {H.~A.}\ \bibnamefont
  {Bechtel}}, \bibinfo {author} {\bibfnamefont {M.~C.}\ \bibnamefont {Martin}},
  \bibinfo {author} {\bibfnamefont {A.}~\bibnamefont {Zettl}}, \bibinfo
  {author} {\bibfnamefont {J.}~\bibnamefont {Analytis}}, \ and\ \bibinfo
  {author} {\bibfnamefont {F.}~\bibnamefont {Wang}},\ }\href {\doibase
  10.1038/nature14364} {\bibfield  {journal} {\bibinfo  {journal} {Nature}\
  }\textbf {\bibinfo {volume} {520}},\ \bibinfo {pages} {650} (\bibinfo {year}
  {2015})}\BibitemShut {NoStop}%
\bibitem [{\citenamefont {Zeng}\ \emph {et~al.}(2020)\citenamefont {Zeng},
  \citenamefont {Chattopadhyay}, \citenamefont {Zhu}, \citenamefont {Qiang},
  \citenamefont {Li}, \citenamefont {Jin}, \citenamefont {Li}, \citenamefont
  {Davies}, \citenamefont {Linfield}, \citenamefont {Zhang}, \citenamefont
  {Chong},\ and\ \citenamefont {Wang}}]{zeng_electrically_2020}%
  \BibitemOpen
  \bibfield  {author} {\bibinfo {author} {\bibfnamefont {Y.}~\bibnamefont
  {Zeng}}, \bibinfo {author} {\bibfnamefont {U.}~\bibnamefont {Chattopadhyay}},
  \bibinfo {author} {\bibfnamefont {B.}~\bibnamefont {Zhu}}, \bibinfo {author}
  {\bibfnamefont {B.}~\bibnamefont {Qiang}}, \bibinfo {author} {\bibfnamefont
  {J.}~\bibnamefont {Li}}, \bibinfo {author} {\bibfnamefont {Y.}~\bibnamefont
  {Jin}}, \bibinfo {author} {\bibfnamefont {L.}~\bibnamefont {Li}}, \bibinfo
  {author} {\bibfnamefont {A.~G.}\ \bibnamefont {Davies}}, \bibinfo {author}
  {\bibfnamefont {E.~H.}\ \bibnamefont {Linfield}}, \bibinfo {author}
  {\bibfnamefont {B.}~\bibnamefont {Zhang}}, \bibinfo {author} {\bibfnamefont
  {Y.}~\bibnamefont {Chong}}, \ and\ \bibinfo {author} {\bibfnamefont {Q.~J.}\
  \bibnamefont {Wang}},\ }\href {\doibase 10.1038/s41586-020-1981-x} {\bibfield
   {journal} {\bibinfo  {journal} {Nature}\ }\textbf {\bibinfo {volume}
  {578}},\ \bibinfo {pages} {246} (\bibinfo {year} {2020})}\BibitemShut
  {NoStop}%
\bibitem [{\citenamefont {Habraken}\ \emph {et~al.}(2012)\citenamefont
  {Habraken}, \citenamefont {Stannigel}, \citenamefont {Lukin}, \citenamefont
  {Zoller},\ and\ \citenamefont {Rabl}}]{Habraken2012}%
  \BibitemOpen
  \bibfield  {author} {\bibinfo {author} {\bibfnamefont {S.~J.~M.}\
  \bibnamefont {Habraken}}, \bibinfo {author} {\bibfnamefont {K.}~\bibnamefont
  {Stannigel}}, \bibinfo {author} {\bibfnamefont {M.~D.}\ \bibnamefont
  {Lukin}}, \bibinfo {author} {\bibfnamefont {P.}~\bibnamefont {Zoller}}, \
  and\ \bibinfo {author} {\bibfnamefont {P.}~\bibnamefont {Rabl}},\ }\href@noop
  {} {\bibfield  {journal} {\bibinfo  {journal} {New J.\ Phys.}\ ,\ \bibinfo
  {pages} {115004}} (\bibinfo {year} {2012})}\BibitemShut {NoStop}%
\bibitem [{\citenamefont {Peano}\ \emph {et~al.}(2016)\citenamefont {Peano},
  \citenamefont {Houde}, \citenamefont {Marquardt},\ and\ \citenamefont
  {Clerk}}]{peano2016topological}%
  \BibitemOpen
  \bibfield  {author} {\bibinfo {author} {\bibfnamefont {V.}~\bibnamefont
  {Peano}}, \bibinfo {author} {\bibfnamefont {M.}~\bibnamefont {Houde}},
  \bibinfo {author} {\bibfnamefont {F.}~\bibnamefont {Marquardt}}, \ and\
  \bibinfo {author} {\bibfnamefont {A.~A.}\ \bibnamefont {Clerk}},\ }\href
  {\doibase 10.1103/PhysRevX.6.041026} {\bibfield  {journal} {\bibinfo
  {journal} {Phys. Rev. X}\ }\textbf {\bibinfo {volume} {6}},\ \bibinfo {pages}
  {041026} (\bibinfo {year} {2016})}\BibitemShut {NoStop}%
\bibitem [{\citenamefont {Mittal}\ \emph {et~al.}(2018)\citenamefont {Mittal},
  \citenamefont {Goldschmidt},\ and\ \citenamefont
  {Hafezi}}]{mittal2018topological}%
  \BibitemOpen
  \bibfield  {author} {\bibinfo {author} {\bibfnamefont {S.}~\bibnamefont
  {Mittal}}, \bibinfo {author} {\bibfnamefont {E.~A.}\ \bibnamefont
  {Goldschmidt}}, \ and\ \bibinfo {author} {\bibfnamefont {M.}~\bibnamefont
  {Hafezi}},\ }\href@noop {} {\bibfield  {journal} {\bibinfo  {journal}
  {Nature}\ }\textbf {\bibinfo {volume} {561}},\ \bibinfo {pages} {502}
  (\bibinfo {year} {2018})}\BibitemShut {NoStop}%
\bibitem [{\citenamefont {Bandres}\ \emph {et~al.}(2018)\citenamefont
  {Bandres}, \citenamefont {Wittek}, \citenamefont {Harari}, \citenamefont
  {Parto}, \citenamefont {Ren}, \citenamefont {Segev}, \citenamefont
  {Christodoulides},\ and\ \citenamefont
  {Khajavikhan}}]{bandres2018topological}%
  \BibitemOpen
  \bibfield  {author} {\bibinfo {author} {\bibfnamefont {M.~A.}\ \bibnamefont
  {Bandres}}, \bibinfo {author} {\bibfnamefont {S.}~\bibnamefont {Wittek}},
  \bibinfo {author} {\bibfnamefont {G.}~\bibnamefont {Harari}}, \bibinfo
  {author} {\bibfnamefont {M.}~\bibnamefont {Parto}}, \bibinfo {author}
  {\bibfnamefont {J.}~\bibnamefont {Ren}}, \bibinfo {author} {\bibfnamefont
  {M.}~\bibnamefont {Segev}}, \bibinfo {author} {\bibfnamefont {D.~N.}\
  \bibnamefont {Christodoulides}}, \ and\ \bibinfo {author} {\bibfnamefont
  {M.}~\bibnamefont {Khajavikhan}},\ }\href@noop {} {\bibfield  {journal}
  {\bibinfo  {journal} {Science}\ }\textbf {\bibinfo {volume} {359}},\ \bibinfo
  {pages} {eaar4005} (\bibinfo {year} {2018})}\BibitemShut {NoStop}%
\bibitem [{\citenamefont {Zhang}\ \emph {et~al.}(2013)\citenamefont {Zhang},
  \citenamefont {MacDonald},\ and\ \citenamefont {Mele}}]{zhang_valley_2013}%
  \BibitemOpen
  \bibfield  {author} {\bibinfo {author} {\bibfnamefont {F.}~\bibnamefont
  {Zhang}}, \bibinfo {author} {\bibfnamefont {A.~H.}\ \bibnamefont
  {MacDonald}}, \ and\ \bibinfo {author} {\bibfnamefont {E.~J.}\ \bibnamefont
  {Mele}},\ }\href {\doibase 10.1073/pnas.1308853110} {\bibfield  {journal}
  {\bibinfo  {journal} {Proceedings of the National Academy of Sciences}\
  }\textbf {\bibinfo {volume} {110}},\ \bibinfo {pages} {10546} (\bibinfo
  {year} {2013})}\BibitemShut {NoStop}%
\bibitem [{\citenamefont {Gao}\ \emph {et~al.}(2018)\citenamefont {Gao},
  \citenamefont {Xue}, \citenamefont {Yang}, \citenamefont {Lai}, \citenamefont
  {Yu}, \citenamefont {Lin}, \citenamefont {Chong}, \citenamefont {Shvets},\
  and\ \citenamefont {Zhang}}]{gao_topologically_2018}%
  \BibitemOpen
  \bibfield  {author} {\bibinfo {author} {\bibfnamefont {F.}~\bibnamefont
  {Gao}}, \bibinfo {author} {\bibfnamefont {H.}~\bibnamefont {Xue}}, \bibinfo
  {author} {\bibfnamefont {Z.}~\bibnamefont {Yang}}, \bibinfo {author}
  {\bibfnamefont {K.}~\bibnamefont {Lai}}, \bibinfo {author} {\bibfnamefont
  {Y.}~\bibnamefont {Yu}}, \bibinfo {author} {\bibfnamefont {X.}~\bibnamefont
  {Lin}}, \bibinfo {author} {\bibfnamefont {Y.}~\bibnamefont {Chong}}, \bibinfo
  {author} {\bibfnamefont {G.}~\bibnamefont {Shvets}}, \ and\ \bibinfo {author}
  {\bibfnamefont {B.}~\bibnamefont {Zhang}},\ }\href {\doibase
  10.1038/nphys4304} {\bibfield  {journal} {\bibinfo  {journal} {Nature
  Physics}\ }\textbf {\bibinfo {volume} {14}},\ \bibinfo {pages} {140}
  (\bibinfo {year} {2018})}\BibitemShut {NoStop}%
\bibitem [{\citenamefont {Schaibley}\ \emph {et~al.}(2016)\citenamefont
  {Schaibley}, \citenamefont {Yu}, \citenamefont {Clark}, \citenamefont
  {Rivera}, \citenamefont {Ross}, \citenamefont {Seyler}, \citenamefont {Yao},\
  and\ \citenamefont {Xu}}]{schaibley_valleytronics_2016}%
  \BibitemOpen
  \bibfield  {author} {\bibinfo {author} {\bibfnamefont {J.~R.}\ \bibnamefont
  {Schaibley}}, \bibinfo {author} {\bibfnamefont {H.}~\bibnamefont {Yu}},
  \bibinfo {author} {\bibfnamefont {G.}~\bibnamefont {Clark}}, \bibinfo
  {author} {\bibfnamefont {P.}~\bibnamefont {Rivera}}, \bibinfo {author}
  {\bibfnamefont {J.~S.}\ \bibnamefont {Ross}}, \bibinfo {author}
  {\bibfnamefont {K.~L.}\ \bibnamefont {Seyler}}, \bibinfo {author}
  {\bibfnamefont {W.}~\bibnamefont {Yao}}, \ and\ \bibinfo {author}
  {\bibfnamefont {X.}~\bibnamefont {Xu}},\ }\href {\doibase
  10.1038/natrevmats.2016.55} {\bibfield  {journal} {\bibinfo  {journal}
  {Nature Reviews Materials}\ }\textbf {\bibinfo {volume} {1}},\ \bibinfo
  {pages} {16055} (\bibinfo {year} {2016})}\BibitemShut {NoStop}%
\bibitem [{\citenamefont {Asb{\'o}th}\ \emph {et~al.}(2016)\citenamefont
  {Asb{\'o}th}, \citenamefont {Oroszl{\'a}ny},\ and\ \citenamefont
  {P{\'a}lyi}}]{asboth_short_2016}%
  \BibitemOpen
  \bibfield  {author} {\bibinfo {author} {\bibfnamefont {J.~K.}\ \bibnamefont
  {Asb{\'o}th}}, \bibinfo {author} {\bibfnamefont {L.}~\bibnamefont
  {Oroszl{\'a}ny}}, \ and\ \bibinfo {author} {\bibfnamefont {A.}~\bibnamefont
  {P{\'a}lyi}},\ }\href {\doibase 10.1007/978-3-319-25607-8} {\emph {\bibinfo
  {title} {A Short Course on Topological Insulators: Band Structure and Edge
  States in One and Two Dimensions}}},\ Lecture Notes in Physics\ (\bibinfo
  {publisher} {Springer International Publishing},\ \bibinfo {year}
  {2016})\BibitemShut {NoStop}%
\bibitem [{COM()}]{COMSOL}%
  \BibitemOpen
  \href@noop {} {}\bibinfo {note} {COMSOL Multiphysics 5.3a,
  http://www.comsol.com/}\BibitemShut {NoStop}%
\bibitem [{\citenamefont {Sekoguchi}\ \emph {et~al.}(2014)\citenamefont
  {Sekoguchi}, \citenamefont {Takahashi}, \citenamefont {Asano},\ and\
  \citenamefont {Noda}}]{Sekoguchi14}%
  \BibitemOpen
  \bibfield  {author} {\bibinfo {author} {\bibfnamefont {H.}~\bibnamefont
  {Sekoguchi}}, \bibinfo {author} {\bibfnamefont {Y.}~\bibnamefont
  {Takahashi}}, \bibinfo {author} {\bibfnamefont {T.}~\bibnamefont {Asano}}, \
  and\ \bibinfo {author} {\bibfnamefont {S.}~\bibnamefont {Noda}},\ }\href@noop
  {} {\bibfield  {journal} {\bibinfo  {journal} {Opt. Express}\ }\textbf
  {\bibinfo {volume} {22}},\ \bibinfo {pages} {916} (\bibinfo {year}
  {2014})}\BibitemShut {NoStop}%
\bibitem [{\citenamefont {Johnson}\ \emph {et~al.}(2000)\citenamefont
  {Johnson}, \citenamefont {Villeneuve}, \citenamefont {Fan},\ and\
  \citenamefont {Joannopoulos}}]{johnson2000linear}%
  \BibitemOpen
  \bibfield  {author} {\bibinfo {author} {\bibfnamefont {S.~G.}\ \bibnamefont
  {Johnson}}, \bibinfo {author} {\bibfnamefont {P.~R.}\ \bibnamefont
  {Villeneuve}}, \bibinfo {author} {\bibfnamefont {S.}~\bibnamefont {Fan}}, \
  and\ \bibinfo {author} {\bibfnamefont {J.~D.}\ \bibnamefont {Joannopoulos}},\
  }\href@noop {} {\bibfield  {journal} {\bibinfo  {journal} {Physical Review
  B}\ }\textbf {\bibinfo {volume} {62}},\ \bibinfo {pages} {8212} (\bibinfo
  {year} {2000})}\BibitemShut {NoStop}%
\bibitem [{\citenamefont {Chutinan}\ and\ \citenamefont
  {Noda}(2000)}]{chutinan2000waveguides}%
  \BibitemOpen
  \bibfield  {author} {\bibinfo {author} {\bibfnamefont {A.}~\bibnamefont
  {Chutinan}}\ and\ \bibinfo {author} {\bibfnamefont {S.}~\bibnamefont
  {Noda}},\ }\href@noop {} {\bibfield  {journal} {\bibinfo  {journal} {Physical
  review B}\ }\textbf {\bibinfo {volume} {62}},\ \bibinfo {pages} {4488}
  (\bibinfo {year} {2000})}\BibitemShut {NoStop}%
\bibitem [{\citenamefont {Huang}\ \emph {et~al.}(2016)\citenamefont {Huang},
  \citenamefont {Zhou},\ and\ \citenamefont {Duan}}]{Huang_2016_type_II}%
  \BibitemOpen
  \bibfield  {author} {\bibinfo {author} {\bibfnamefont {H.}~\bibnamefont
  {Huang}}, \bibinfo {author} {\bibfnamefont {S.}~\bibnamefont {Zhou}}, \ and\
  \bibinfo {author} {\bibfnamefont {W.}~\bibnamefont {Duan}},\ }\href {\doibase
  10.1103/PhysRevB.94.121117} {\bibfield  {journal} {\bibinfo  {journal} {Phys.
  Rev. B}\ }\textbf {\bibinfo {volume} {94}},\ \bibinfo {pages} {121117}
  (\bibinfo {year} {2016})}\BibitemShut {NoStop}%
\bibitem [{\citenamefont {{Hopcroft}}\ \emph {et~al.}(2010)\citenamefont
  {{Hopcroft}}, \citenamefont {{Nix}},\ and\ \citenamefont
  {{Kenny}}}]{hopcroft_2010}%
  \BibitemOpen
  \bibfield  {author} {\bibinfo {author} {\bibfnamefont {M.~A.}\ \bibnamefont
  {{Hopcroft}}}, \bibinfo {author} {\bibfnamefont {W.~D.}\ \bibnamefont
  {{Nix}}}, \ and\ \bibinfo {author} {\bibfnamefont {T.~W.}\ \bibnamefont
  {{Kenny}}},\ }\href@noop {} {\bibfield  {journal} {\bibinfo  {journal}
  {Journal of Microelectromechanical Systems}\ }\textbf {\bibinfo {volume}
  {19}},\ \bibinfo {pages} {229} (\bibinfo {year} {2010})}\BibitemShut
  {NoStop}%
\bibitem [{\citenamefont {Safavi-Naeini}\ and\ \citenamefont
  {Painter}(2014)}]{Safavi-Naeini2014}%
  \BibitemOpen
  \bibfield  {author} {\bibinfo {author} {\bibfnamefont {A.~H.}\ \bibnamefont
  {Safavi-Naeini}}\ and\ \bibinfo {author} {\bibfnamefont {O.}~\bibnamefont
  {Painter}},\ }in\ \href {\doibase 10.1007/978-3-642-55312-7_10} {\emph
  {\bibinfo {booktitle} {Cavity Optomechanics}}},\ \bibinfo {series and number}
  {Quantum Science and Technology},\ \bibinfo {editor} {edited by\ \bibinfo
  {editor} {\bibfnamefont {M.}~\bibnamefont {Aspelmeyer}}, \bibinfo {editor}
  {\bibfnamefont {T.~J.}\ \bibnamefont {Kippenberg}}, \ and\ \bibinfo {editor}
  {\bibfnamefont {F.}~\bibnamefont {Marquardt}}}\ (\bibinfo  {publisher}
  {Springer Berlin Heidelberg},\ \bibinfo {year} {2014})\ pp.\ \bibinfo {pages}
  {195--231}\BibitemShut {NoStop}%
\end{thebibliography}

%%%%%%%%%%%%%%%%%%%%%%%%%%%%%%%%%%%%%%%%%%%%%%%%%%%%%%%%%%%%%%%%%%%%%%%%%%%%%%%%%%
\end{document}